République Tunisienne
Ministère de l'Enseignement Supérieur
et de la Recherche Scientifique

Université de Sfax
Faculté des Sciences Economiques
et de Gestion
Département d'Informatique

**Mémoire en vue de l'obtention du diplôme en Mastère en Systèmes d'Information et Nouvelles Technologies**

*Présenté Par* : **Mouti HAMMAMI**

# Maintenance de l'outil Wr2fdr de traduction de Wright vers CSP

***Soutenu le 9 août 2011 devant le jury composé par :***

| | | |
|---|---|---|
| **Mr.** | Ahmed HAJ KACEM | Président |
| **Mr.** | Bilel GARGOURI | Membre |
| **Mr.** | Mohamed Tahar BHIRI | Directeur de recherche |

**Année universitaire : 2010-2011**

# Liste des figures

# *Introduction*

L'ADL formel Wright [**22**], [**23**] et [**24**] permet de décrire et de raisonner sur des architectures logicielles. En effet, Wright supporte des concepts structuraux tels que composant, connecteur et configuration permettant de décrire les aspects structuraux d'une architecture logicielle.

En outre, les aspects comportementaux d'une architecture logicielle sont spécifiés en utilisant CSP de Hoare [**3**]. De plus l'ADL Wright définit onze propriétés standards [**24**] relatives à la cohérence d'architectures logicielles décrites en Wright. Parmi ces propriétés, quatre propriétés sont automatisables par l'outil Wr2fdr [**26**] qui accompagne l'ADL Wright. Ces quatre propriétés sont liées à la cohérence d'un composant, à la cohérence d'un connecteur et à la compatibilité Port/Rôle.

L'outil Wr2fdr développé par l'université de Carnegie Mellon accepte en entrée une architecture logicielle décrite en Wright et produit en sortie une spécification acceptable par le Model-checker FDR [**29**]. Hormis les fonctionnalités lexico-syntaxique de Wright et la traduction de Wright vers CSP, l'outil Wr2fdr automatise les quatre propriétés citées ci-dessus en utilisant la technique de raffinement CSP. Suite à des expérimentations avec Wr2fdr, nous avons constaté des défaillances liées à l'implémentation de ces propriétés. Vu l'importance de cet outil, nous avons contacté les auteurs de Wright, expliqué les problèmes rencontrés, récupérer le code source écrit en C++ afin de le corriger et le faire évoluer. Dans un premier temps nous avons suivi une approche de test fonctionnel – Wr2fdr est assimilé à une boite noire – orientée tests syntaxiques afin de détecter les défaillances de l'outil Wr2fdr. Dans un deuxième temps, nous avons suivi une démarche de rétro-conception (ou encore rétro-ingénierie) afin d'extraire une vue générale de l'outil Wr2fdr : diagramme de classe en UML, méthodes essentielles, fonctionnement général de Wr2fdr et algorithmes et structures de données fondamentaux utilisés par Wr2fdr. Ceci nous a permis de **localiser** et **corriger** les défaillances identifiées précédemment et de **faire évoluer** l'outil Wr2fdr.

Ce mémoire comporte trois chapitres. Le premier chapitre présente les aspects structuraux et comportementaux de Wright, la sémantique formelle de Wright basée sur celle de CSP de Hoare et les propriétés standards de Wright en focalisant sur celles censées être automatisables par l'outil Wr2fdr. En outre, il propose une activité de vérification de l'outil Wr2fdr basée sur le test fonctionnel orienté tests syntaxiques.

Le second chapitre présente la démarche de rétro-conception que nous avons appliqué sur le code source de Wr2fdr et les artefacts produits notamment le diagramme de classes extrait.

Enfin, le troisième chapitre présente les modifications que nous avons apportées à l'outil Wr2fdr.

Une conclusion et des perspectives terminent ce mémoire. De plus, ce mémoire comporte des annexes décrivant des extraits des modifications apportées à l'outil Wr2fdr.

# Chapitre 1
# L'ADL formel Wright

Dans ce chapitre, nous allons présenter d’une façon détaillée l’ADL formel Wright et l’outil Wr2fdr qui l’accompagne. Dans un premier temps, nous abordons les aspects structuraux et comportementaux de l’ADL Wright. Dans un deuxième temps, nous étudions la sémantique formelle de Wright basée sur celle de CSP de Hoare. Dans un troisième temps, nous analysons les propriétés standards de Wright liées à la cohérence et complétude des architectures logicielles décrites en Wright. Dans un quatrième temps nous discutons les techniques de vérification de certaines propriétés standards de Wright basées sur la technique du raffinement de CSP. Enfin, nous présentons et évaluons l’outil Wr2fdr permettant d’automatiser certaines propriétés standards définies par Wright à savoir cohérence d’un Composant, cohérence d’un Connecteur et compatibilité Port/Rôle d’un assemblage de Composants Wright.

## 1. Les aspects structuraux

Wright [**22**], [**23**] et [**24**] repose sur quatre concepts qui sont le composant, le connecteur, la configuration et le style. Pour bien illustrer comment ces concepts peuvent être utilisés pour définir une architecture pour un système donné, nous allons analyser le système fourni ci-dessous inspiré de [**18**].

Nous voulons développer une application *Index* qui prenne en entrée une suite de caractères et renvoie un ensemble de couples (mot, numéro(s) de ligne) tel que :

- Un mot correspond à une séquence de lettres et/ou de chiffres. Ces mots sont délimités par des séparateurs. Dès qu'un mot a été trouvé, il est transcrit en minuscule.
- La numérotation des lignes commence à 1 et son incrémentation a lieu lorsque nous rencontrons le caractère spécial CR (retour chariot).
- Les mots identiques sont fusionnés et leurs numéros de ligne sont soit également fusionnés s'ils sont identiques, soit stockés par ordre croissant s'ils sont différents. Ainsi, à la sortie de notre index, il ne peut pas exister deux couples ayant le même mot.
- Les couples doivent être ordonnés par un tri alphabétique au niveau des mots.
- Illustrons notre *Index* par un petit exemple (voir Figure 1.1).

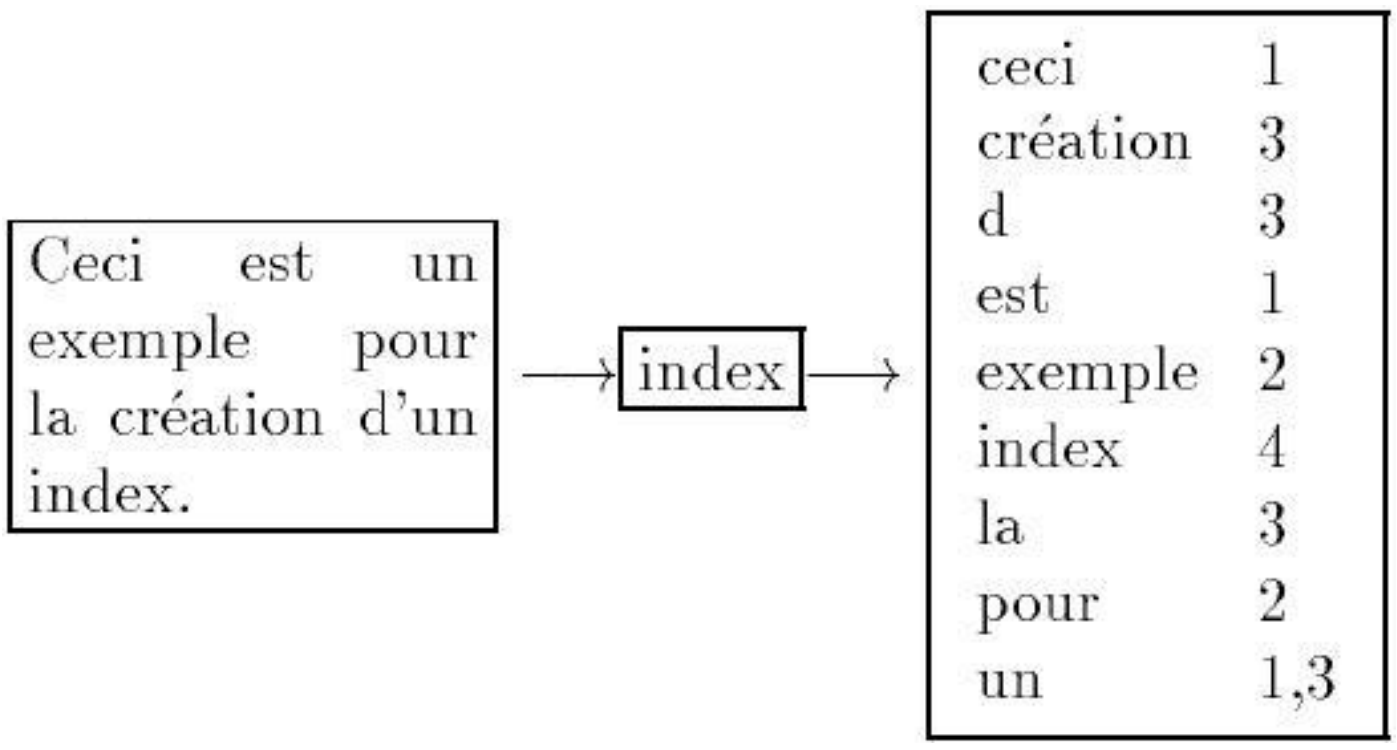

**Figure 1.1.** *Exemple de l'application Index*

Nous avons choisi de définir notre architecture en adaptant le style Filtre/Pipe représentée par trois composants : Filtre Texte, Filtre Tri et Filtre Fusion ; et par deux connecteurs de type pipe : Pipe1 et Pipe2 (voir Figure 1.2).

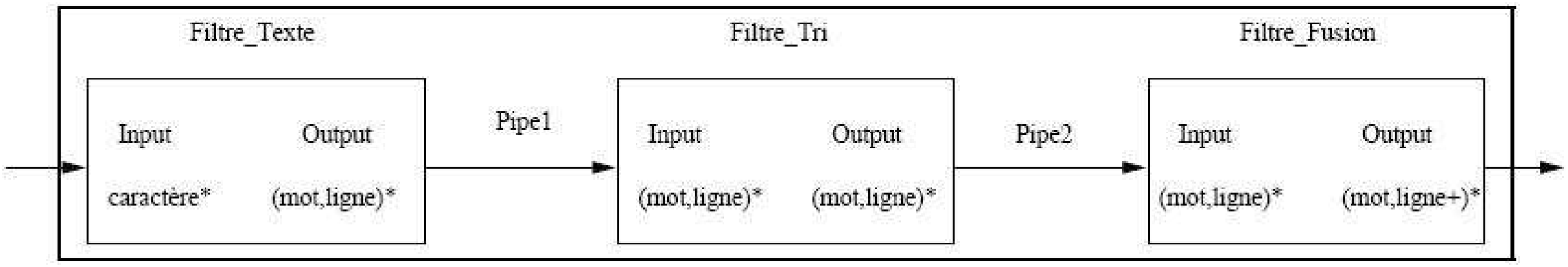

**Figure 1.2.** *Architecture de l'application Index*

Le composant Filtre_Texte prend en entrée une suite de caractères. Si le caractère qu'il lit est :

- Une lettre ou un chiffre alors il concatène ce caractère au mot qu'il est en train de construire puis il lit le caractère suivant.
- Le caractère spécial CR alors il incrémente le compteur de lignes puis lit le caractère suivant.
- Un séparateur alors si le caractère précédent n'était pas un séparateur ou si ce séparateur n'est pas le premier caractère, il transcrit le mot en minuscule et il envoie le couple (mot, numéro de ligne). Ensuite il lit le caractère suivant.

Lorsqu'il a terminé de lire la suite de caractères, il envoie le dernier couple s'il lui reste des données et puis il s'arrête.

En ce qui concerne le composant Filtre_Tri, il doit lire tous les couples (mot, numéro de ligne) que lui envoie le composant Filtre_Texte par l'intermédiaire du connecteur Pipe1. Ensuite il trie tous ces couples par ordre alphabétique pour les mots et par ordre croissant pour les numéros de ligne lorsque les mots sont identiques. Ensuite il envoie ces couples triés au connecteur Pipe2 par son port de sortie un à un.

Quant au composant Tri_Fusion, il reçoit ces couples (mot, numéro de ligne) triés en entrée par l'intermédiaire du connecteur Pipe2. Dès qu'il récupère un couple :

- Il l'oublie si ce couple est identique au précédent (cela évite les doublons).
- Il fusionne son numéro de ligne avec celui du couple précédent si les mots des deux couples sont identiques.
- Il envoie le couple (mot, numéro(s) de ligne) du couple précédent sur son port de sortie et stocke ce nouveau couple si les mots des deux couples sont différents.

### *1.1. Le concept composant*

Un composant en Wright est une unité abstraite localisée et indépendante. La description d'un composant contient deux parties importantes qui sont l'interface (interface) et la partie calcul (computation).

L'interface consiste en un ensemble de ports, chacun représente une interaction avec l'extérieur à laquelle le composant peut participer. La spécification d'un port décrit deux aspects de l'interface d'un composant :

- Elle décrit partiellement le comportement attendu d'un composant : le comportement du composant est vu par la « lorgnette » de ce port particulier.
- Elle décrit ce que le composant attend du système dans lequel il va interagir.

Le calcul décrit ce que le composant fait de point de vue comportemental. Il mène à bien les interactions décrites par les ports et montre comment elles se combinent pour former un tout cohérent. C'est sur cette spécification que l'analyse des propriétés du composant va être basée. Un composant peut avoir plusieurs ports (et donc de multiples interfaces). Les spécifications fournissent plus que des signatures statiques de l'interface. Elles indiquent aussi des interactions dynamiques. La figure 1.3 fournit les aspects structuraux du composant Filtre_Texte de l'application *Index*. Les aspects comportementaux de ce composant sont décrits d'une façon informelle.

**Component** Filtre_Texte
    **Port** Input = *Lire les données, s'arrêter au signal end-of-data*
    **Port** Output = *Envoyer les données, signaler la terminaison par close*
    **Computation** = *Lire les données du port Input, suivant les cas continuer à lire les données du port Input ou envoyer le couple (mot,numéro de ligne) sur le port Output*

**Figure 1.3.** *Composant Filtre_Texte*

### *1.2. Le concept connecteur*

Un connecteur représente une interaction entre une collection de composants. Il possède un type. Il spécifie le patron d'une interaction de manière explicite et abstraite. Ce patron peut être réutilisé dans différentes architectures.

Il contient deux parties importantes qui sont un ensemble de rôles et la glu :

- chaque rôle indique comment se comporte un composant qui participe à l'interaction.
- La glu décrit comment les rôles travaillent ensemble pour créer cette interaction.

Comme pour le calcul (au niveau du composant), la spécification de la glu (au niveau du connecteur) définit l'entière spécification du comportement. Elle coordonne le comportement des composants dans le cas où ceux-ci obéissent effectivement aux comportements indiqués par les rôles. La figure 1.4 fournit les aspects structuraux du connecteur Pipe de l'application *Index*. Les aspects comportementaux de ce connecteur sont décrits d'une façon informelle.

**Connector** Pipe
    **Role** Source = *Lire les données, s'arrêter au signal end-of-data*
    **Role** Sink = *Envoyer les données, signaler la terminaison par close*
    **Glue** = *Sink va recevoir les données dans l'ordre délivré par Source*

**Figure 1.4.** *Connecteur Pipe*

### *1.3. Le concept configuration*

La configuration permet de décrire l'architecture d'un système en regroupant des instances de composants et des instances de connecteurs. La description d'une configuration est composée de trois parties qui sont la déclaration des composants et des connecteurs utilisés dans l'architecture, la déclaration des instances de composants et de connecteurs, les descriptions des liens entre les instances de composants et de connecteurs. La figure 1.5 fournit la configuration de l'application *Index*.

```
Configuration Index
    Component Filtre_Texte
      ...
    Connecteur Pipe
      ...
    ...
    Instances
      Texte   :  Filtre_Texte
      Tri     :  Filtre_Tri
      Fusion  :  Filtre_Fusion
      P1,P2   :  Pipe
    Attachements
      Texte.Output   as   P1.Source
      Tri.Iutput     as   P1.Sink
      Tri.Output     as   P2.Source
      Fusion.Input   as   P2.Sink

End Configuration
```

**Figure 1.5.** *Configuration de l'application Index*

Wright supporte la composition hiérarchique. Ainsi, un composant peut être composé d'un ensemble de composants. Il en va de même pour un connecteur. Lorsqu'un composant représente un sous-ensemble de l'architecture, ce sous-ensemble est décrit sous forme de configuration dans la partie calcul du composant.

### *1.4. Le concept style*

Le style d'architecture permet de décrire un ensemble de propriétés communes à une famille de système comme, par exemple, les systèmes temps réel ou les systèmes de gestion de paye. Il permet de décrire un vocabulaire commun en définissant un ensemble de types de connecteur et de composant et un ensemble de propriétés et de contraintes partagés par toutes les configurations appartenant à ce style.

## 2. Les aspects comportementaux

Les aspects comportementaux d'une architecture logicielle décrite en Wright sont spécifiés en utilisant CSP de Hoare [**3**] et [**4**]. Les aspects fondamentaux de CSP sont : évènement et processus.

### *2.1. Les événements*

Dans le modèle CSP, tout est représenté par des événements. Un événement correspond à un moment ou une action qui présente un intérêt. Pour les caractériser, nous leur choisissons un nom qui doit être unique pour pouvoir les identifier et les utiliser au moment de la spécification. Dans le cas d'un composant qui ne fait qu'envoyer des messages jusqu'à un message d'arrêt, nous avons les deux événements suivants : write et end-of-data. Comme nous le verrons plus tard, nous avons besoin de faire la différence entre un événement initialisé et un événement observé. Grâce à cette distinction, il nous est possible de mieux contrôler l'interaction en sachant quel composant a initialisé l'événement de ceux qui l'observent. Comme CSP ne fait pas cette distinction, CSP pour Wright le fait :

- Un événement qui est initialisé possède en plus une barre au dessus de lui. Par exemple, le port Output utilisera l'événement $\overline{\text{write}}$ pour indiquer que c'est lui qui l'initialise.
- Un événement observé sera noté comme d'habitude. Le rôle Serveur utilise l'événement write pour indiquer qu'il observe cet événement.
- Une dernière propriété importante des événements est qu'il est possible de décrire des événements qui transmettent des données et d'autre qui reçoivent des données :
    - un événement peut fournir une donnée. Par exemple le port Output pourra spécifier le fait qu'il veut envoyer la donnée x par l'événement write!x.
    - Un événement peut recevoir des données. Comme par exemple pour le rôle Source qui veut recevoir une donnée x utilisera l'événement write?x.

### *2.2. Les processus*

Pour définir un comportement, il faut pouvoir combiner les événements. Un processus correspond à la modélisation de comportement d'un objet par une combinaison d'évènements et d'autres processus simples. L'ensemble des évènements que le processus P peut exécuter est appelé son alphabet (ou interface) est dénoté αP. Le comportement le plus simple est de rien faire : un tel processus est dénoté par STOP.

Pour décrire des comportements plus élaborés, CSP offre les opérateurs donnés ci-dessous.

#### 2.2.1. *Préfixe*

Le préfixe noté → permet de définir un processus en explicitant les événements qu'il peut exécuter. Si a est un événement et P est un processus, alors :
a → P est le processus qui peut exécuter a et se comporte ensuite comme le processus P. l'opérateur de préfixe est toujours utilisé sous cette forme avec un événement à la gauche de → et un processus à droite. Il est possible d'enchaîner les préfixes. Par exemple a → b → Q correspond à l'expression de processus a → (b → Q) en appliquant l'associativité à droite. Le processus § est défini comme suit : √ → STOP.

#### 2.2.2. *Récursivité*

Les définitions récursives permettent de décrire en CSP des processus qui agissent sans fin. Par exemple : Alternative = On → Off → Alternative est un processus défini de manière récursive. Il exécute les événements On et Off en alternance. La récursivité croisée (ou mutuelle récursion) est également autorisée en CSP.

#### 2.2.3. *Opérateur de choix*

Il existe en CSP trois opérateurs de choix sur les processus :

- Le plus simple, dénoté par « | », agit sur les processus de la forme « a → P ». Le processus « a → P | b →Q » peut soit exécuter l'événement a et se comporter ensuite comme le processus P, soit exécuter l'événement b et se comporte ensuite comme le processus Q. les préfixes sont nécessairement distincts (a ≠ b). Il est possible d'utiliser | avec plus de deux choix a → P | b →Q | …| z → R.
- Le choix déterministe ou externe noté □: nous le nommerons ainsi car c'est l'environnement qui choisit le comportement du processus. Si nous avons le processus : e → P □ f → Q et que l'environnement s'engage dans l'événement f alors le processus s'engagera dans cet événement et se comportera comme le processus Q. Ce choix est typiquement utilisé entre des événements observés.

- Le choix non déterministe ou interne noté Π. A l'inverse du choix déterministe, c'est le processus qui choisit de façon non déterministe le comportement à choisir parmi plusieurs. Le processus : e → P Π f → Q va choisir entre initialiser l'événement e et continuer comme P ou initialiser l'événement f et continuer comme Q il décide lui-même de ce choix sans se préoccuper de l'environnement. Ce choix est typiquement utilisé entre des événements initialisés.

2.2.4. *Evènements cachés*

Il est possible à partir d'un processus P de cacher certains évènements de son alphabet. Si A est un ensemble d'événements et si P est un processus, alors : P\A est le processus dont les événements appartenant à A devient des événements internes du processus.

2.2.5. *Composition parallèle*

Pour décrire plusieurs processus en concurrence, CSP introduit l'opérateur de composition parallèle entre processus. P || Q est la composition parallèle de P et de Q. Dans ce cas, tous les événements de P et Q sont exécutés en synchronisation. Par exemple, supposons que les processus P et Q sont définis par :

P = (e → f → P) □ (g → P)

Q = e → (f → Q □ g → Q)

Dans ce cas le processus R = P || Q a le comportement suivant :

- Comme P ne peut s'engager que dans e ou g et Q que dans e, R s'engagera dans e.
- A ce moment P ne peut que continuer dans f alors que Q peut continuer dans f ou g. D'où, R s'engagera dans f.
- A ce point les deux processus ne peuvent que revenir qu'à leur état initial et donc la séquence <e, f> se répétera indéfiniment.

Nous pouvons donc décrire le comportement de R ainsi : R = e → f → R.

## 3. Sémantique de Wright

Wright étant entièrement basé sur CSP au niveau de la spécification des interactions, sa sémantique l'est aussi. Il faut donc comprendre en premier lieu la sémantique de CSP.

### *3.1. Modélisation mathématique des processus CSP*

Un processus CSP est triplet (A, F, D) où A représente l'alphabet du processus, F représente ses échecs et D représente ses divergences.

#### 3.1.1. *Alphabet*

L'alphabet représente l'ensemble des événements sur lequel le processus a une influence. De ce fait, si un évènement n'est pas dans l'alphabet d'un processus, alors ce processus ne le connaît pas, ne le traite pas et donc l'ignore.

#### 3.1.2. *Echecs*

Les échecs d'un processus sont une paire de traces et de refus. Une trace est une séquence d'événements permise par le processus. L'ensemble des traces possibles d'un processus P est noté *traces(P)*. Un refus correspond à un ensemble d'événements proposés pour lequel le processus refuse de s'engager. Cet ensemble d'événements refusés par un processus P est noté *refus(P)*. Cette notion de refus permet une distinction formelle entre processus déterministe et non déterministe. En effet un processus déterministe ne peut jamais refuser un événement qu'il peut entamer alors qu'un processus non déterministe le peut.

#### 3.1.3. *Divergences*

Une divergence d'un processus est définie comme une de ses traces quelconque après laquelle il y a un comportement chaotique. Ce comportement chaotique est représenté par le processus CHAOS : $CHAOS_A = STOP \sqcap (\forall x : A \bullet x \rightarrow CHAOS_A)$

Ce processus peut se comporter comme n'importe quel autre. C'est le plus non déterministe, le plus imprévisible, le plus incontrôlable de tous. La divergence est donc utilisée pour représenter des situations catastrophiques ou des programmes complètement imprédictibles (comme des boucles infinies).

#### 3.1.4. *Modèles Sémantiques*

Les trois principaux modèles sémantiques [**29**] sont les traces, les échecs stables et les traces-divergences.

- Le modèle de traces associe à chaque processus les séquences finies d'événements admises par ce processus. Ce modèle permet donc de représenter les comportements possibles de processus sous forme de traces. Les traces du processus *P* sont dénotées par *traces(P)*.

- Le modèle des échecs stables associe à chaque processus *P* les couples de la forme *(t, E)*, où *t* est une trace finie admise par *P* et *E* est l'ensemble des évènements que le processus ne peut pas exécuter après avoir exécuté les événements de *t*. L'ensemble de ces couples est noté *failures(P)*. Ce modèle permet de caractériser les blocages de *P*. En effet, si *E* est égal à l'ensemble des évènements exécutables par *P*, alors *P* se retrouve bloqué.
- Enfin, le modèle des échecs-divergences associe à chaque processus *P* l'ensemble de ses échecs stables et l'ensemble de ses divergences. Un processus *P* n'est divergent que s'il se trouve dans un état dans lequel les seuls événements possibles sont les événements internes. Cet état est dit divergent. L'ensemble des divergences de *P* noté *divergences(P)*, est l'ensemble des traces *t* telles que le processus se trove dans un état divergent après avoir exécuté *t*. Si le processus est déterministe, alors *divergences(P)* est vide.

## 4. Le raffinement CSP

Le raffinement consiste à calculer et à comparer les modèles sémantiques de deux processus. Le raffinement dépend donc du modèle considéré. Par exemple, dans le cas du modèle des échecs-divergences, si *P* et *Q* sont deux processus alors *Q* raffine *P*, noté :

$$P \sqsubseteq_{FD} Q \text{ si : } failures\ (Q) \subseteq failures(P) \wedge divergence(Q) \subseteq divergences(P)$$

Dans cet exemple il n'est pas utile de comparer *traces*(P) et *traces*(Q), car par définition :

$$failures(Q) \subseteq failures(P) \Rightarrow traces(Q) \subseteq traces(P)$$

Intuitivement, *Q* est égal ou meilleur que *P* dans le sens où il à moins de risque d'échec et divergence. *Q* est plus prévisible et plus contrôlable que *P*, car si *Q* peut faire quelque chose d'indésirable ou refuser quelque chose, *P* peut le faire aussi. Concrètement cela signifie qu'un observateur ne peut pas distinguer si un processus a été substitué par un autre.

## 5. Utilisation de la sémantique de CSP dans Wright

La sémantique de Wright est entièrement basée sur CSP. En effet, chaque partie d'une architecture logicielle Wright est modélisée par un processus CSP. Pour analyser l'interaction de ces processus, il faut les combiner par l'opérateur « || » de CSP. Mais un problème

émerge : la sémantique de CSP utilise des noms d'événements globaux. En effet pour décrire que deux processus CSP interagissent, il suffit qu'ils partagent un nom d'événement identique. Tandis que les noms d'événements en Wright agissent comme des noms locaux propres aux composants et que l'interaction a lieu par connecteurs.

Afin de réaliser cette correspondance d'évènements locaux de Wright en événements globaux de CSP, Wright identifie et résout systématiquement les deux problèmes suivants :

- Au niveau des instances :

Il peut exister plusieurs instances d'un même type. Ainsi des interactions indésirées peuvent avoir lieu en introduisant de multiples copies d'un même processus. Pour résoudre ce premier problème, il suffit de préfixer chaque nom d'évènement par le nom de son instance ainsi un évènement a :

  - 3 niveaux : *N.P.e* (nom du composant, nom du port, nom de l'évènement), si le Calcul (Computation) utilise un évènement du port P,
  - 2 niveaux : *N.e* (nom du composant, nom de l'évènement), si le Calcul (Computation) utilise un évènement interne (non associé à un port).

- Au niveau des liens :

Comme les noms d'événements utilisés dans les Composants et les Connecteurs sont locaux, pour assurer une synchronisation entre un Composant et un Connecteur (indiqué par un lien) il faut pouvoir changer les noms d'événements afin de respecter le fait qu'en CSP deux processus ne communiquent que s'ils partagent un même événement. Pour résoudre ce deuxième problème, il suffit de renommer les événements du Connecteur par les noms des événements des Composants correspondants : si nous avons un nom d'événement pour un connecteur *Conn.Role.e* et que le rôle de ce Connecteur est lié à un Port d'un Composant *Comp.Port* alors nous voulons que le nom de l'événement du Connecteur *Conn.Role.e* soit renommé *Comp.Port.e*.

## 6. Validations de la description

Nous avons vu comment Wright peut être utilisé pour exprimer la structure et le comportement d'une architecture logicielle ou d'un style d'architecture. Mais ces descriptions ne sont réellement intéressantes que si elles nous apportent suffisamment d'informations pour pouvoir analyser et fournir des propriétés sur l'architecture logicielle en question.

Wright permet l'étude de deux critères fondamentaux sur une architecture logicielle qui sont la cohérence et la complétude.

Informellement, la cohérence correspond au fait que la description a un sens, c'est-à-dire qu'aucune partie n'en contredit une autre.

La complétude permet de vérifier que la description contient suffisamment d'informations pour prétendre à une analyse, c'est-à-dire que la description n'a pas omis de détails pour réaliser notre analyse.

Il est important de vérifier que le système résultant de sa description est cohérent car s'il ne l'est pas, tout raffinement ou implémentation restera incohérent. La complétude est tout aussi importante car une analyse ne peut être basée que sur ce que nous connaissons du système. Si nous analysons la communication d'un Composant, mais qu'une partie de l'interface est laissée de côté, comment est-il possible de faire une bonne analyse ? Par exemple, une analyse sur une ressource partagée ne peut être faite que si tous les participants qui accèdent à cette ressource sont connus.

Le problème de la complétude est spécialement critique pour un architecte à cause de l'importance de son abstraction à ce niveau de conception. Il y a toujours une tension entre le besoin d'inclure des informations critiques nécessaires pour garantir des propriétés importantes du système et le risque de mettre du désordre dans l'architecture en mettant des contraintes et des détails qui peuvent rendre l'architecture difficile à manier et à exploiter.

Nous allons maintenant décrire la liste des tests ou propriétés élaborés par les auteurs de Wright liés à la cohérence et la complétude d'une architecture logicielle en Wright un soin particulier sera accordé aux propriétés automatisables grâce au raffinement CSP.

### *6.1. Description informelle des propriétés Wright*

Nous allons expliciter informellement l'ensemble des propriétés intégrées dans Wright sur la cohérence et la complétude.

#### 6.1.1. *Cohérence*

La cohérence consiste à vérifier que tous les éléments décrivant l'architecture logicielle (les Composants, les Connecteurs et la Configuration) sont cohérents.

##### 6.1.1.1. Cohérence d'un composant

Nous avons vu dans la description d'un Composant qu'il est constitué par deux parties, la partie interface décrivant les Ports et la partie Calcul, ainsi nous vérifions la cohérence d'un Composant en nous assurant que le Calcul obéit aux règles d'interaction définies par les Ports.

Le premier aspect décrit par le Port correspond au comportement attendu du Composant. Il faut s’assurer que la cohérence entre le comportement des ports et celui du Calcul (Computation) par la notion de la projection.

Un Port est une projection d’un Composant si ce dernier agit de la même manière que le Port quand nous ignorons tous les événements n’appartenant pas à l’alphabet de ce Port.

Illustrons la notion de projection en prenant le Composant suivant (voir Figure 1.6).

**Component** Double

**Port** Input = $read?x \rightarrow Input \; \Box \; close \rightarrow \S$

**Port** Output = $\overline{write!x} \rightarrow Output \sqcap \overline{close} \rightarrow \S$

**Computation** =

$$Input.read?x \rightarrow \overline{Output.write!(2 * x)} \rightarrow Computation$$
$$\Box$$
$$Input.close \rightarrow \overline{Output.close} \rightarrow \S$$

**Figure 1.6.** *Cohérence d’un Composant*

Si nous ignorons tous les événements qui n’appartiennent pas à l’alphabet du Port *Input*, nous obtenons :

$$Computation = Input.read?x \rightarrow Computation \; \Box \; Input.close \rightarrow \S$$

Le port *Input* est bien une projection du Calcul. De même manière, il est facile de montrer que le port *Output* est une projection du Calcul.

Le second aspect du Port correspond à l’interaction avec l’environnement. Si la spécification des Ports ne tient pas compte d’un événement, alors le Composant (partie Calcul) n’a pas à s’en occuper. Mais d’un autre côté, il y a des situations dans lesquelles il peut être approprié d’avoir des spécifications d’un Composant qui décrivent des comportements qui n’auront pas forcément lieu. Ceci est notamment le cas lors de la réutilisation d’une spécification d’un Calcul plus général que nos besoins. Pour toutes ces raisons, la spécification des Ports peut ne couvrir qu’un sous ensemble des situations que le Composant peut effectivement gérer. Illustrons nos propos par l’exemple du Composant *Double* qui décrit le comportement d’échecs indiqué par l’évènement *fail* si nous souhaitons utiliser ce Composant sans se préoccuper de cet événement, nous le spécifions par la figure 1.7.

**Component** Double

$$\textbf{Port}\ \text{Input} = read?x \rightarrow Input \,\square\, close \rightarrow \S$$

$$\textbf{Port}\ \text{Output} = write!x \rightarrow Output \sqcap \overline{close} \rightarrow \S$$

$$\textbf{Computation} =$$

$$Input.read?x \rightarrow \overline{Output.write!(2 * x)} \rightarrow Computation$$
$$\square$$
$$Input.close \rightarrow \overline{Output.close} \rightarrow \S$$
$$\square$$
$$In.fail \rightarrow \S$$

**Figure 1.7.** *Cohérence d'un Composant réutilisé*

Si nous ignorons l'hypothèse du port Input qui est que l'événement fail ne va pas avoir lieu et nous projetons le port Output, nous obtenons :

Computation = $\overline{\text{Ouput.write}}$ !(2*x) → Computation □ $\overline{\text{Output.close}}$ → § □ §

Le port Output n'est plus une projection du Calcul. Ainsi la propriété permettant de vérifier la cohérence d'un composant Wright est formulée comme suit :

Propriété 1: cohérence Port / Calcul

*La spécification d'un port doit être une projection du Calcul, sous l'hypothèse que l'environnement obéisse à la spécification de tous les autres ports.*

Intuitivement, la propriété 1 stipule que le composant ne se préoccupe pas des évènements non traités par les ports (ici évènement *fail*).

6.1.1.2. Cohérence d'un connecteur

La description d'un connecteur doit vérifier que la coordination des rôles par la glu cohérente avec le comportement attendu des composants. Sachant qu'un processus CSP est dit en situation d'interblocage quand il peut refuser de participer à tout évènement, mais n'a pas pour autant terminé correctement (en participant à l'évènement §). Inversement, un processus est sans interblocage s'il ne peut jamais être en situation d'interblocage. Ainsi Wright propose la propriété :

Propriété 2: Connecteur sans interblocage

*La glu d'un connecteur interagissant avec les rôles doit être sans interblocage.*

Une autre catégorie d'incohérence est détectable comme une situation d'interblocage, lorsque la spécification d'un rôle est elle-même incohérente. Dans une spécification d'un rôle complexe, il peut y avoir des erreurs qui mènent à une situation dans laquelle aucun évènement n'est possible pour ce participant même si la Glu était prête à prendre tout évènement.

Propriété 3: Rôle sans interblocage

*Chaque rôle d'un connecteur doit être sans interblocage.*

Pour empêcher le conflit de contrôle, un évènement ne doit être initialisé que par un unique processus, tous les autres processus ne faisant que l'observer.

Propriété 4: Un initialiseur unique

*Dans une spécification de connecteur, tout évènement ne doit être initialisé que par un rôle ou la glu. Tous les autres processus doivent soit l'observer, soit l'oublier (grâce à leur alphabet).*

La dernière propriété vérifie que les notions pour l'initialisation et l'observation des événements sont utilisées correctement.

Propriété 5: Engagement de l'initialiseur

*Si un processus initialise un évènement alors il doit s'engager dans cet évènement sans être influencé par l'environnement.*

6.1.1.3. Cohérence d'une configuration

Au niveau de la déclaration d'instances, la cohérence s'applique aux deux points suivants :

- Le nom de l'instance est-il unique?
- Des paramètres raisonnables ont-il été donnés ?

Dans l'exemple ci-dessous nous avons paramétré le nombre d'instances du port Output du composant Filtre_Texte.

**Component** Filtre_Texte (nout : 1..)

**Port** Input = DataInput

**Port** $Output_{1..nout}$ = DataOutput

**Computation** = lire des données du port Input. Envoyer ces données successivement sur les ports $Output_1$, $Output_2$, …,$Output_{nout}$

Le nombre d'instances du port Output sera déterminé au moment de l'instanciation du filtre.

Propriété 6: Substitution des paramètres

*Une déclaration d'instance paramétrant un type doit résulter d'une validation de ce type après avoir substitué tous les paramètres formels manquant.*

Dans le cas de paramètre numérique, il faut s'assurer que les paramètres entrent dans les bonnes données dans la description du type.

Propriété 7: Test des valeurs sur leur intervalle donné

*Un paramètre numérique ne doit pas être plus petit que la limite inférieure (si elle est déclarée), et pas plus grande que la limite supérieure (si elle est déclarée).*

Au niveau de liens la question suivante se pose : Quels ports peuvent être utilisés pour ce rôle? La vérification sur le fait que les protocoles du port et du rôle soient identiques n'est pas suffisante. En effet, nous voulons avoir la possibilité d'attacher un port qui n'a pas un protocole identique au rôle.

Considérons le rôle suivant :

**Role** Source = $\overline{\text{write !x}} \rightarrow \text{Source} \Box \overline{\text{close}} \rightarrow §$

Le rôle Source peut être attaché au port suivant :

**Port** Output3 = $\overline{\text{write! 1}} \rightarrow \overline{\text{write! 2}} \rightarrow \overline{\text{write! 3}} \rightarrow \overline{\text{close}} \rightarrow §$

Le rôle Source et le port Output3 ne sont pas identiques. Le rôle Source qui émet des suites de x a une description plus générale que le port Ouput3 qui émet la suite 1 2 3.

D'autre part il faut toujours vérifier qu'il n'existe pas une incompatibilité entre le rôle et le port qui lui est attaché. Par exemple, nous ne voulons pas accepter le fait qu'un port BadOutput (sans l'évènement close) puisse être attaché au rôle Source.

**Port** BadOutput = $\overline{\text{write !x}} \rightarrow \text{BadOutput} \sqcap §$

Propriété 8: compatibilité port / rôle

*Tout port attaché à un rôle doit toujours continuer son protocole dans une direction que le rôle peut avoir.*

6.1.1.4. Cohérence d'un style

Les deux propriétés suivantes concernent le concept de Style d'architecture. Une configuration d'un système est cohérente avec ses styles déclarés si elle obéit à chacune de leurs contraintes.

Propriété 9: Contraintes de style

*Les prédicats d'un style doivent être vrais pour une configuration déclarée être dans ce style. Les contraintes d'un Style doivent être cohérentes entre elles.*

Exemple :

$\forall$ c : Component ; p : Ports (c) $\bullet$ Type (p) = DataOutput

$\exists$ c : Component ; p : Ports (c) $\bullet$ Type (p) = DataOutput

Ces deux contraintes sont en contradiction, donc le Style contenant ces deux contraintes est incohérent.

<u>Propriété 10: Cohérence de Style</u>

*Au moins une configuration doit satisfaire les contraintes de style*

6.1.2. *Complétude*

Une catégorie de complétude importante que vérifie Wright concerne la configuration :

- Au niveau des liens, si un lien est omis alors un composant va dépendre des évènements qui ne vont jamais avoir lieu, ou une interaction va échouer car il manque un participant,
- D’autre part, il existe des ports de composants qui n’ont pas besoin d’être attachés et il y a des interactions qui peuvent continuer même si un participant manque.

Pour ces deux raisons, il n’est pas suffisant de contrôler que tous les ports et rôles soient bien attachés.

<u>Propriété 11: la complétude des liens</u>

*Chaque port (respectivement rôle) non attaché dans la configuration doit être compatible avec le rôle (respectivement port) §.*

Pour résumer, on peut donner la liste des propriétés effectuées par Wright sur la cohérence et la complétude.

1. Cohérence des ports avec le Calcul (composant)
2. Absence d’interblocage sur les connecteurs (connecteur)
3. Absence d’interblocage sur les rôles (rôle)
4. Initialiseur unique (connecteur)
5. Engagement de l’initialiseur (n’importe quel processus)
6. Substitution des paramètres (instance)
7. Borne d’intervalle (instance)
8. Compatibilité port / rôle (lien)
9. Contraintes pour les styles (configuration)
10. Cohérence de style (style)
11. Complétude des liens (configuration)

## 7. Techniques de vérification des propriétés Wright

Dans cette section, nous nous interrogeons sur les techniques potentielles permettant de prouver les propriétés Wright présentées précédemment.

### *7.1. Utilisation du raffinement CSP*

Le raffinement CSP permet le développement incrémental des systèmes CSP. Un processus CSP peut être raffiné progressivement jusqu'à son implémentation (raffinement ultime). Par exemple si une composition parallèle de deux processus P et Q raffine une spécification abstraite décrite par le processus S, alors nous écrivons :

$$S \sqsubseteq P \| Q$$

Ensuite, nous pouvons développer la spécification S en raffinant d'une façon séparée P et Q :si $P \sqsubseteq P'$ et $Q \sqsubseteq Q'$, alors la composition de P' et Q' raffine aussi S : $S \sqsubseteq P' \| Q'$.

Egalement le raffinement CSP peut être utilisé pour vérifier des propriétés de sûreté ou de vivacité. En effet, des propriétés Wright sont formalisées grâce au raffinement CSP. Ces propriétés sont les suivantes :

- Propriété 1: Cohérence des ports avec le calcul.
- Propriété 2: Absence d'interblocage sur les connecteurs.
- Propriété 3: Absence d'interblocage sur les rôles.
- Propriété 8: Compatibilité port / rôle.

### *7.2. Formalisation*

Afin de formaliser les propriétés Wright en utilisant le raffinement CSP, nous définissons les ensembles suivants :

$\alpha P$ : l'alphabet du processus P.

$\alpha_i P$ : le sous –ensemble de $\alpha P$ correspondant au évènements initialisés.

$\alpha o P$ : le sous –ensemble de $\alpha P$ correspondant au évènements observés.

- **Propriété 1: Cohérence Port / Calcul**

Comme nous l'avons noté, la spécification d'un port a deux aspects :

- Des exigences sur le comportement du composant (le composant accomplit le comportement décrit par le port).

- Des suppositions sur l’environnement (qu’est-ce que l’environnement, c’est-à-dire les rôles des connecteurs auxquels le composant peut être attaché, va exiger pendant l’interaction).

Ainsi, pour modéliser le processus du Calcul dans l’environnement indiqué par les ports :

**a.** Nous devons prendre les ports et construire un processus qui est restreint aux évènements observés (se qui extrait les suppositions de l’environnement).

**Définition 1**

Pour tout processus P= (A, F, D) et un ensemble d’événements E, $P \lceil E = (A \cap E, F', D')$

*où* $F' = \{(t', r') \mid (t, r) \in F \mid t' = t \lceil E \wedge \forall \; r' = r \cap E\}$ *et* $D' = \{t' \mid \exists \; t \in D \mid t' = t \lceil E\}$.

La projection d’une trace (t ⌈ E) est une trace qui contient tous les éléments de t qui sont dans E, dans le même ordre, sans tous les éléments qui ne sont pas dans E.

Exemple

$<acadbcabc> \lceil \{a, b\} = <aabab>$

**b.** Nous devons rendre le nouveau processus déterministe. Ainsi, nous assurons que les décisions prises dans l’interaction sont faites par le Calcul et non pas par les ports.

**Définition 2**

Pour tout processus $P = (A, F, D)$, $det\,(p) = (A, F', \emptyset)$

*où* $F' = \{(t, r) \mid t \in Traces(P) \wedge \forall \; e : r \bullet t \,\hat{}\, < e > \notin Traces(P)\}$.

La fonction *det*(P) a les mêmes traces que P, mais avec moins de refus. Ainsi, n’importe quel événement qui a lieu à tout point est entièrement contrôlable par l’environnement : *det*(P) est déterministe.

**c.** Il nous reste plus qu’à faire interagir ce nouveau processus déterministe (*det*(P)) avec celui du Calcul (C) en les mettant en parallèle : C || *det*(P). Nous avons donc les traces de P mais où les décisions sont prises par C.

En utilisant le raffinement, nous pouvons vérifier que le Calcul respecte bien les exigences de ports.

<u>Propriété 1: Cohérence Port / Calcul</u>

*Pour un composant avec un processus de Calcul C et des ports P,* $P_1, \dots P_n$ *; C est cohérent avec P si* $P \sqsubseteq (C \parallel \forall \; i : 1..n \parallel det(Pi \lceil \alpha oPi)) \lceil \alpha P$.

- **Propriété 2 et Propriété 3: Absence d’interblocage sur les connecteurs et Absence d’interblocage sur les rôles**

Ces deux propriétés reviennent à vérifier si un processus est sans interblocage. D’une façon formelle, un processus P =(A, F, D) est sans interblocage si pour toute trace t telle que (t, A)∈ F, last(t) = √. Mais ceci peut être exprimé par une relation de raffinement entre le processus $DF_A$ et P: $DF_A \sqsubseteq P$ avec $DF_A$ est défini comme suit :

$DF_A = (\sqcap\, e : A \bullet e \rightarrow DF_A) \sqcap \S.$

Le processus $DF_A$ permet toutes les traces possibles sur l'alphabet A mais sans jamais avoir la possibilité de refuser tous les événements : il s'agit d'un processus sans interblocage.

- **Propriété 8: compatibilité port / rôle**

La distinction entre un port et un rôle est que le port décrit un comportement spécifique alors qu'un rôle décrit un pattern de comportement permettant le lien de plusieurs ports.
Par contre le lien d'un port à un rôle doit toujours respecter les contraintes de spécification de ce rôle. Ainsi, le comportement d'un port attaché à un rôle est le comportement de ce processus port restreint aux traces de ce processus rôle.

Comme la restriction d'une trace est effectuée par la version déterministe d'un processus, nous testons donc le processus P || det(R) pour exprimer cette restriction au processus rôle.
Pour pouvoir utiliser le raffinement dans le test de compatibilité, il faut que les alphabets des deux processus port et rôle soient identiques. Pour cela nous définissons comment augmenter l'alphabet d'un processus.

**Définition 3**

Pour tout processus P et un ensemble d'évènement A, $P_{+A} = P \parallel STOP_A$

Propriété 8: compatibilité

Un port P est compatible avec un rôle R, noté P compat R, si

$R_{+(\alpha P - \alpha R)} \sqsubseteq P_{+(\alpha R - \alpha P)} \parallel det(R)$.

## 8. Automatisation

Les auteurs de Wright proposent un outil appelé Wr2fdr [**2**], [**24**] censé d'automatiser les quatre propriétés décrites précédemment. Pour y parvenir, l'outil Wr2fdr traduit la spécification Wright en une spécification CSP dotée des relations de raffinement à vérifier. La spécification CSP engendrée pour l'outil Wr2fdr est soumise à l'outil de Model checking - FDR (Failure-Divergence Refinement) [**28**]. Dans la suite nous présentons successivement FDR, Wr2fdr et une vérification de l'outil wr2fdr.

### *8.1. FDR*

FDR permet de vérifier de nombreuses propriétés sur des systèmes d’état finis. FDR s’appuie sur la technique de « model checking » [**29**]. Celle-ci effectue la vérification d’un modèle d’un système par rapport aux propriétés qui sont attendues sur ce modèle. Cette vérification est entièrement automatisée et consiste à explorer tous les cas possibles. Le résultat de cette analyse est soit la confirmation que chaque propriété est maintenue par le modèle, soit qu’elle ne l’est pas. Dans le dernier cas, le « model checker » (outil) renvoie un contre-exemple qui montre comment la propriété n’est pas maintenue.

FDR est basé sur la théorie de CSP et précisément sur la sémantique opérationnelle de CSP : modèle de traces, modèle d’échecs stables et modèle d’échecs/divergences (voir chapitre 2 section 3). En FDR, la méthode pour établir qu’une propriété vérifiée revient à réaliser un raffinement entre deux processus P (processus abstrait) et Q (processus raffiné) représentés par deux machines d’états finis. Sachant que la propriété à vérifier est traduite par la relation de raffinement entre P et Q.

Le problème de FDR réside dans l’exploration de tous les états à considérer pour vérifier la propriété. Très vite les limites de la machine sont atteintes. Cependant sur des espaces d’état raisonnables, FDR permet de vérifier la propriété ou, dans le cas contraire, de donner une trace correspondant à l’ensemble de transitions et des états qui conduise au contre-exemple. Mais dans le cas où un contre-exemple ne serait pas trouvé, il est difficile voire impossible de statuer sur la validité ou non de la propriété. Cela dépend de l’espace d’état considéré et de la finitude de la machine d’états représentant le processus CSP. Ce dernier problème est non décidable [**28**].

La dernière version de FDR2 améliore le passage à l’échelle de l’outil en proposant des nouvelles techniques permettant de combattre l’explosion combinatoire de la taille des machines d’états finis en utilisant des méthodes de compression. Par exemple, la composition parallèle de deux processus P et Q ayant chacun 1000 états nécessite une machine de 1000000 d’état mais la composition compressée proposée par FDR2 nécessite uniquement 10000 états.

### *8.2. L’outil Wr2fdr*

Wr2fdr est un outil développé par l’université de Carnegie Mellon. Il accompagne l’ADL Wright. Il permet de traduire une spécification Wright en une spécifiacation CSP acceptée par l’outil FDR. L’outil Wr2fdr est censé assurer les fonctionnalités suivantes :

- Analyse lexico-syntaxique d'une spécification Wright,
- Génération de code CSP,
- Déterminisation d'un processus CSP : det (P), ceci permet de traiter l'opération non déterministe (⊓) de CSP,
- Calcul de l'alphabet d'un processus CSP : αP. En effet, FDR exige explicitement lors de la composition parallèle d'un processus (||) leurs alphabets,
- Correspondance entreles événements locaux de Wright et les événements globaux de CSP,
- Calcul des relations de raffinement liées aux proprités 1, 2, 3 et 8 permettant de vérifier respectivement la cohérence Port / Calcul, l'absence d'interblocage sur les connecteurs, l'absence d'interblocage sur les rôles et la compatibilité Port / Rôle (voir chapitre 2 section 6 et 7).

La version actuelle de l'outil Wr2fdr ne fait pas la distinction entre les événements initialisés et observés. De plus, ces événements ne doivent pas porter des informations ni d'entrée ni de sortie.

### *8.3. Vérification de l'outil Wr2fdr*

L'outil Wr2fdr (voir Figure 1.8) accepte en entrée un fichier contenant une spécification Wright et produit en sortie un fichier content une spécification CSP acceptable par l'outil de model-cheking FDR afin de vérifier les propriétés 1, 2, 3 et 8 (voir 2.6.1). En effet, l'outil Wr2fdr est censé automatiser ces propriétés en utilisant le concept de raffinement de CSP.

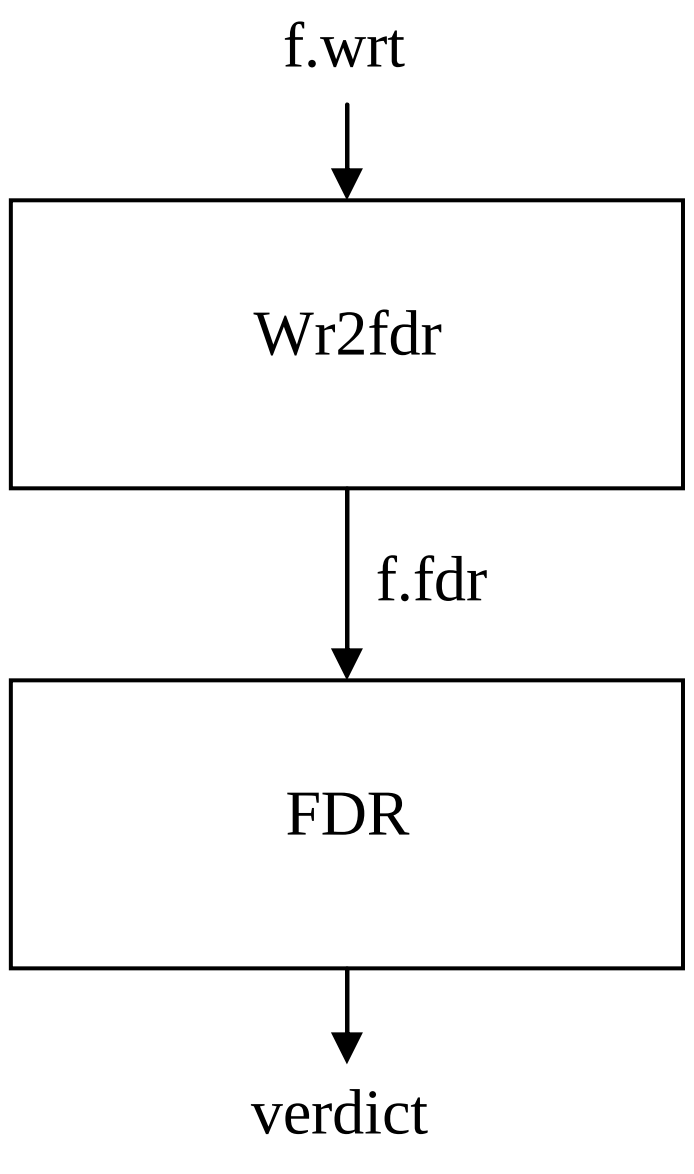

**Figure 1.8.** *Fonctionnement de l'outil Wr2fdr*

8.3.1. *Tests syntaxiques*

Nous considérons le programme Wr2fdr comme boîte noire. Ainsi, nous nous plaçons dans le cadre de test fonctionnel [**25**]. Le programme Wr2fdr nécessite des données d'entrée (des spécifications ou des descriptions en Wright) respectant une syntaxe rigide et bien définie : un sous-ensemble de la syntaxe BNF de Wright [**24**]. Afin de couvrir l'espace de données du programme Wr2fdr, nous retenons les deux critères de couverture suivants :

**Critère 1**: couverture des symboles terminaux, tels que « configuration », « composant », « as », « instances », « [] », « → », « |~| », etc.

**Critère 2**: couverture des règles de production permettant de définir les conditions syntaxiques offertes par Wright.

Nous avons suivi une approche de prédiction des sorties attendues afin de tester le programme Wr2fdr. La fonction d'oracle permettant de comparer la sortie observée par rapport à la sortie attendue pour une DT (Donnée de Test) fournie est actuellement manuelle. Une automatisation de celle-ci peut être envisagée en s'inspirant de la commande *diff* offerte par un système d'exploitation de type Unix.

En menant une activité de test fonctionnel orientée tests syntaxiques, nous avons constaté des écarts entre la spécification et l'implémentation de l'outil Wr2fdr. Ainsi nous pouvons dire que l'implémentation de l'outil Wr2fdr n'est pas conforme à sa spécification. En fait l'outil Wr2fdr peut produire des spécifications CSP non acceptables par FDR. En outre il peut s'arrêter brutalement en signalant une erreur à l'exécution.

Dans la suite, nous allons détailler les défaillances détectées lors du test de l'outil Wr2fdr. Ces défaillances concernent principalement le calcul des relations de raffinement liées aux propriétés 1, 2, 3 et 8 (voir chapitre 2 section 6).

8.3.2. *Défaillances liées aux propriétés 2 et 3 : cohérence du Connecteur*

En Wright, la cohérence d'un connecteur est obtenue par la vérification des deux propriétés 2 et 3. Pour tester le comportement de l'outil Wr2fdr vis-à-vis de ces deux propriétés, nous avons soumis l'entrée appelée *PipeConn.wrt* fournie par la Figure 1.9.

```
Style PipeConn
Connector Pipe
   Role Writer = write -> Writer |~| close -> TICK

   Role Reader = DoRead |~| ExitOnly
   where {
         DoRead = read -> Reader [] readEOF -> ExitOnly
         ExitOnly = close -> TICK
   }

   Glue = Writer.write -> Glue [] Reader.read -> Glue
         [] Writer.close -> ReadOnly [] Reader.close -> WriteOnly
   where {
         ReadOnly = Reader.read -> ReadOnly
                  [] Reader.readEOF -> Reader.close -> TICK
                  [] Reader.close -> TICK
         WriteOnly = Writer.write -> WriteOnly [] Writer.close -> TICK
   }

Constraints
   // no constraints
End Style
```

**Figure 1.9.** *Cas de test pour les propriétés 2 et 3*

L'outil Wr2fdr génère la spécification CSP PipeConn.fdr2 (voir figure 1.10). Lors de la vérification des trois relations de raffinement signalées par assert, l'outil FDR rencontre des problèmes (voir figure 1.11) visiblement d'ordre syntaxique. En effet un examen du fichier PipeConn.fdr2 montre que les identificateurs colorés (ou marqués) ne sont pas définis.

```
-- FDR
compression functions
transparent diamond
transparent normalise
-- Wright defined processes
channel abstractEvent
DFA = abstractEvent -> DFA |~| SKIP
quant_semi({},_) = SKIP
quant_semi(S,PARAM) = |~| i:S @ PARAM(i) ; quant_semi(diff(S,{i}),PARAM)
power_set({}) = {{}}
power_set(S) = { union(y,{x}) | x <- S, y <- power_set(diff(S,{x}))}
-- Style PipeConn
-- events for abstract specification
channel readEOF, read, close, write
-- Connector Pipe
  -- generated definitions (to split long sets)
  ALPHA_Pipe = {|Reader.readEOF, Reader.read, Reader.close, Writer.write
     , Writer.close|}

  ReadOnly = ((Reader.read -> ReadOnly) [] ((Reader.readEOF -> (Reader.close
        -> SKIP)) [] (Reader.close -> SKIP)))
  WriteOnly = ((Writer.write -> WriteOnly) [] (Writer.close -> SKIP))
  Glue = ((Writer.write -> Glue) [] ((Reader.read -> Glue) [] ((Writer.close
        -> ReadOnly) [] (Reader.close -> WriteOnly))))
-- Rôle Writer
  ALPHA_Writer = {close, write}
  ROLEWriter = ((write -> Writer) |~| (close -> SKIP))
  WriterA = ROLEWriter [[ x <- abstractEvent | x <- ALPHA_Writer ]]
  assert DFA [FD= WriterA
-- Rôle Reader
  ALPHA_Reader = {readEOF, read, close}
  DoRead = ((read -> Reader) [] (readEOF -> ExitOnly))
  ExitOnly = (close -> SKIP)
  ROLEReader = (DoRead |~| ExitOnly)
  ReaderA = ROLEReader [[ x <- abstractEvent | x <- ALPHA_Reader ]]
  assert DFA [FD= ReaderA
channel Writer: {close, write}
channel Reader: {readEOF, read, close}
Pipe = ( (ROLEWriter[[ x <- Writer.x | x <- {close, write } ]]
      [| diff({|Writer|}, {}) |]
    (ROLEReader[[ x <- Reader.x | x <- {readEOF, read, close } ]]
      [| diff({|Reader|}, {}) |]
    Glue)) )
PipeA = Pipe [[ x <- abstractEvent | x <- ALPHA_Glue ]]
assert DFA [FD= PipeA
-- End Style
```

**Figure 1.10.** *Fichier CSP PipeConn.fdr2*

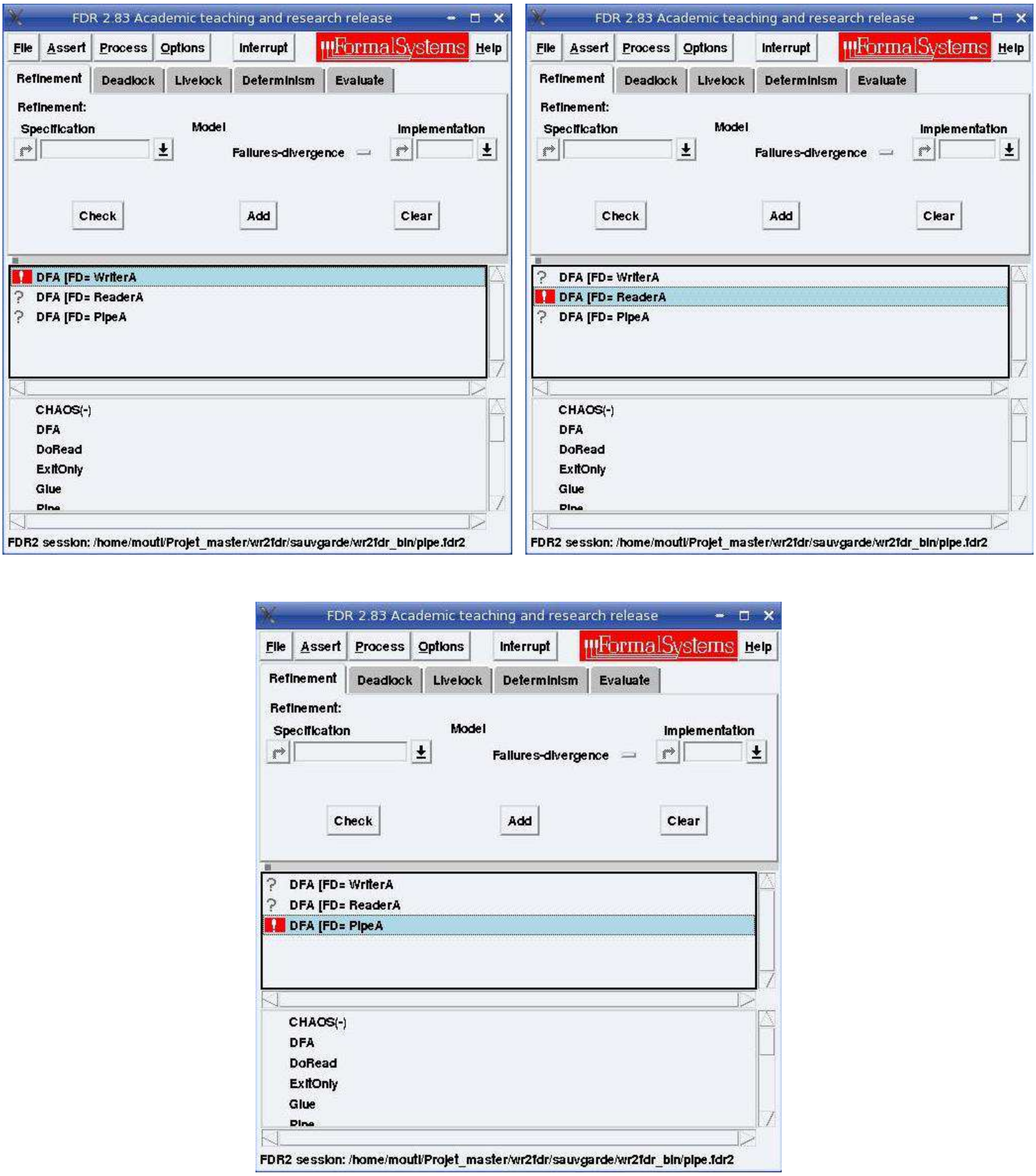

**Figure 1.11.** *Problèmes rencontrés par FDR*

### 8.3.3. *Défaillances liées à la propriété 1: cohérence Port/Calcul*

Nous avons exécuté le programme Wr2fdr avec le cas de test fourni par la figure 1.12. Un tel cas comporte un seul composant appelé *Double* et par conséquent la propriété visée est évidement la propriété 1. L'exécution de Wr2fdr sur ce cas de test entraîne une erreur à l'exécution (voir Figure 1.13) : erreur de segmentation traduisant souvent l'utilisation d'un pointeur qui ne pointe nulle part ceci est plausible car Wr2fdr est écrit en C++.

```
Style Double
  Component Double
    Port In = read -> In [] close -> TICK
    Port Out = _write -> Out |~| _close -> TICK
    Computation = In.read -> _Out.write -> Computation [] In.close -> _Out.close -> TICK
constraints
    //no constraints
End Style
```

**Figure 1.12.** *Cas de test pour la propriété 1*

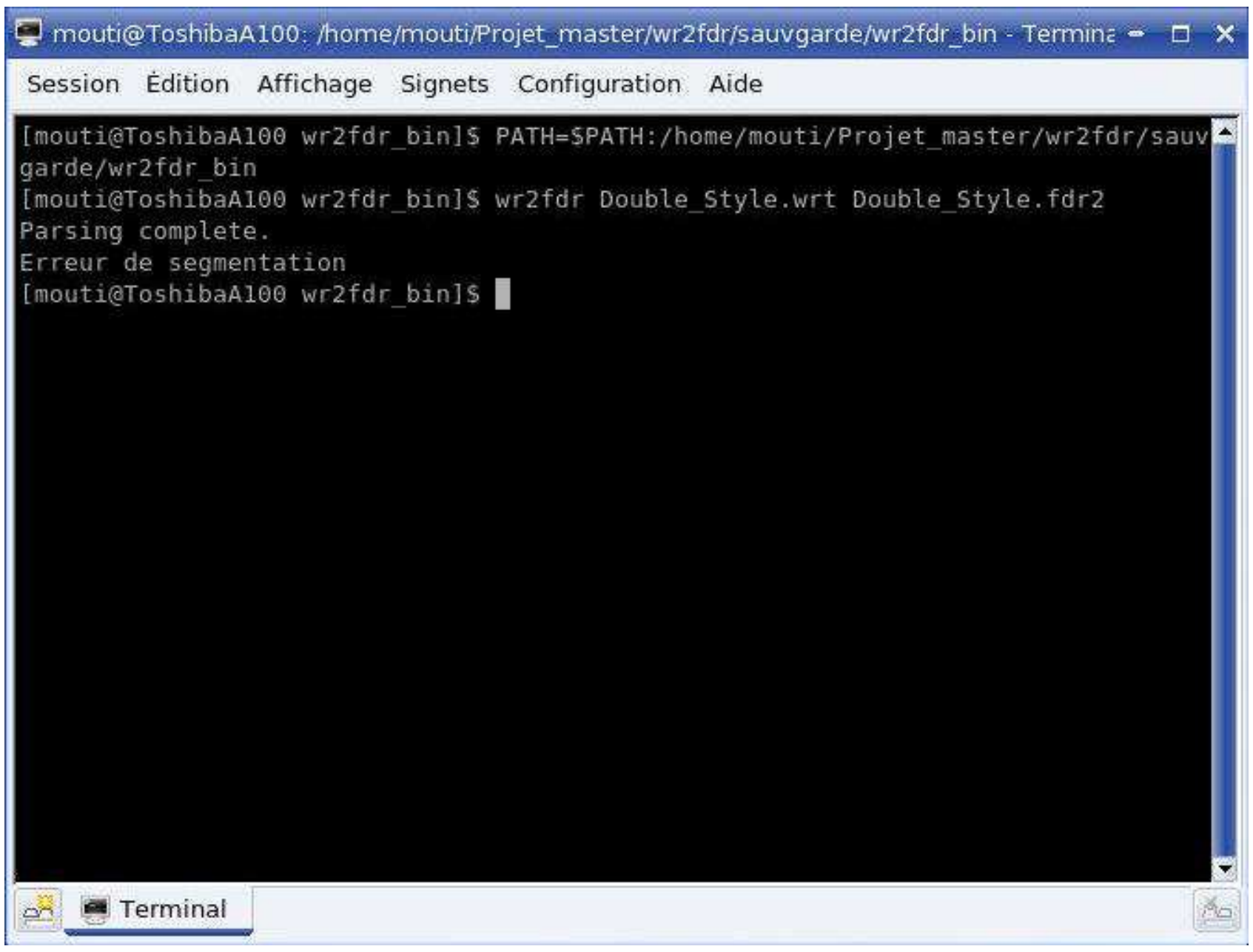

**Figure 1.13.** *Arrêt brutal de l'outil Wr2fdr*

### 8.3.4. *Défaillances liées à la propriété 8: compatibilité Port/Rôle*

Pour pouvoir tester le comportement de l'outil Wr2fdr vis-à-vis de la propriété 8, il faut faire appel à la construction *Configuration* avec notamment les clauses *Instances* et *Attachments*.

L'exécution du programme Wr2fdr avec le cas de test fourni par la figure 1.14 entraîne le même arrêt brutal rencontré précédemment.

```
Configuration ABC
  Component Atype
    Port Output = _a -> Output |~| TICK
    Computation = _Output.a -> Computation |~| TICK
  Component Btype
    Port Input = c -> Input [] TICK
    Computation = Input.c -> _b -> Computation [] TICK
  connector Ctype
    Role Origin = _a -> Origin |~| TICK
    Role Target = c -> Target [] TICK
    Glue = Origin.a -> _Target.c -> Glue [] TICK

  Instances
    A : Atype
    B : Btype
    C : Ctype
  Attachments
    A.Output As C.Origin
    B.Input As C.Target
End Configuration
```

**Figure 1.14.** *Cas de test pour la propriété 8*

## 9. Conclusion

Dans ce chapitre, nous avons présenté d'une façon approfondie tous les éléments fondamentaux de l'ADL formel Wright : aspects structuraux, aspects comportementaux basés sur ceux de CSP de Hoare, sémantique formelle de Wright basée sur celle de CSP, propriétés standards liées à la cohérence et la complétude d'assemblages de composants Wright et formalisation de certaines propriétés en utilisant le concept raffinement de CSP.

Ensuite, nous avons présenté les fonctionnalités de l'outil Wr2fdr. Celui-ci est censé de traduire de Wright vers CSP en automatisant certaines propriétés standards définies par Wright. Enfin, nous avons vérifié l'outil Wr2fdr en suivant une activité de test fonctionnel orientée tests syntaxiques. Ainsi, nous avons constaté des défaillances lors du test de l'outil Wr2fdr. Ces défaillances concernent principalement le calcul des relations de raffinement liées aux propriétés 1, 2, 3 et 8 (voir chapitre 2 section 6). Vu l'importance de cet outil, nous avons contacté les auteurs de Wright - en passant par le professeur Jean-Pierre Giraudin de l'équipe SIGMA de LIG, Grenoble – expliqué les problèmes rencontrés, récupéré le code source afin de le corriger et le compléter.

Dans la suite de ce mémoire, nous allons décrire l'activité de maintenance corrective et évolutive que nous avons menée sur l'outil Wr2fdr.

# Chapitre 2
# Rétro-ingénierie de l'outil Wr2dfr

L'objet de ce chapitre est l'étude approfondie du code source de l'outil Wr2fdr écrit en C++. Dans un premier temps nous allons énumérer les caractéristiques techniques générales de l'outil Wr2fdr. Dans un deuxième temps nous allons extraire l'architecture à objets sous forme d'un diagramme de classes UML de l'outil Wr2fdr. Pour y parvenir, nous proposons et appliquons des règles de rétro-ingénierie de C++ vers UML. Dans un troisième temps, nous aborderons les techniques d'implémentation retenues par les auteurs de Wr2fdr. Enfin, nous allons évaluer l'outil Wr2fdr en tant que logiciel.

## 1. Caractéristiques techniques générales de l'outil Wr2fdr

L'outil Wr2fdr est écrit en C++. Son code source est réparti physiquement sur plusieurs fichiers : trois fichiers « .hpp » et huit fichiers « .cpp ». La complexité textuelle de l'outil Wr2fdr est de l'ordre de 16000 lignes C++. L'outil Wr2fdr englobe un analyseur lexico-syntaxique de Wright développé en utilisant les deux générateurs d'analyseurs lexicaux et syntaxique célèbres Lex et Yacc [**10**], [**12**] et [**17**]. Le fonctionnement général de l'outil Wr2fdr est traduit par une séquence d'opérations. Dans un premier temps, l'opération *parse_result* est exécutée afin d'analyser syntaxiquement le fichier d'entrée contenant une spécification Wright. En cas de succès, cette opération produit un arbre syntaxique abstrait (structure de données astNode voir section 2.2.1). En cas d'échec, des erreurs lexico-syntaxiques sont signalées. Dans un deuxième temps, l'opération *fdrprint* applicable sur un objet de type *AstNode* est exécutée afin de produire la traduction CSP correspondante.

## 2. Extraction de la partie statique de l'outil Wr2fdr

L'extraction de la partie statique sous forme d'un diagramme de classes UML à partir d'une implémentation C++ est une tâche délicate, assez difficile et notamment laborieuse. Pour y parvenir, nous avons appliqué des règles plus ou moins systématiques de transformation de C++ vers UML. Ainsi, nous avons pu extraire l'architecture à objets de l'outil Wr2fdr. Dans la suite, nous allons décrire les règles utilisées pour la transformation de C++ vers UML [**13**]. En outre, nous allons présenter le diagramme de classes UML extrait en appliquant ces règles.

### *2.1. Règles de rétro-ingénierie de C++ vers UML*

Dans la suite, nous allons proposer des règles de transformation de C++ vers UML illustrées par des exemples issus du code source de l'outil Wr2fdr.

**<u>Règle 1:</u>**

Un fichier d'entête d'extention « .hpp » contenant la déclaration de plusieurs (>1) classes est traduit par un package UML.

*Exemple :*

Le fichier *Wr2fdr.hpp* contient plusieurs classes telles que *AstNode*, *AstList*, *Name*, *Style*, *Configuration* et *Component* est traduit par un package UML.

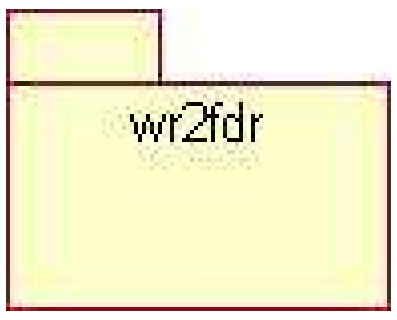

**<u>Règle 2:</u>**

Une classe C++ est traduite par une classe UML en tenant compte de son statut : classe abstraite ou effective.

*Exemple :*

```
class astNode {
  …
};
```

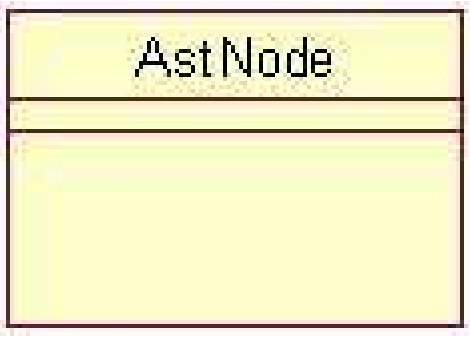

**<u>Règle 3:</u>**

Les attributs publics de type simple en C++ sont traduits par des attributs publics en UML de type équivalent. Egalement, les méthodes publiques en C++ sont traduites par des méthodes publiques en UML en tenant compte des aspects suivants : nature de la méthode (création, consultation, modification), nature logique des paramètres formels, typage des paramètres formels et le statut de la méthode (virtuelle ou effective). Nous utilisons les stéréotypes « constructor », « query » et « update » pour traduire respectivement en UML les méthodes de création, consultation et modification en C++. Egalement, nous utilisons le stéréotype « destructor » pour traduire une méthode de destruction en C++.

*Exemple :*

```
class binaryOp : public astNode {
  private:
    char       *parallel_set;
    void        CheckCalculations(void);
    void        SplitInteractSet(Set *long_set);
  public:
    binaryOp(int, astNode *,astNode *);
    virtual Set *    findValueAgain (void);
…
};
```

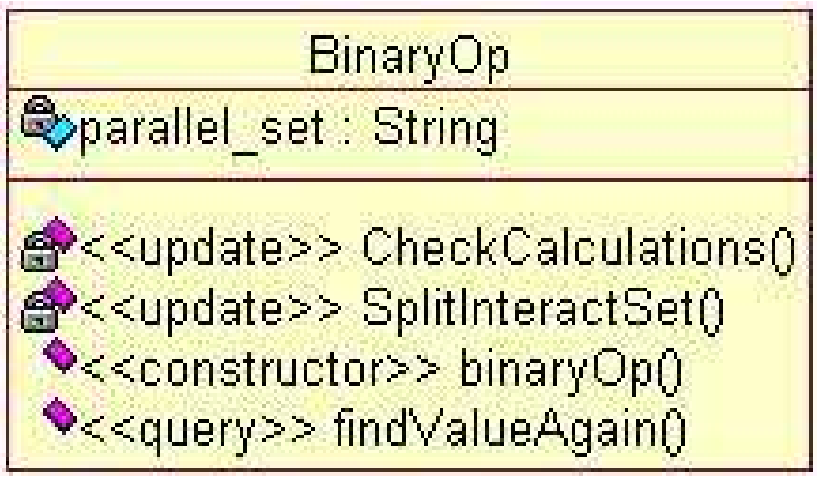

**Règle 4:**

Une relation client C++ matérialisée par un attribut public à base d'une autre classe est traduite par une association unidirectionnelle UML. La multiplicité de cette association (0..1, 0..n et p..n) et déduite de type de base de l'attribut concerné. Pour pouvoir la considérer comme agrégation ou composition UML il faut analyser davantage l'implantation.

*Exemple :*

```
class name {
  public:
....
};

class connector {
  public:
    name        *conn_name;
…
};
```

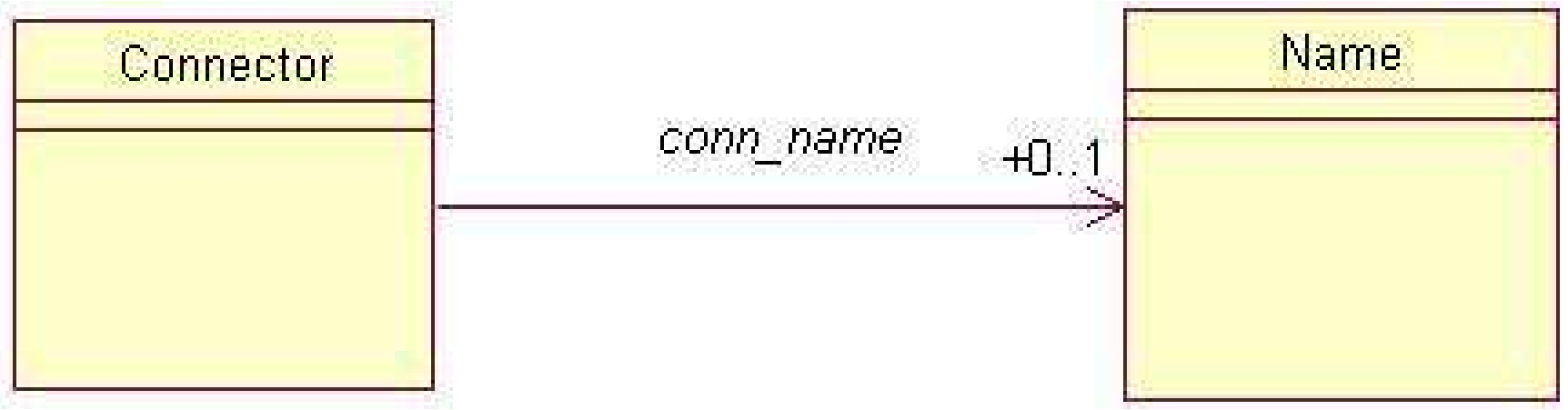

**Règle 5:**

Une relation d'héritage en C++ est traduite par une généralisation UML.

*Exemple :*

```
class component : public astNode {
  public:
…
};
```

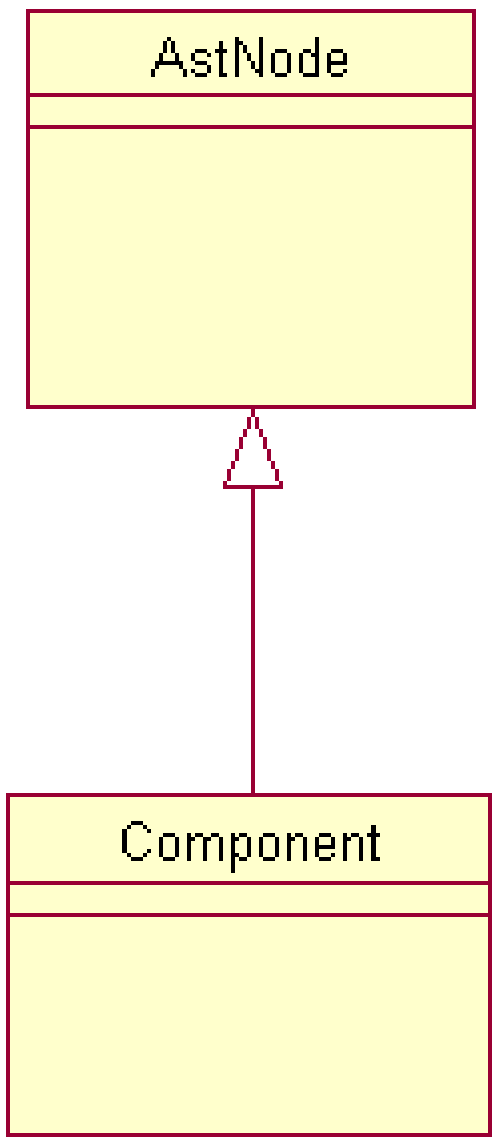

**Règle 6:**

Une classe abstraite en C++ est traduite par une classe UML abstraite.

*Exemple :*

```
Class astNode {
virtual void fdrprint(void);
…
}
```

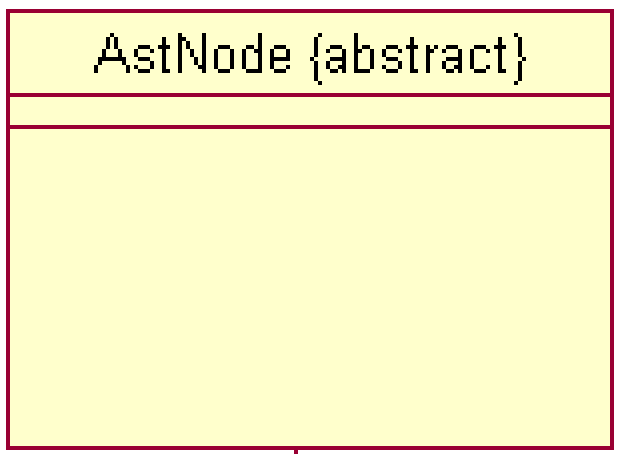

### *2.2. Diagramme de classe de l'outil Wr2fdr*

Dans cette section, nous allons appliquer plus au moins systématiquement les règles de retro-ingénierie de C++ vers UML proposée dans la section précédente afin d'extraire la partie statique de l'outil Wr2fdr sous forme d'un diagramme de classe UML. Pour y parvenir, nous allons suivre une démarche de rétro-conception comportant les deux étapes suivantes :

- Etape1 : Extraire les classes d'implémentation et de conception.
- Etape2 : Extraire les classes d'analyse modélisant les concepts métier de l'ADL Wright.

#### 2.2.1. *Classes d'organisation*

Ces classes ne sont pas issues directement du cahier des charges de l'outil Wr2fdr basé sur l'ADL Wright et FDR2. Elles facilitent l'implémentation efficace et modulaire des fonctionnalités souhaites de Wr2fdr (Voir chapitre 1 section 8.2).

**La classe AstNode**

Il s'agit plutôt d'une classe de conception. Elle permet de regrouper sous forme d'un arbre appelé arbre syntaxique abstrait les constructions syntaxiques offertes par l'ADL Wright telles que composant, connecteur, configuration, style (voir chapitre 1 section 1).

Elle utilise avec profit les services offerts par les classes d'implémentation fournies ci-dessous afin d'implémenter ses constituants.

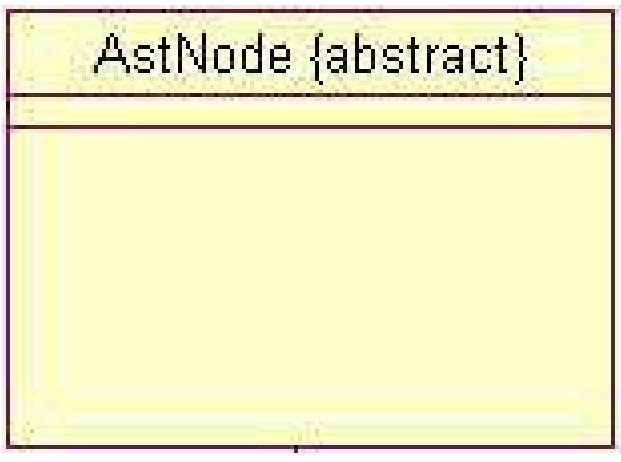

**Classe SList**

Elle modélise la cellule de base de la SD Liste linéaire. Les éléments stockés dans cette structure est de type *AstNode*. Le modèle UML issu de la classe *SList* en C++ est fourni par la figure 2.1. La classe *SList* est dédiée à la classe *Set*.

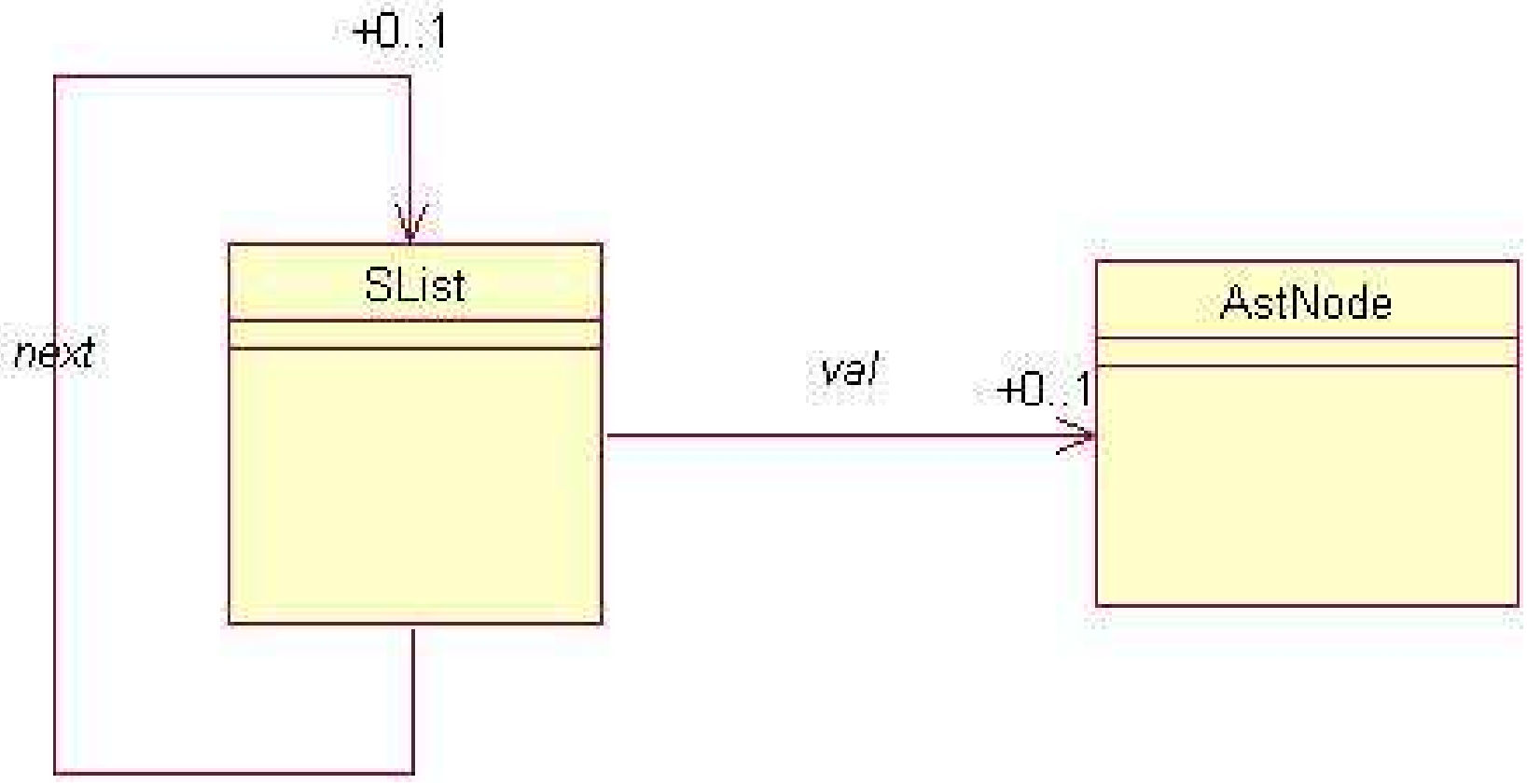

**Figure 2.1.** *Modèle UML issu de la classe SList en C++*

**Classe Set**

Cette classe implémente la structure de donnée *Ensemble* en utilisant la classe *SList*. Ainsi, Chaque élément de la classe *Set* est de type *SList* et par conséquent de type *AstNosde*.

La classe *Set* supporte la notion de curseur permettant de connaître l'élément courant et d'installer un itérateur sur une collection de type *Set*. En outre, la classe *Set* offre la plupart des opérations applicables sur la SD Ensemble telles que : *union*, *intersection*, *ajouter*, *supprimer*, *appartenance*, *cardinalité* et les différents parties d'un ensemble (PowerSet). Le modèle UML issu de la classe *Set* en C++ est fourni par la figure 2.2.

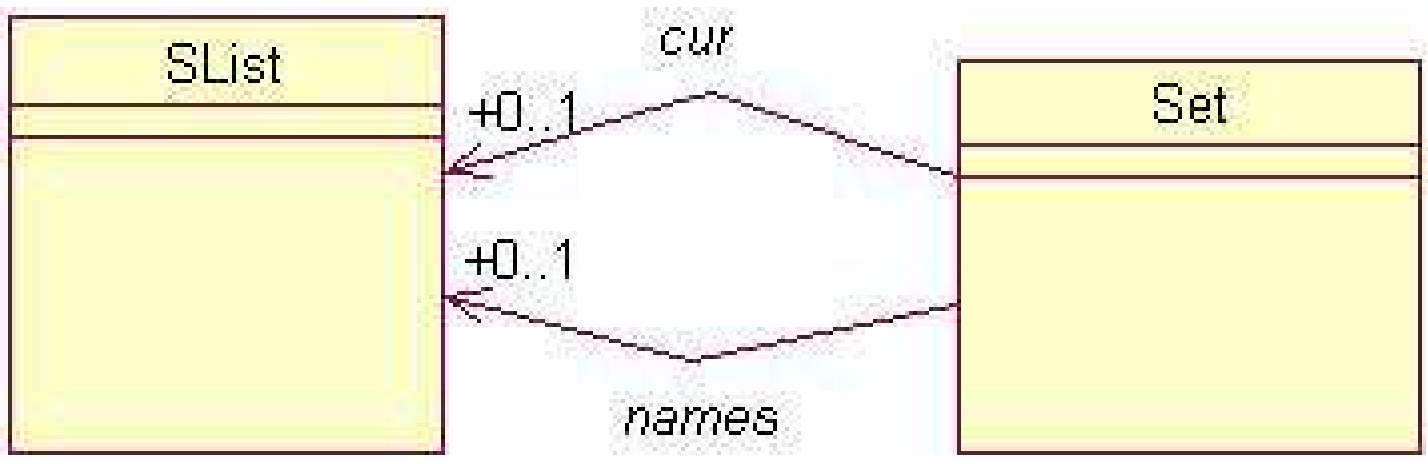

**Figure 2.2.** *Modèle UML issu de la classe Set en C++*

**Classes SymEntry et Relation**

Ces deux classes implémentent le concept relation au sens mathématique dont l'ensemble de départ et d'arrivée est *AstNode*. La classe *SymEntry* propose une implémentation sous forme d'une liste linéaire du concept relation. Chaque élément de cette liste est un couple dont le premier élément est de type *AstNode* et le second est de types *Set*. Tandis que la classe *Relation* détient une entrée de type *SymEntry* et offre la plupart des opérations applicables sur le concept relation au sens mathématique telles que : union, composition séquentielle, fermeture transitive, image relationnelle, ajouter et supprimer. Le modèle UML issu de deux classe *SymEntry* et *Relation* en C++ est fourni par la figure 2.3.

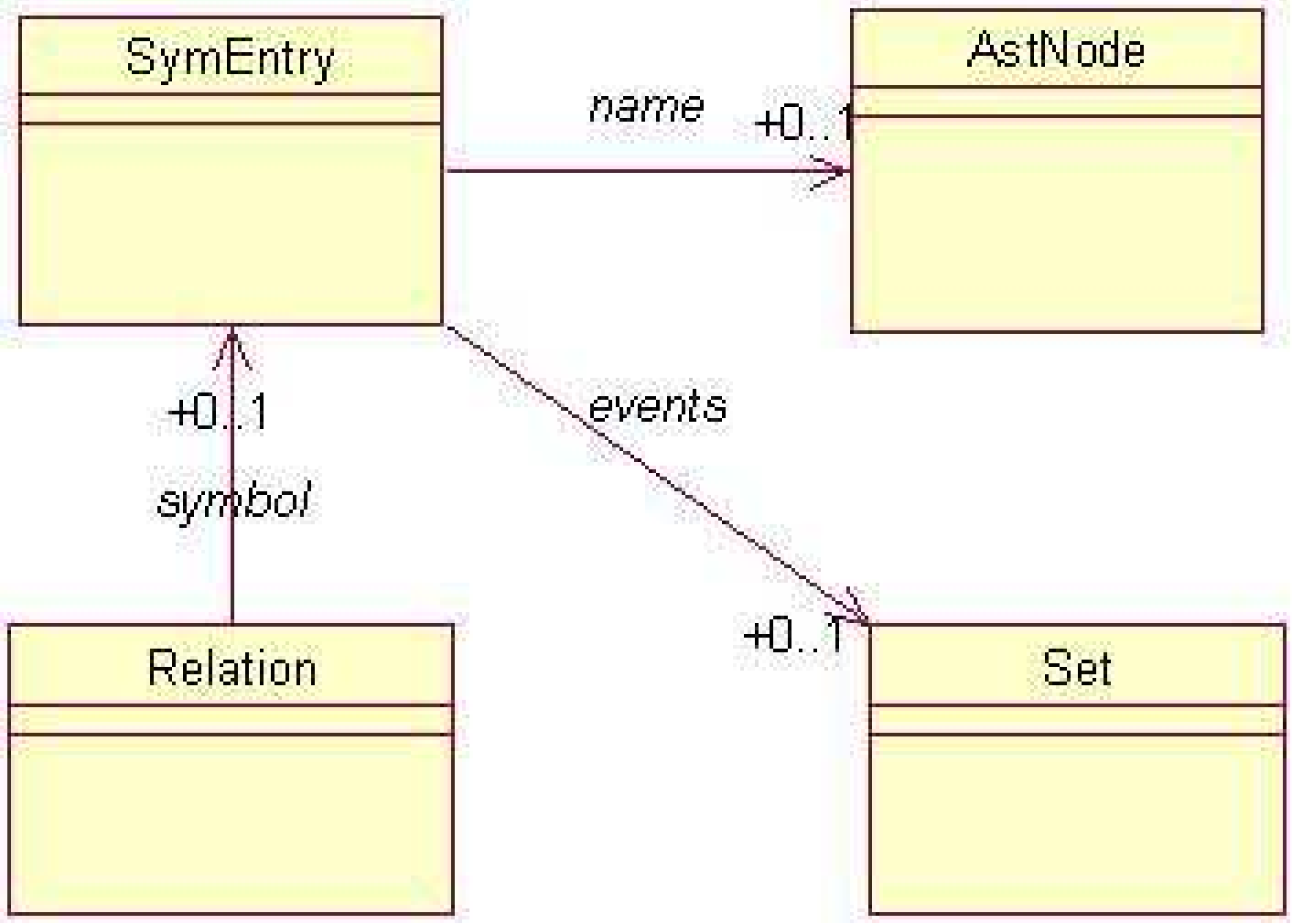

**Figure 2.3.** *Modèle UML issu de deux classes SymEntry et Relation*

**<u>Classe LookUpTable</u>**

Elle permet d'implémenter le concept mathématique fonction partielle dont l'ensemble de départ et d'arrivée est *AstNode*. Une fonction partielle bâtie sur *AstNode* est assimilée à une liste linéaire de maplets dont le premier élément et le second est de type *AstNode*.

La classe *LookUpTable* offre un service permettant de parcourir les couples de fonction partielle et un autre permettant de connaître l'image d'un élément donné. Le modèle UML issu de la classe *LookUpTable* en C++ est fourni par la figure 2.4.

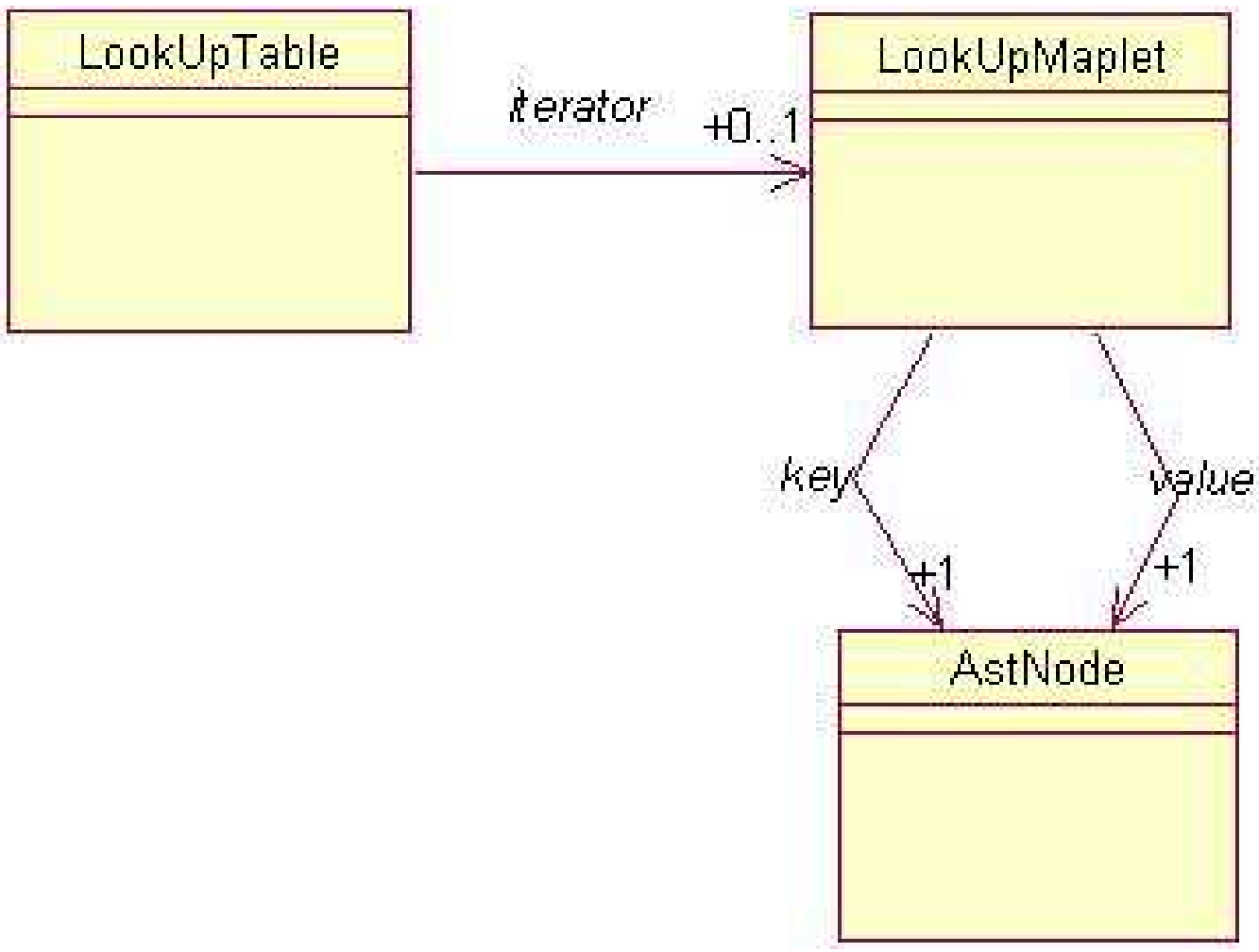

**Figure 2.4.** *Modèle UML issu de la classe LookUpTable en C++*

**Classe AstList**

Elle descend de la classe abstraite *AstNode*. Elle permet de regrouper sous forme d'une liste linéaire des objets appartenant à des classes qui décrivent directement ou indirectement de la classe *AstNode*. La classe *AstList* incarne une liste linéaire polymorphe et offre des opérations standards sur les listes linéaires telles que ajouter, inverser, suivant, copier, et des opérations spécifiques liées à la manipulation des constructions syntaxiques de l'ADL Wright. Le modèle UML issu de la classe *AstList* en C++ est fourni par la figure 2.5.

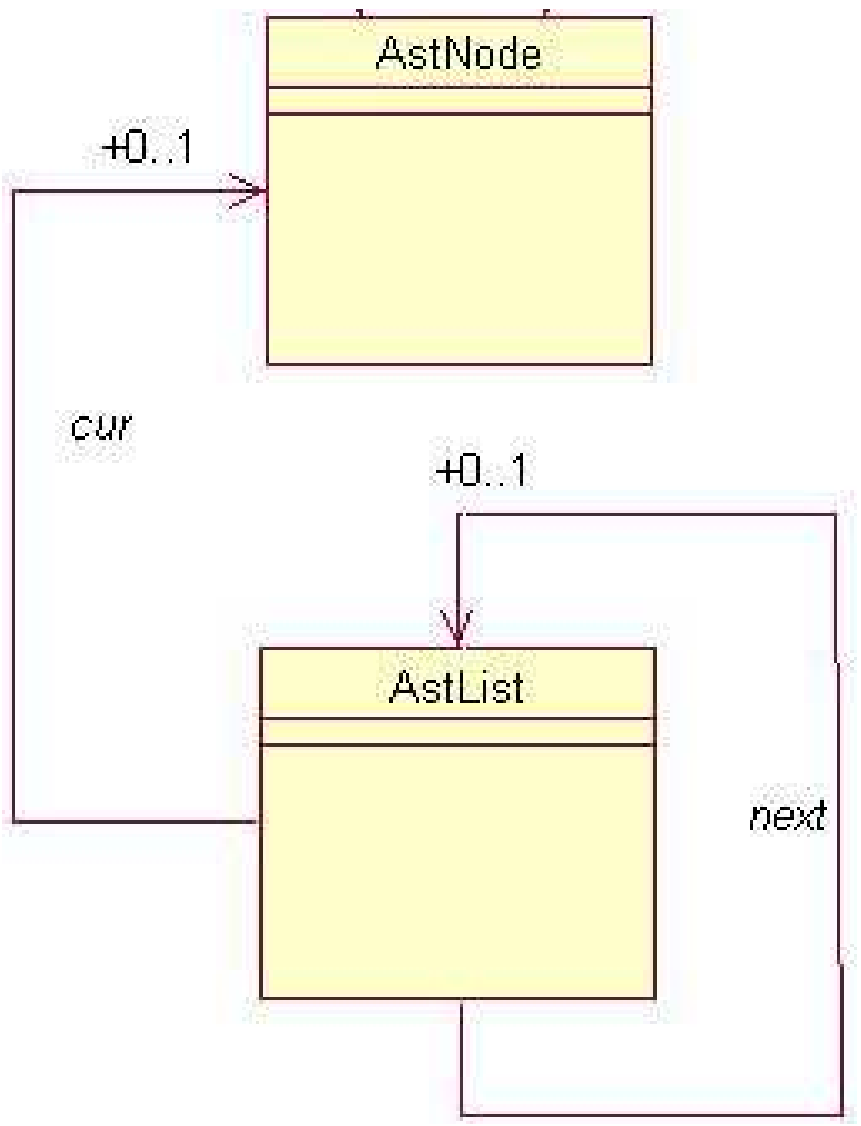

**Figure 2.5.** *Modèle UML issu de la classe AstList en C++*

**La classe Name**

Il s'agit plutôt d'une classe de conception. Elle permet de mémoriser l'identification d'un concept Wright et ses paramètres éventuels. En outre, la classe *Name* offre des services standards et spécifiques permettant de gérer les aspects liés à l'identification d'un concept Wright.

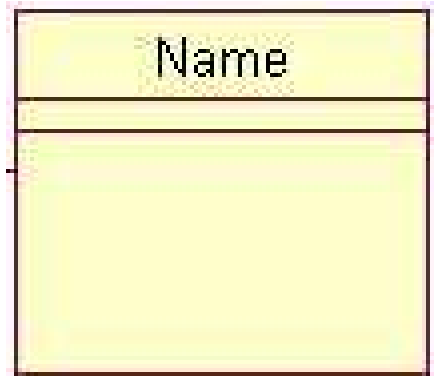

2.2.2. *Classes d'analyse*

Les classes d'analyse modélisent des concepts métier issus de l'ADL Wright. De tels concepts traduisent aussi bien des aspects structuraux que comportementaux.

2.2.2.1. Concepts structuraux

Ils concernent les constructions structurelles offertes par Wright : *Component*, *Connector*, *Configuration* et *Style*. Ces concepts sont modélisés par des classes en C++ qui dérivent directement d'*AstNode* (voir Figure 2.6).

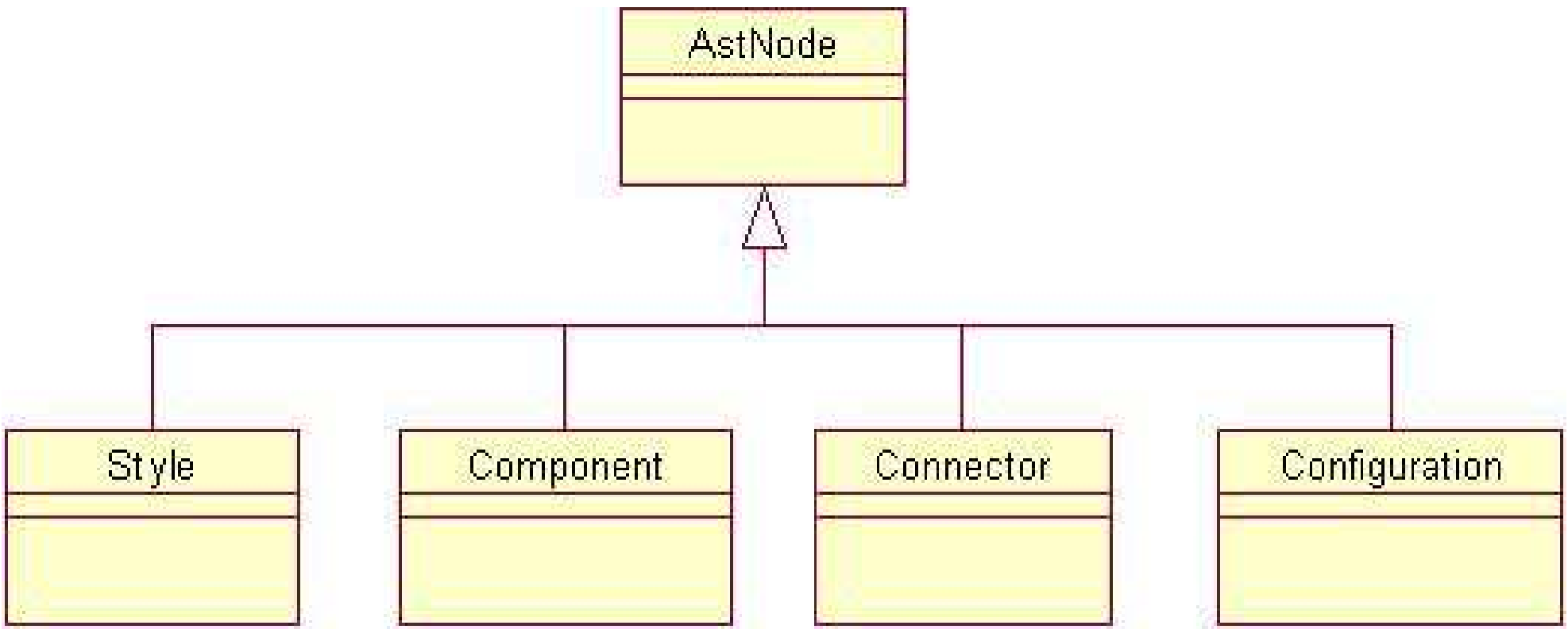

**Figure 2.6.** *Modélisation des concepts structuraux de Wright*

**<u>Classe Component</u>**

La classe *Component* apporte les attributs suivants : *comp_name*, *params*, *ports* et *computation*. Ces attributs mémorisent respectivement des références sur des objets de type *Name*, *AstList* et *AstNode* (voir Figure 2.7).

De plus, Component fournit des implémentations aux méthodes abstraites *copy*, *eq*, *wrprint*, *fdrprint*, *calculationPass*, *postCalculationPass* et *SplitLongSets* venant de la classe ascendante *AstNode*.

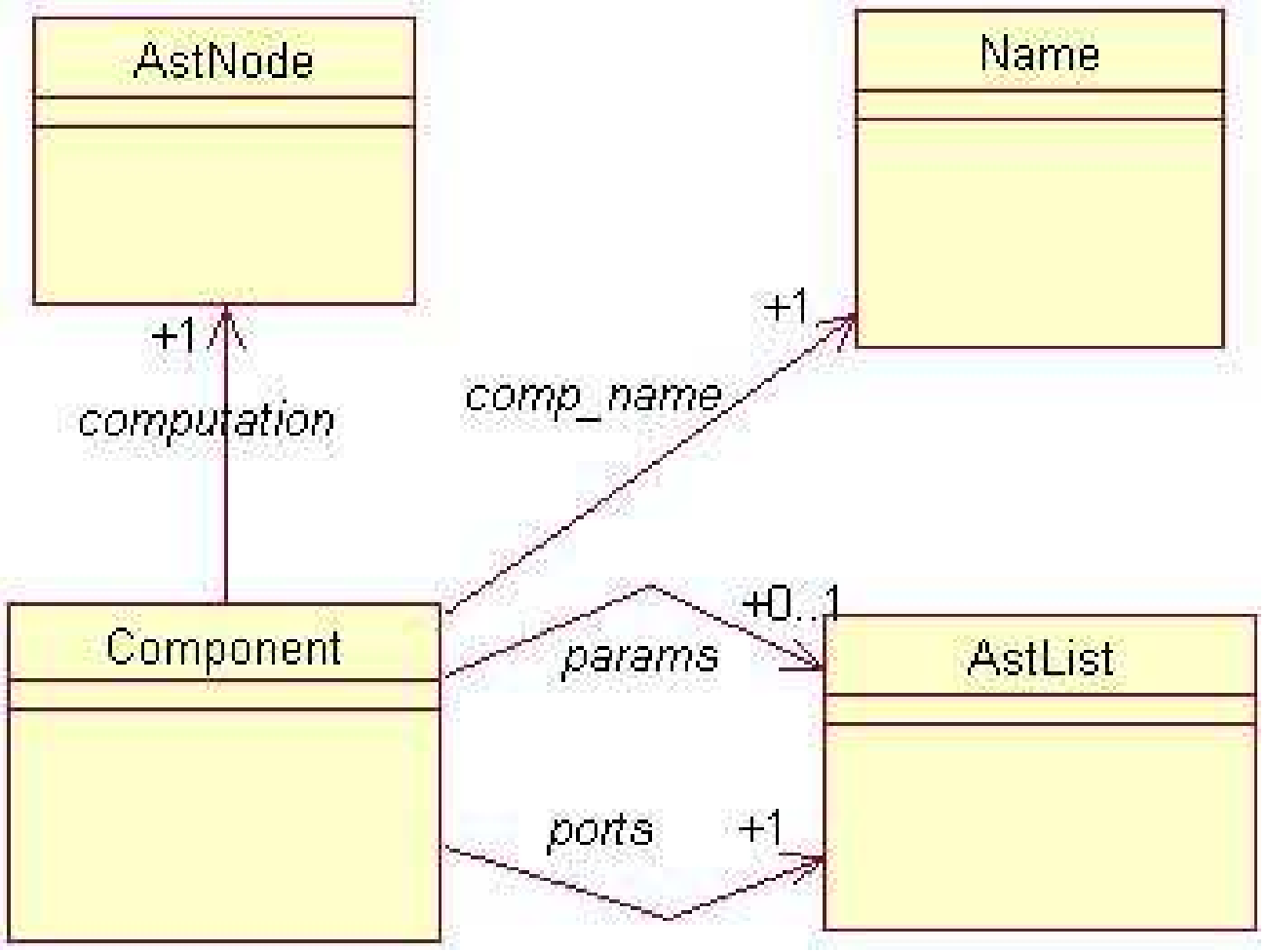

**Figure 2.7.** *Classe Component*

**Classe Connector**

Elle apporte des nouveaux attributs comme *name*, *params*, *roles* et *glue* mémorisant respectivement des références de type *Name*, *AstList*, *AstList* et *Declaration* (voir figure 2.8).

De plus, cette classe propose des implémentations des méthodes telles que *copy*, *eq*, *wrprint*, *fdrprint*, *CalculationPass*, *PostCalculationPass* venant de la classe ascendante *AstNode*.

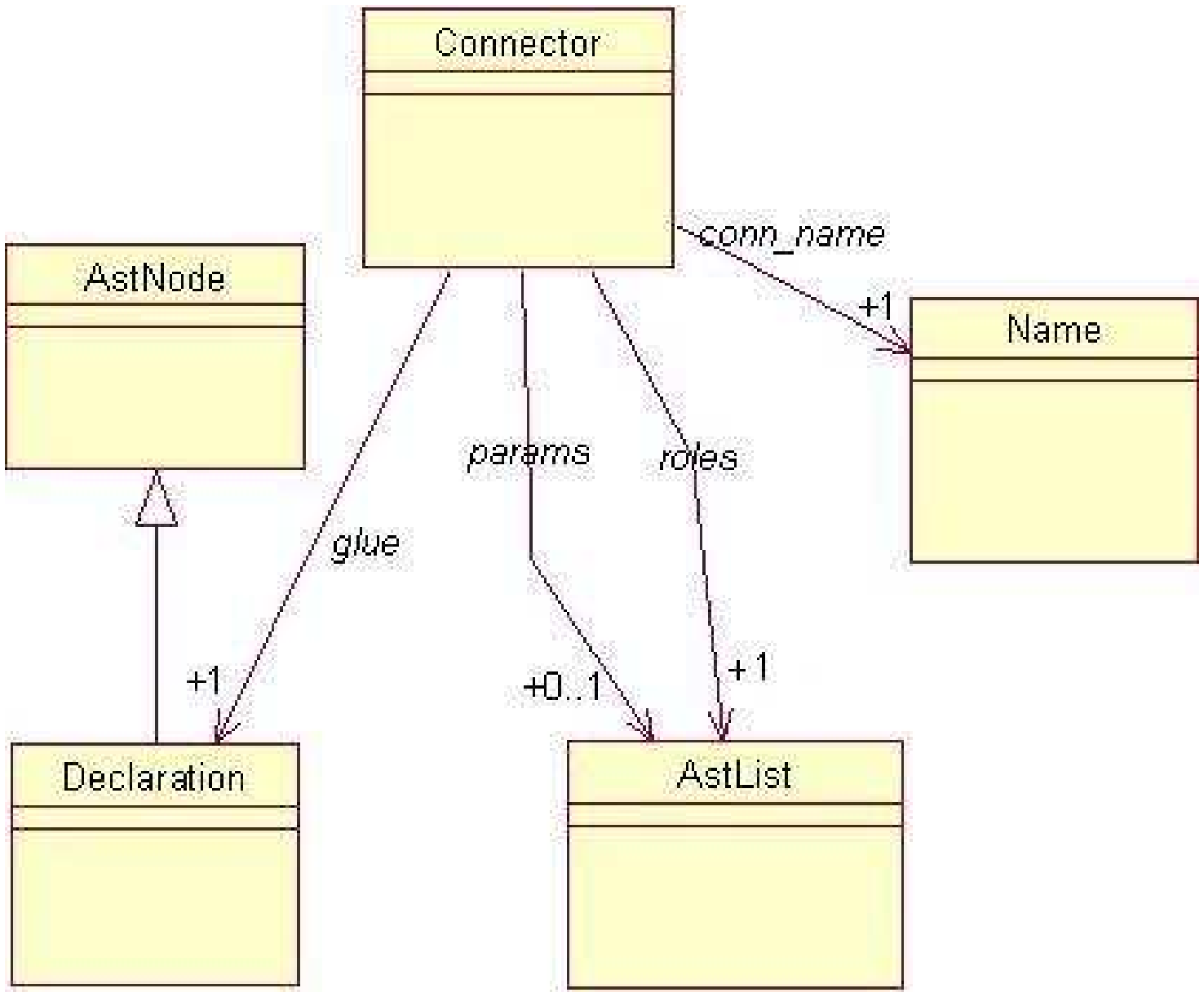

**Figure 2.8.** *Classe Connector*

**Classe Style**

Elle apporte trois nouveaux attributs : *style_name*, *types* et *constraint* mémorisant des références sur des objets de type *Name*, *AstList* et *AstNode* (voir figure 2.9).

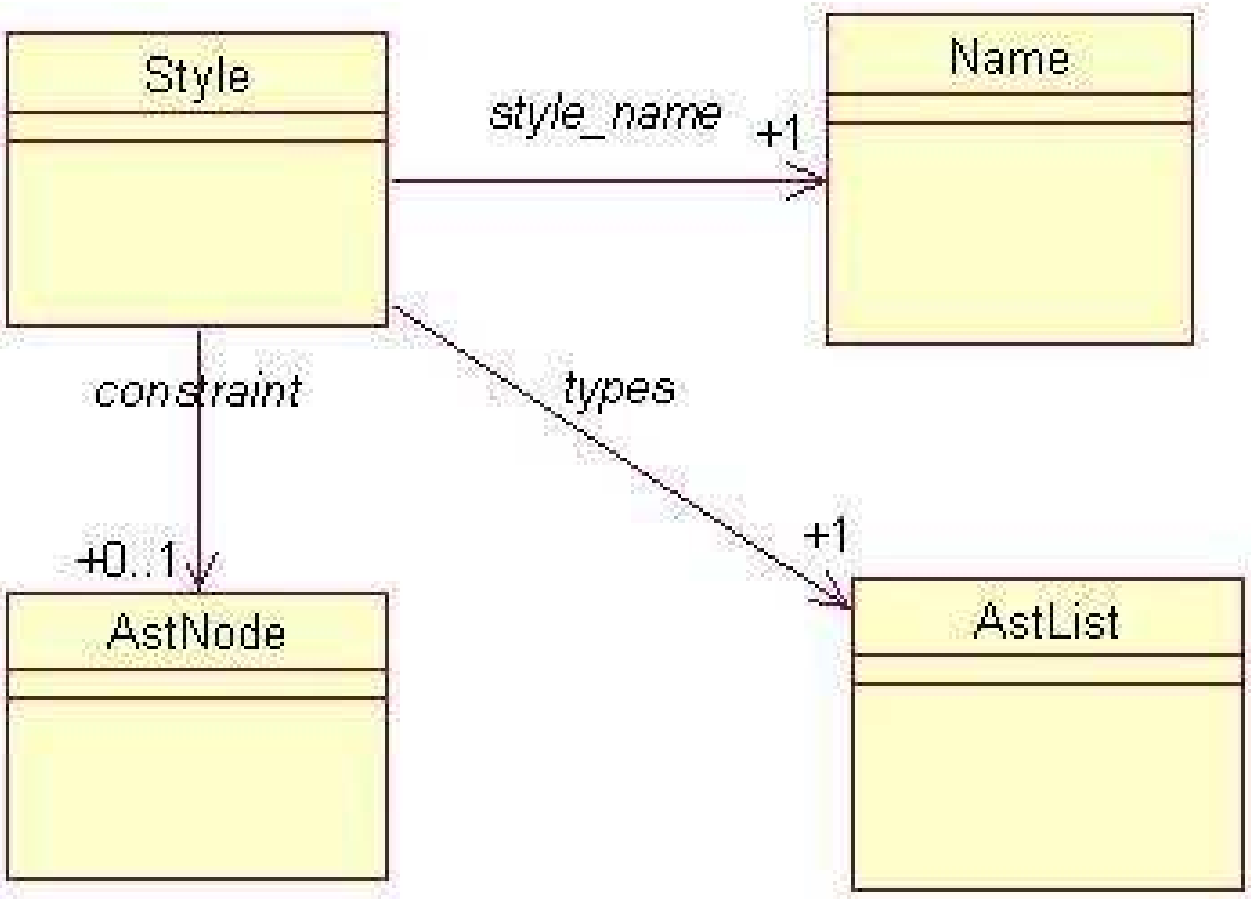

**Figure 2.9.** *Classe Style*

**<u>Classe Configuration</u>**

Elle apporte de nouveaux attributs *conf_name*, *style_name*, *types*, *instances*, *attachments* mémorisant des références sur des objets respectivement de type *Name*, *Name*, *AstList* et *AsList* (voir figure 2.10)

En outre, elle implémente les méthodes abstraites telles que *copy*, *eq*, *wrprint*, *fdrprint*, *CalculationPass*, *PostCalculationPass*. Venant de la classe ascendante AstNode.

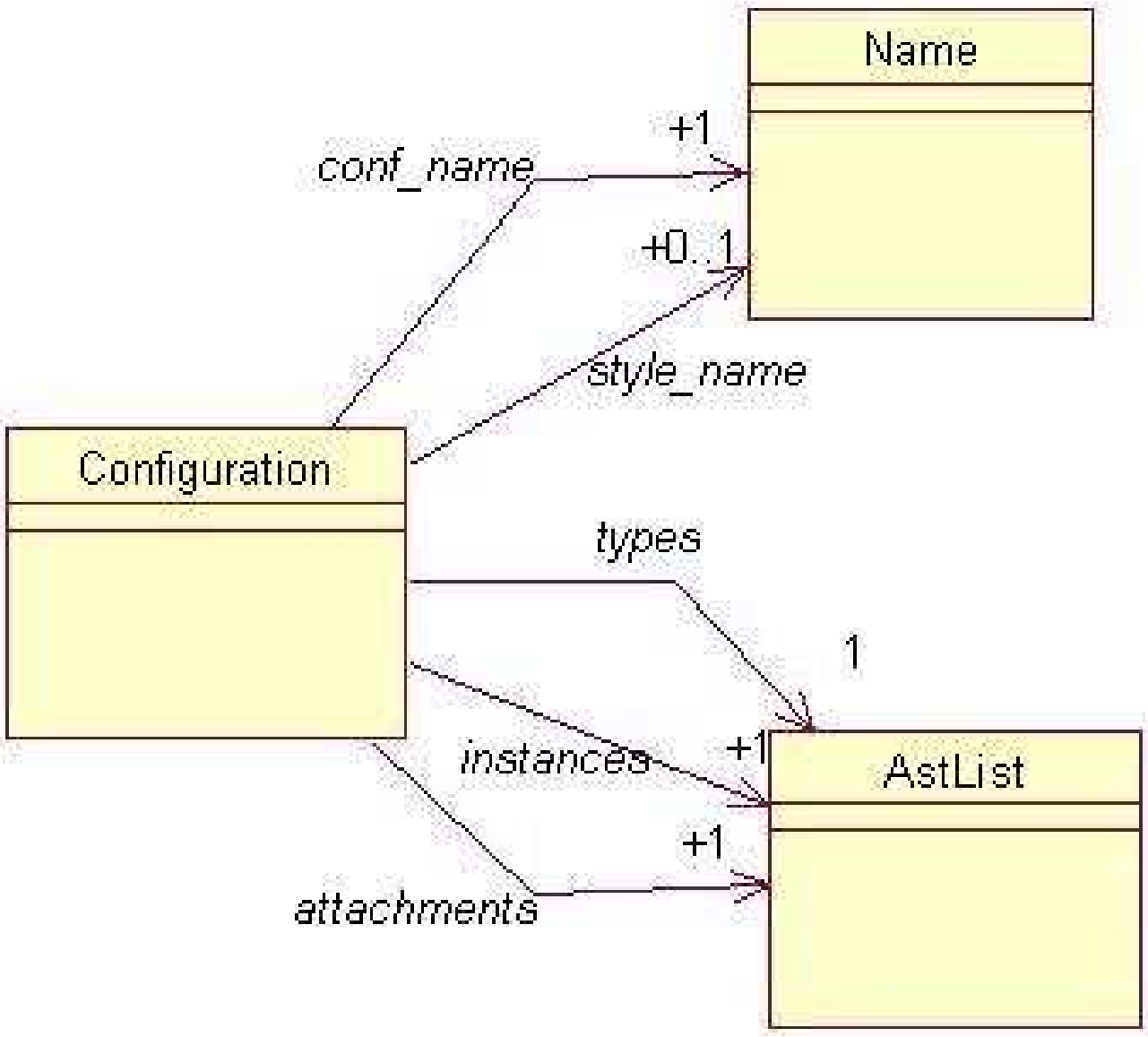

**Figure 2.10.** *Classe Configuration*

La figure 2.11 récapitule sous forme d'un diagramme de classe UML les différents fragments liés à la modélisation des concepts structuraux de l'ADL Wright.

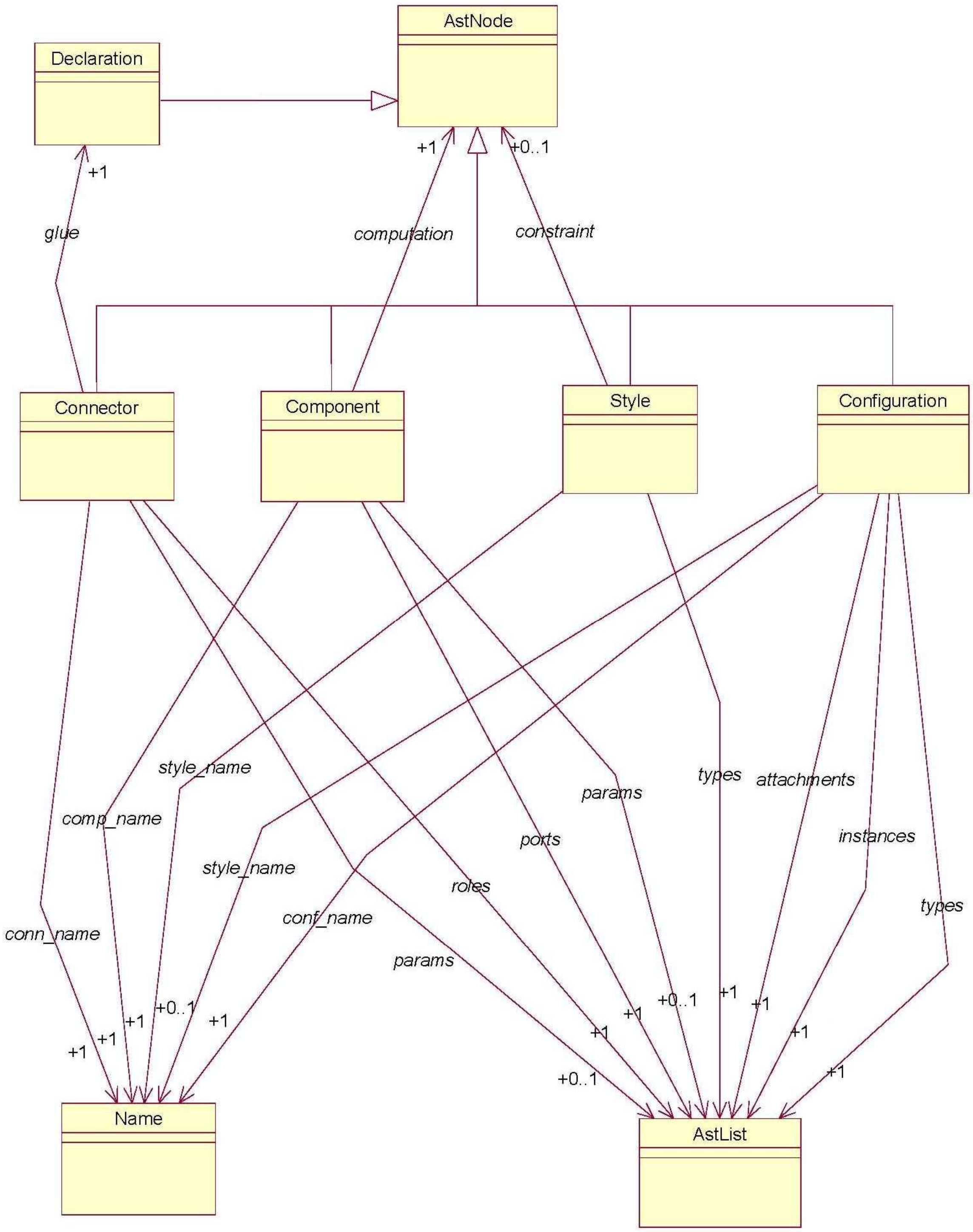

**Figure 2.11.** *Concepts structuraux de Wright en UML*

2.2.2.2. Concepts comportementaux

Ils concernent essentiellement les deux concepts événement et processus supportés par CSP de Wright. Toutes les notions relatives à ces deux concepts sont assimilées à des classes qui descendent de la classe *AstNode* (voir Figure 2.12).

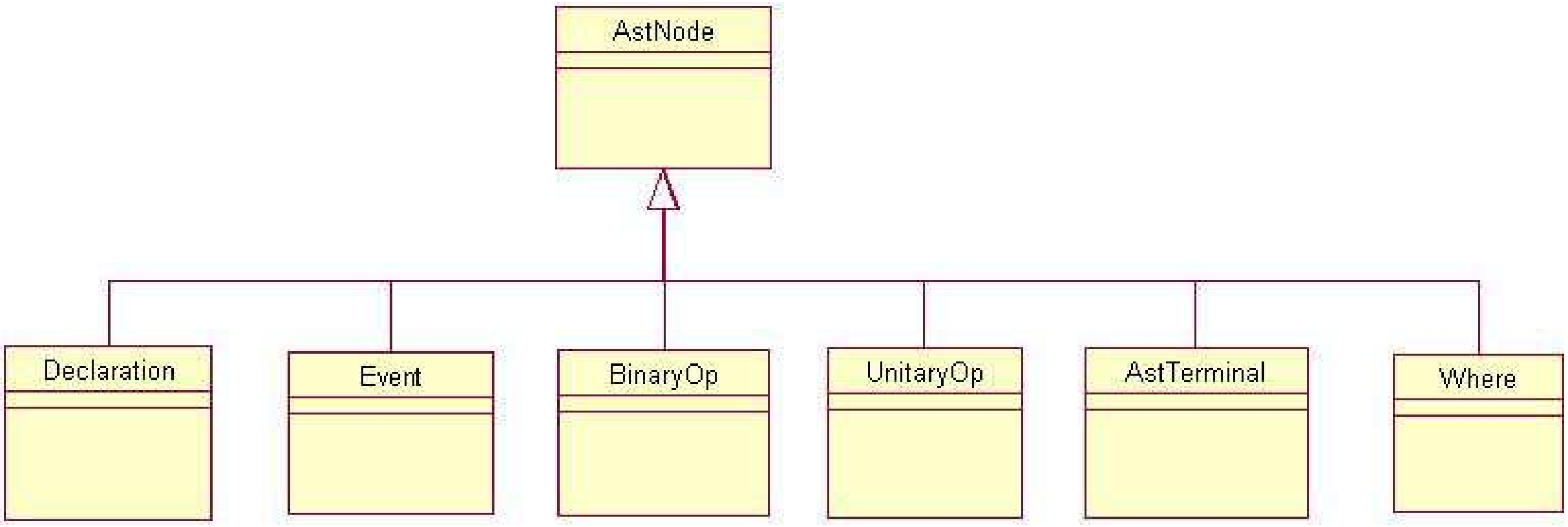

**Figure 2.12.** *Concepts Comportementaux de Wright*

## Classe Declaration

Elle modélise le concept processus de Wright. Cette classe apporte deux attributs principaux *n* et *defn* mémorisant des objets de type *AstNode*. Ces deux objets sont liés à l'identité et à la définition du processus concerné (voir Figure 2.13).

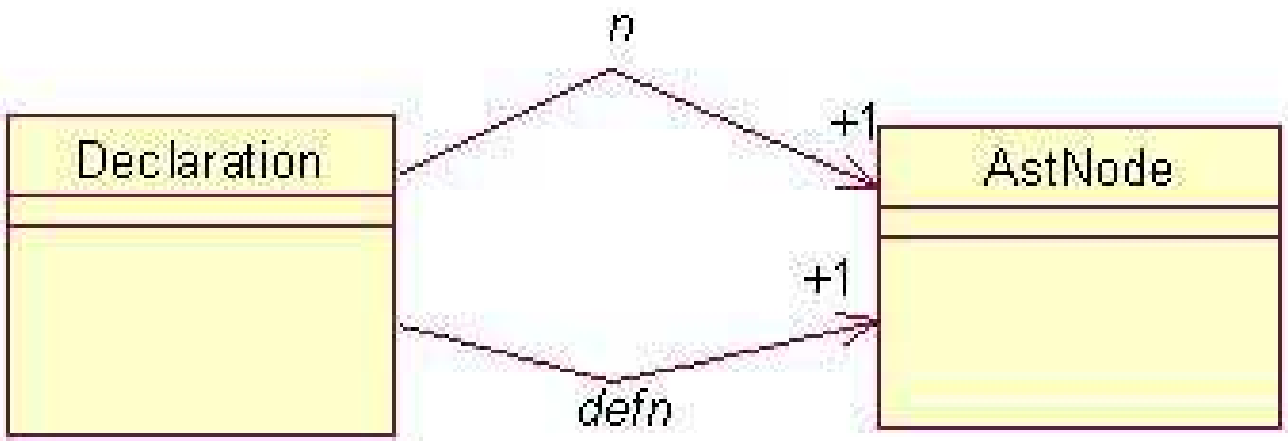

**Figure 2.13.** *Classe Declaration*

## Classe Event

Elle modélise le concept Evénement (voir chapitre 1 section 2.1). Elle apporte un attribut principal mémorisant un objet de type *Name* (voir figure 2.14). En outre, elle apporte un service *AddPrefixToEvent* permettant d'associer l'événement au processus concerné.

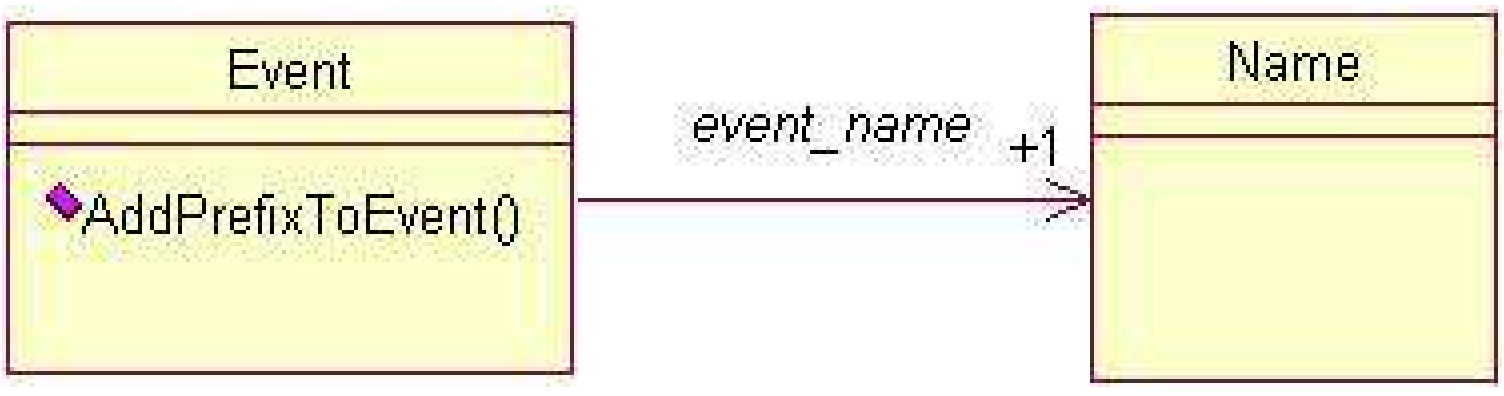

**Figure 2.14.** *Classe Event*

**Classe BinaryOp**

Elle modélise les opérateurs binaires CSP tels que : Choix déterministes (□), choix non déterministe (⊓) et préfixage (→). Elle apporte deux attributs *first* et *second* mémorisant des objets de type *AstNode*. Les deux attributs modélisent respectivement les deux opérandes gauche et droite de l'opérateur binaire concerné (voir figure 2.15).

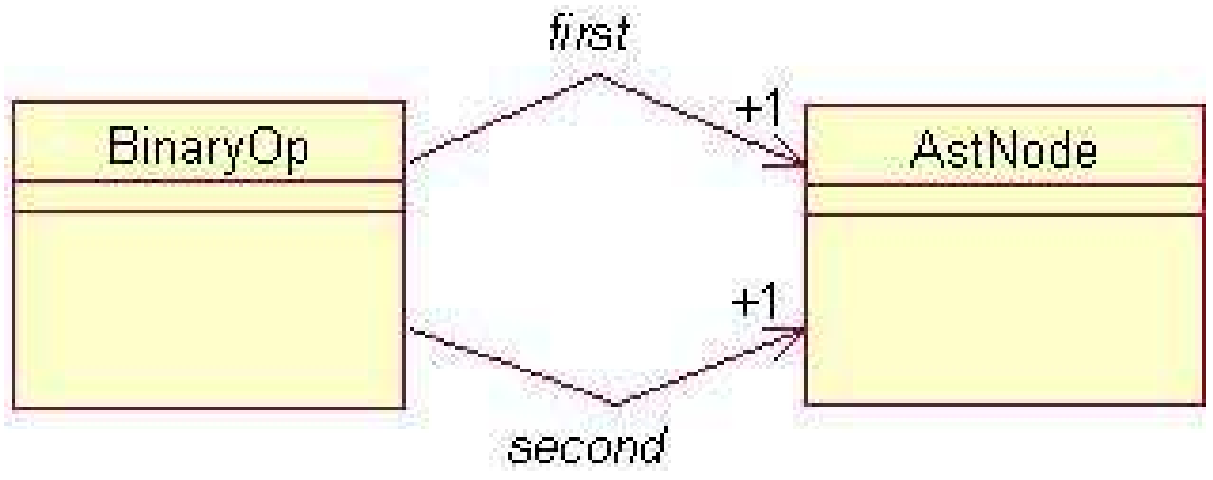

**Figure 2.15.** *Classe BinaryOp*

**Classe UnitaryOp**

Elle modélise la notion d'événement initialisé (voir chapitre 1 section 2.1). Elle apporte un attribut *first* de type *AstNode* mémorisant un événement perçu comme initialisé (voir Figure 2.16).

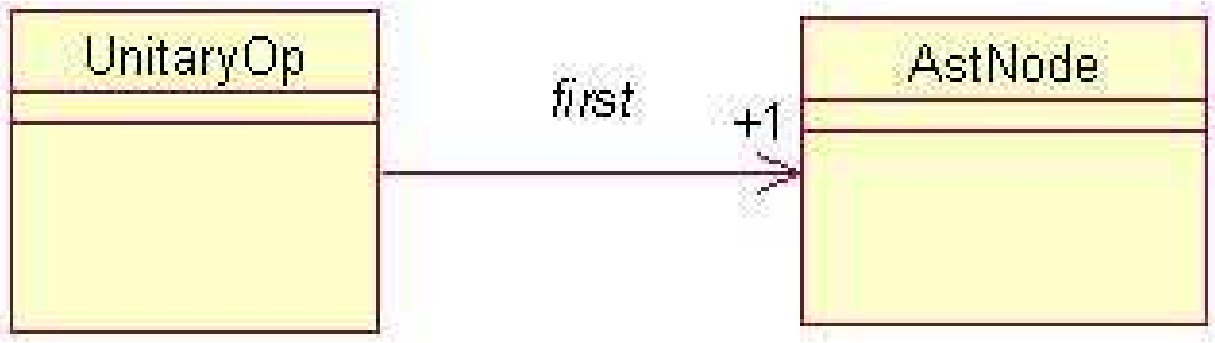

**Figure 2.16.** *Classe UnitaryOp*

**Classe AstTerminal**

Cette classe représente les terminaux dans l'ADL Wright comme le processus *TICK* (√→STOP) indiquant une bonne terminaison du processus.

**Classe Where**

Elle modélise la construction Where de CSP de Hoare. Celle-ci permet de nommer un sous-processus CSP. Elle apporte deux attributs mémorisant deux objets respectivement de type *AstNode* et *AstList* (voir Figure 2.17).

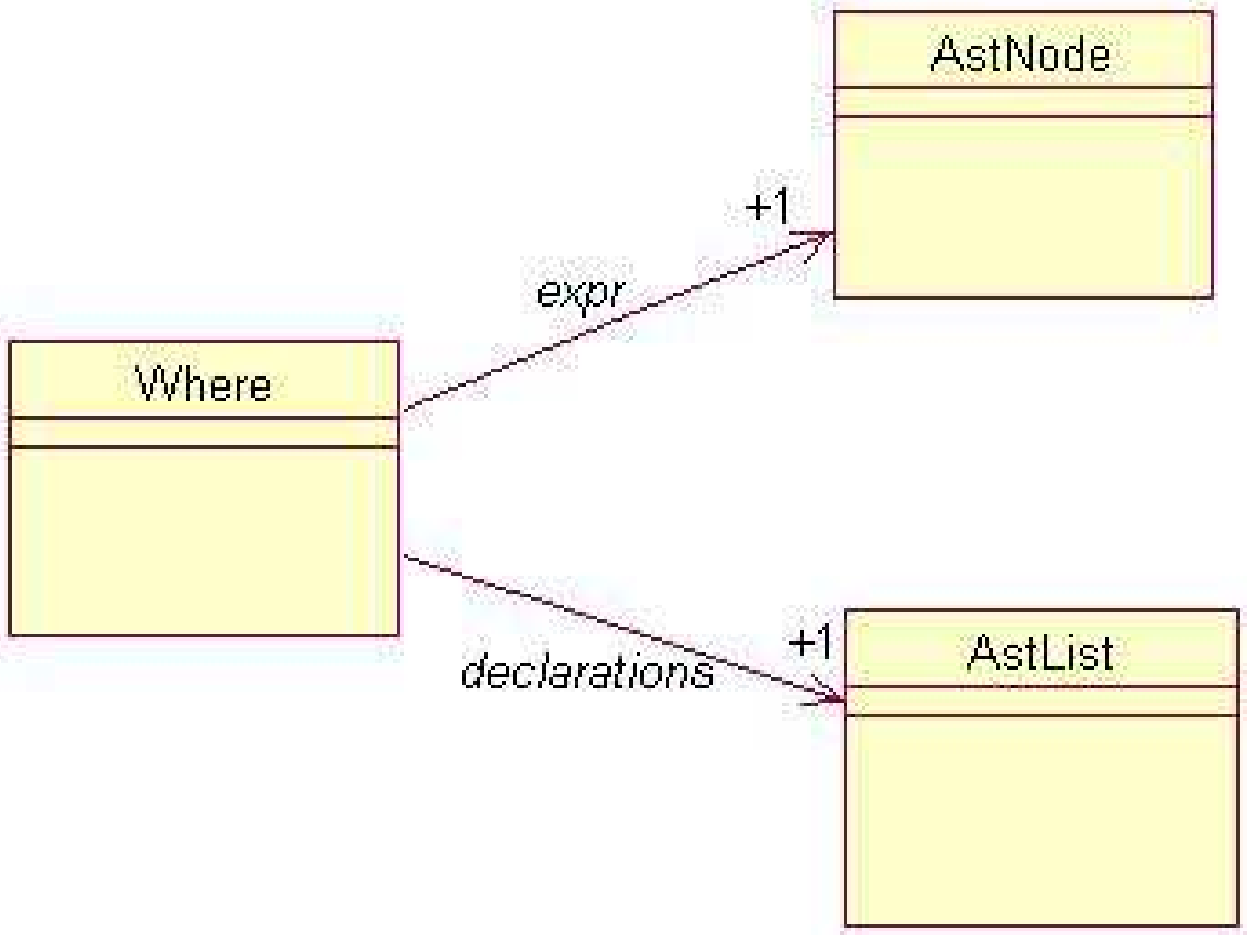

**Figure 2.17.** *Classe Where*

La figure 2.18 récapitule sous-forme d'un diagramme de classe UML les différents fragments liés à la modélisation des concepts comportementaux de l'ADL Wright.

La figure 2.19 regroupe sous forme d'un diagramme de classe UML les aspects structuraux et comportementaux de l'ADL Wright issus du code source de Wr2fdr écrit en C++.

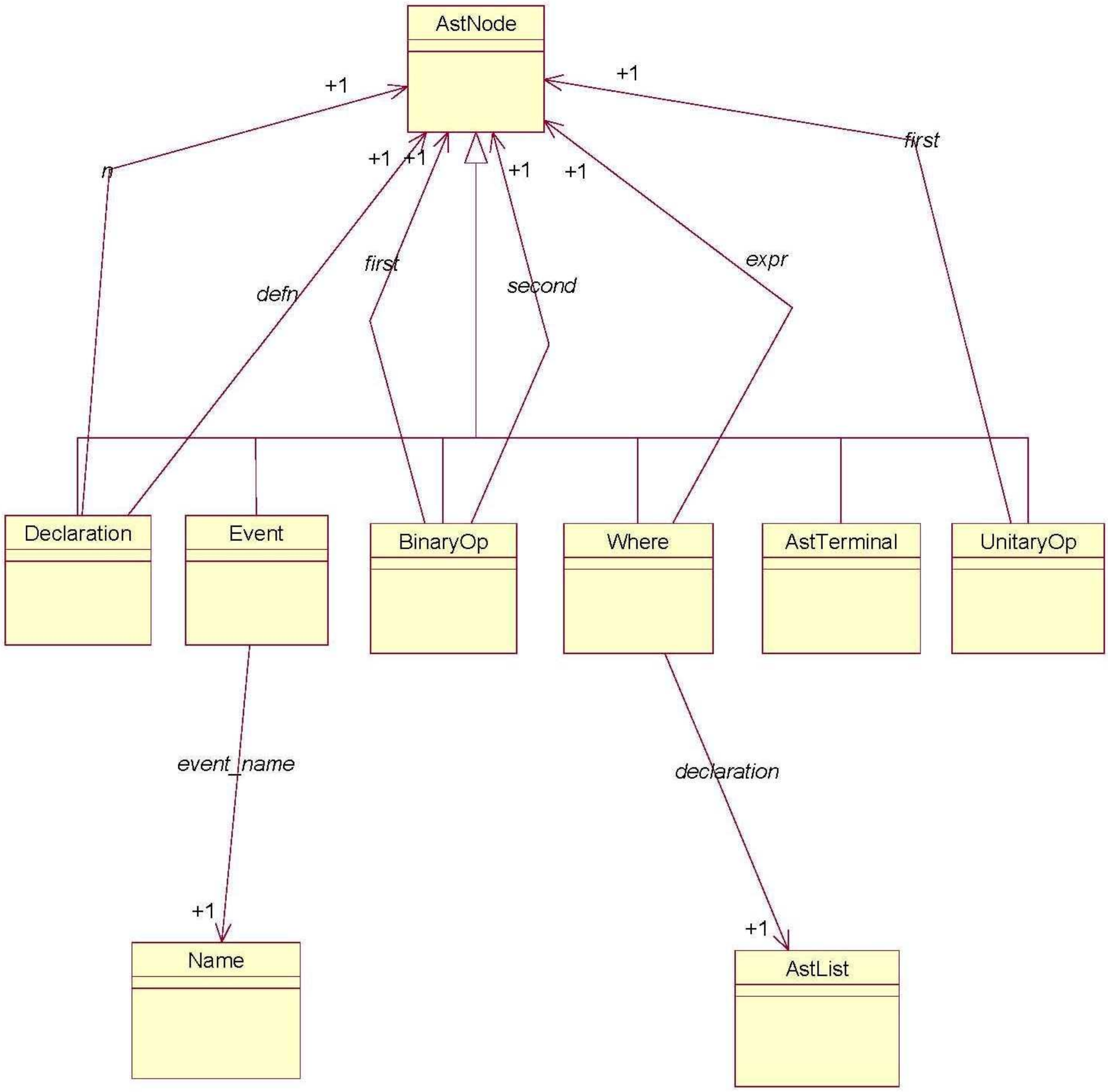

**Figure 2.18.** *Concepts Comportementaux de l'ADL Wright*

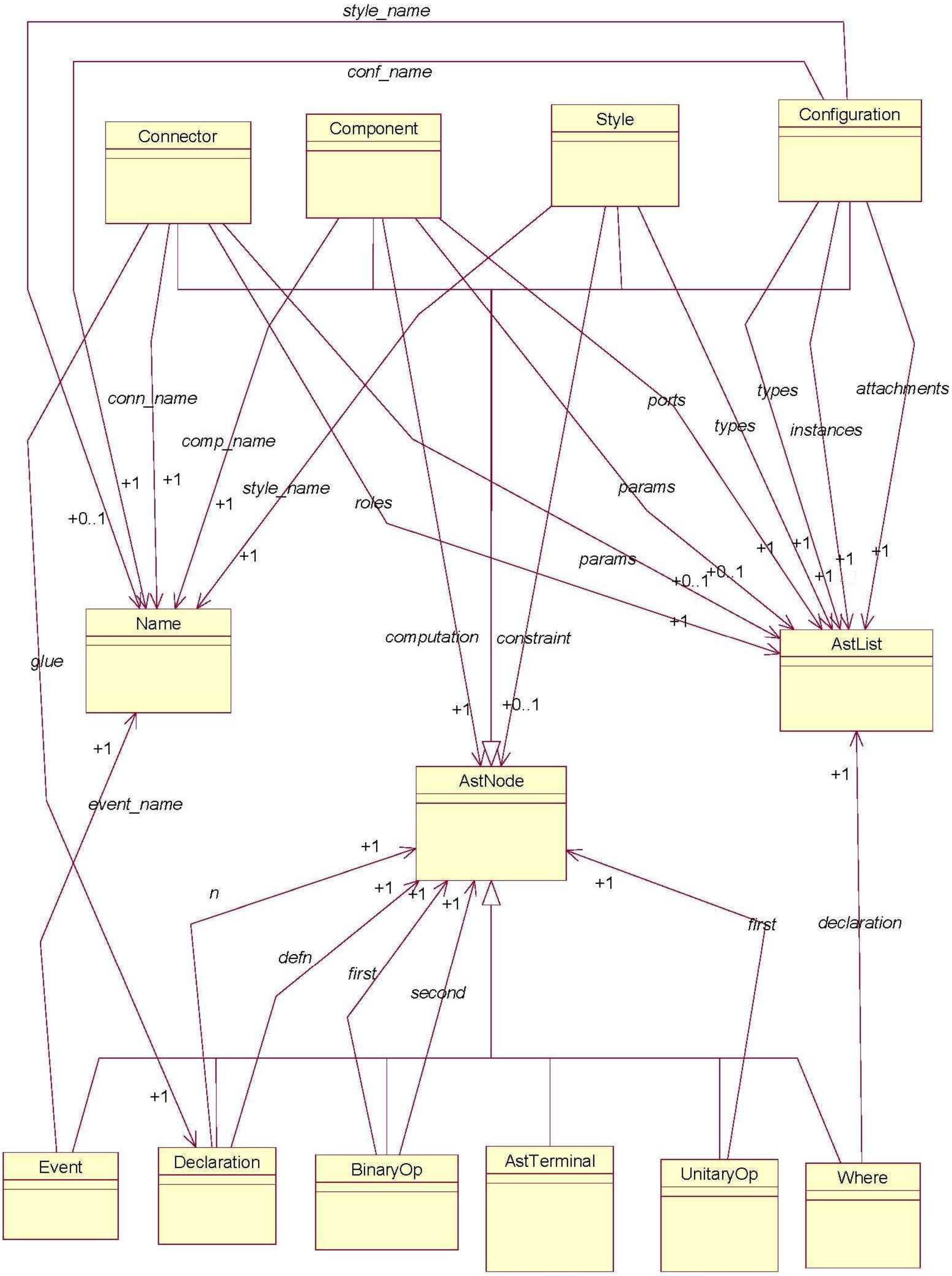

**Figure 2.19.** *Concepts structuraux et comportementaux de l'ADL Wright issus de Wr2fdr*

## 3. Techniques d'implémentation

### *3.1. Analyseur lexico-syntaxique de l'ADL Wright*

L'outil Wr2fdr intègre un analyseur lexico-syntaxique d'une architecture logicielle décrite en Wright. Un tel analyseur est construit en réutilisant les deux générateurs célèbres Lex et Yacc [**10**], [**12**] et [**17**]. Les expressions régulières décrivant les constructions lexicales soumises à Lex sont regroupées dans le fichier *wr2fdr.l*. Tandis que les règles de production relatives à la grammaire de Wright sont regroupées dans le fichier *wr2fdr.y*. Celui-ci est soumis à Yacc. L'analyseur syntaxique produit par Yacc relatif au fichier *wr2fdr.y* perçoit l'analyseur lexical produit par Lex relatif au fichier *wr2fdr.l* comme un sous-programme renvoyant l'unité lexicale courante *int yytext (void).*

En cas de succès, l'analyseur lexico-syntaxique de Wr2fdr produit l'arbre syntaxique abstrait matérialisé par la classe *AstNode* et ses classes descendantes telles que *Component, Connector, Name, Configuration, Style, Event, Declaration, Where, BinaryOp* (voir figure 2.19).

### *3.2. Structures des données fondamentales*

L'outil Wr2fdr utilise des structures de données classiques telles que *Set* et *AstList* (voir section 2.2.1). Ces deux classes permettent de regrouper des objets à base d'*AstNode*. En outre, la classe *AstNode* et les classes qui descendent de celle-ci telles que *Component*, *Name*, *AstTerminal, Event*, *BinaryOp* modélisent la structure de données Arbre selon l'approche par objets. Par exemple, le processus CSP associé au port *OutPut* :

OutPut = _a → OutPut |~| Tick est traduit par l'arbre syntaxique abstrait fourni par la figure 2.20.

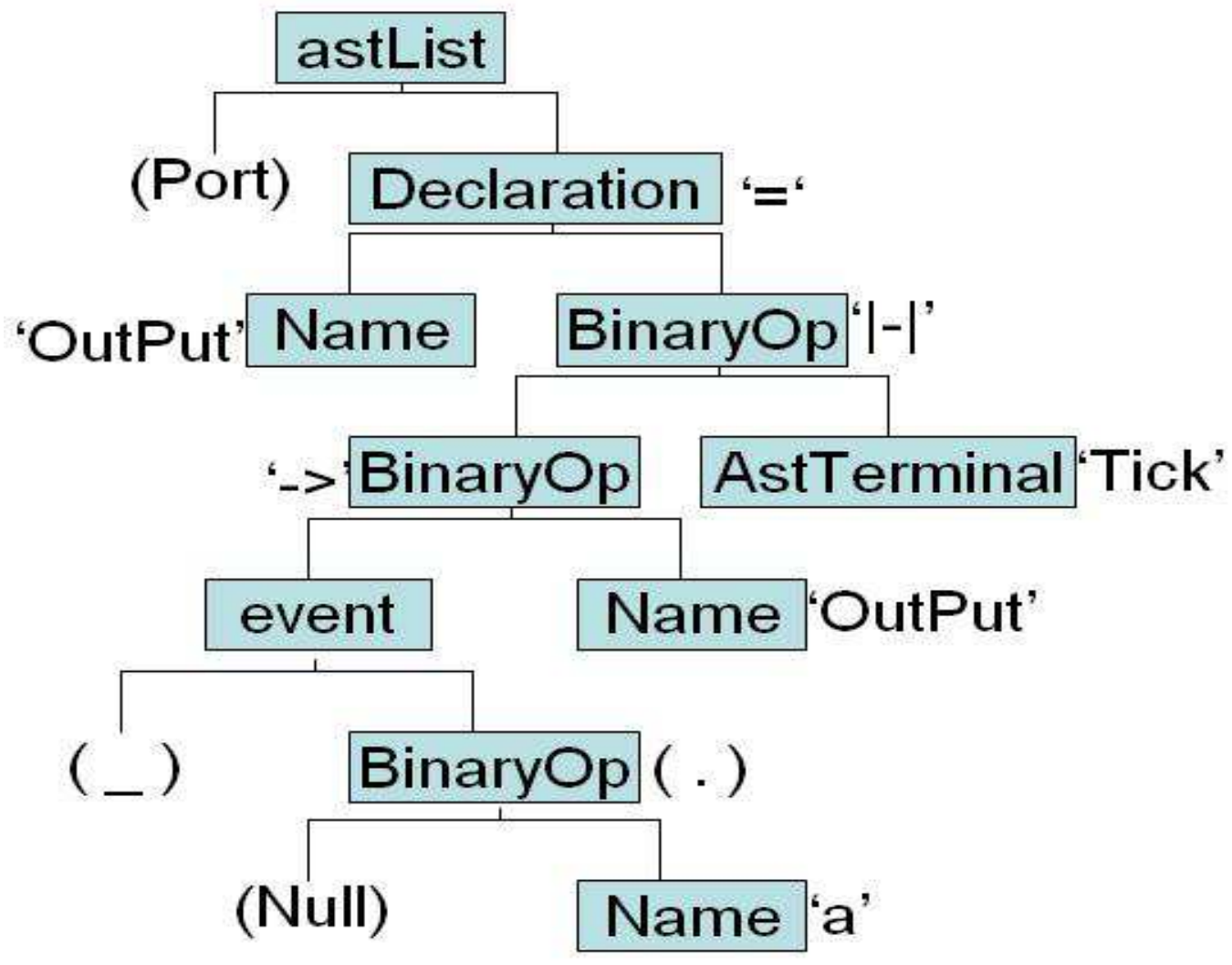

**Figure 2.20.** *Arbre syntaxique abstrait correspondant au port Output*

### *3.3. Concepts mathématiques*

L'outil Wr2fdr propose une implémentation orientée objet aux deux concepts mathématiques relation et fonction partielle : *Relation* et *LookupTable* (voir section 2.2.1). Par exemple, La classe *Relation* est utilisée pour mettre en association les processus et leurs événements (voir Figure 2.21).

*Exemple :*

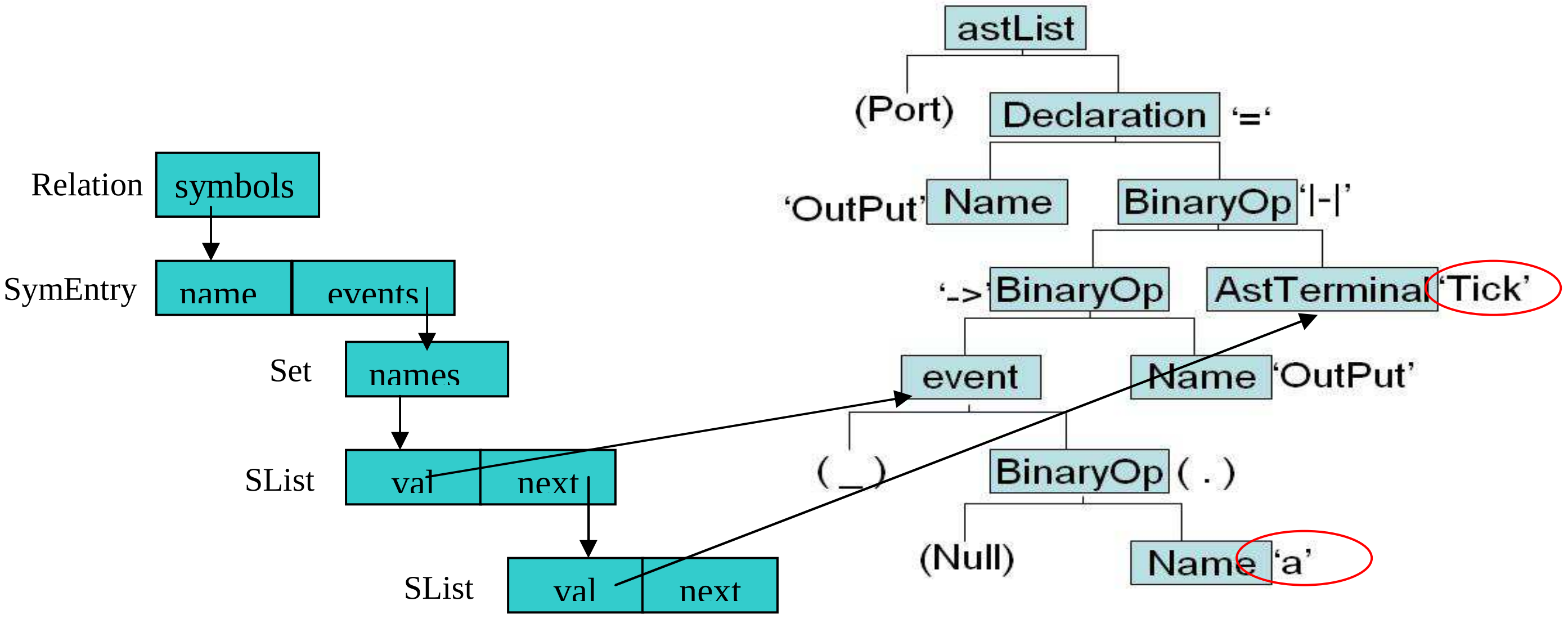

**Figure 2.21.** *Calcul de l'alphabet du port Output*

## *3.4. Quelques algorithmes*

### 3.4.1. *Fonction Principale Main*

La fonction principale *Main* regroupe les différentes étapes d'analyse et de génération. Elle commence par l'ouverture du fichier contenant la spécification d'architecture Wright (fichier .wrt) ensuite elle lance l'analyseur lexico-syntaxique (fonction *yyparse*) pour préparer le modèle d'analyse : arbre syntaxique abstrait en cas de succès. Ce modèle est parcouru pour calculer les constituants de chacun des processus utilisés (méthode *CalculationPass*). En plus, tout événement utilisé dans un port ou un rôle doit être reconnu comme événement du composant ou du connecteur (méthode *PostCalulationPass*). Enfin, la fonction main utilise la méthode *fdrpint* pour générer la sortie CSP acceptable par FDR correspondant à la spécification Wright soumise.

```
int main(int argc,char **argv) {
    astNode    *root;  // pointe au résultat de l'analyse (un noeud)
    astNode    *tmp;   // temp var utilisé pour inverser l'ordre de root
    int         error; // utilisé pour vérifier les erreurs dans le code

    // effectuer des manipulations de fichiers de base, en vérifiant que
    // les fichiers peuvent être ouverts avec succès et que le bon nombre
    // de paramètres est utilisé comme arguments de ligne de commande
    // jgrivera: Envoie des messages d'erreur à stderr au lieu d'un fichier
    // efile = fopen("wr2fdr.log","w");

    efile = stderr;

    // établit une fonction qui est appelée à chaque fois que le renvoie de
    // new est 0. Cela signifie que les appels individuels à new n'ont pas
    // besoin différemment. d'être explicitement vérifiée à moins qu'ils
    // doivent être traités Cela affecte tous les appels à new, même ceux
    // des autres fichiers.
    set_new_handler (OutOfMemoryError);

    if (argc <3) {
      fprintf(efile,"usage: %s <infile> <fdrfile>\n",argv[0]);
      exit(1);
    }

    // ouvrir le fichier d'entrée et pointe l'analyseur sur lui
    if ((ifile = fopen(argv[1],"r")) == NULL) {
      fprintf(efile,"can't open file for input: %s.\n",argv[1]);
      exit(1);
    } else {
      yyin = ifile;
    }

    // appel à l'analyseur pour opérer sur le fichier d'entrée
    error = yyparse();
    tmp = parse_result;
    fprintf(efile,"Parsing complete.\n");

    // effectuer un premier passage sur l'arbre pour dériver les alphabets
    if (!error && tmp!=NULL) {
```

```
    // avant de faire le test de calcul, nous avons besoin d'inverser
    // l'ordre de la liste (l'analyseur recueille les spécifications dans
      //l'ordre inverse)
      root = ((astList *)tmp)->reverse(); // pas d'espace alloué
      tmp = NULL;

      root->CalculationPass();
      root->PostCalculationPass();
      fprintf(efile,"Done with the calculation pass.\n");

      PrintFDRHeaders ();
      root->fdrprint ();
      declaration::fdrPrintGeneratedDeclarations ();

    // face à une tentative d'analyse échouée, effacer et sortir
    } else {
      fclose(ofile);
      if (error) {
          fprintf(efile,"a problem occured in the parsing stage.\n");
      } else {
          fprintf(efile,"null root.\n");
      }
      fclose(efile);
      exit(1);
    }

    fprintf(efile,"wr2fdr done.\n\n");
    return 0;

} // fin main
```

Les trois méthodes *calculationPass*, *postCalculationPass* et *fdrprint* sont des méthodes virtuelles (ou abstraites) appartenant à la classe fondatrice *AstNode*. Elles sont implémentées par les classes qui descendent d'*AstNode*. La méthode *calculationPass* permet de regrouper l'alphabet de chaque concept Wright.la méthode *postCalculationPass* permet d'augmenter les alphabets des composants et des connecteurs par les événements respectivement de leurs ports et rôles. Elle est également utilisée dans la résolution et le calcul des alphabets des processus imbriqués. Enfin la méthode *fdrprint* génère la sortie CSP correspondant à l'élément architectural Wright concerné.

### 3.4.2. *Méthode calculationPAss*

La méthode c*alculationPass* de la classe *Declaration* réutilise la classe *Relation* (voir section 3.3) pour regrouper l'alphabet correspondant à chaque concept Wright.

```
void declaration::CalculationPass(void) {
   Relation *totSymP;
   Relation *totSymE;
   symEntry *allEvents;

       // Si ce n'est pas une déclaration de la clause where
   if ((gtype==INTDECL_T || gtype==ROLE_T || gtype==PORT_T || gtype==GLUE_T
         || gtype==COMPUTATION_T) && defn && n->gtype==NAME_T ) {
```

```
      // Produire le premier tableau de symbole; Cela peut être fait
      // de la même manière pour tous les types de déclarations
      findValue();
      findImmediate();
      if (!processesToProcesses) {
        fprintf(efile,"declaration without processesToProcesses.\n");
        exit(1);
      }
      // Produire la deuxième table de symboles:
      //   processAlphabet = (processesToProcesses+ ; processesToEvents)

      totSymP = processesToProcesses->closure();
      if (!totSymP) {
          fprintf(efile,"Problem:  calculation totSymP failed.\n");
          exit(1);
      }
      if (!processesToEvents) {
          fprintf(efile,"declaration without processesToEvents.\n");
          exit(1);
      }
      totSymE = ScopedCompose(totSymP, processesToEvents);
      processAlphabet = totSymE;
      processAlphabet->afficheRelation();

      // Maintenant nous ajoutons notre alphabet à la table
      allEvents = processAlphabet->find(n);
      if (allEvents) {
          higherScope->processAlphabet->add (n, allEvents->events);
      }
      // Nous avons besoin d'étendre la deuxième table à tous
      // les nœuds subordonnés
      // même si elle ne changera pas ce que ce nœud pense totEvents
      findTotal();
   } // fin si
} // déclaration::CalculationPass
```

### 3.4.3. *Méthode PostCalculationPass*

Dans la suite, nous donnons l'implémentation de la méthode *postCalulationPass* de la classe *Declaration.*

```
void declaration::PostCalculationPass (void) {
    astNode *current_event = NULL;
    astNode *current_resolved_event = NULL;
    Set           *resolved_events = NULL;
    // Actuellement, seule les expressions susceptibles d'être non résolus
    // sont des événements dans totEvents.
    // (dans le cas des références circulaires).
    if (totEvents) {
      current_event = totEvents->start();
      while (current_event) {
          if (current_event->gtype != EVENT_T) {
            ResolveAlphabet (totEvents);
            break;           // il a terminé la boucle
          }
          current_event = totEvents->next();
      } // end while
    }
    defn->PostCalculationPass ();
}
```

### 3.4.4. *Le sous-proframme PrintFDRHeaders*

Le sous-proframme *PrintFDRHeaders* injecte dans le fichier de sortie de l'outil Wr2fdr, l'entête du model-cheker FDR l'entête. De plus il définit le processus *DeadlockFreeAbstraction* $DF_A$ pour la vérification de la propriété 2 (voir chapitre 1 section 7.2).

```
void PrintFDRHeaders (void) {
    // Inclure un processus DeadlockFreeAbstraction afin que nous puissions
    // effectuer les tests de cohérence de connecteur.
    doPrint("-- FDR compression functions");
    newLine();
    doPrint("transparent diamond");
    newLine();
    doPrint("transparent normalise");
    newLine();
    newLine();
    newLine();
    doPrint("-- Wright defined processes");
    newLine();
    doPrint("channel abstractEvent");
    newLine();
    doPrint("DFA = abstractEvent -> DFA |~| SKIP");
    newLine();
    newLine();
    doPrint("quant_semi({},_) = SKIP");
    newLine();
    doPrint("quant_semi(S,PARAM) = |~| i:S @ PARAM(i) ;
quant_semi(diff(S,{i}),PARAM)");
    newLine();
    newLine();
    doPrint("power_set({}) = {{}}");
    newLine();
    doPrint("power_set(S) = { union(y,{x}) | x <- S, y <-
power_set(diff(S,{x}))}");
    newLine();
    newLine();
    newLine();

}
```

### 3.4.5. *Méthode fdrprint*

L'implémentation de la méthode *fdrprint* de la classe *Connecteur* doit générer la propriété 3. Pour y parvenir, elle doit générer le processus décrivant le comportement du connecteur (glu). En plus elle doit invoquer la méthode *fdrprint* de la classe *Declaration* pour préparer la relation de raffinement de la propriété 3. Enfin la méthode *fdrprint* génère la relation de raffinement correspondant à la propriété 2.

```
void connector:: fdrprint (void) {
    declaration       *curRole;
    int                i;
    Set               *role_internal_events = NULL;

    event::BindEventData (NULL);

    // Première sortie des processus fondamentaux du Role et Glue qui
    //seront utilisées pour construire la définition du connecteur
    doPrint("-- Connector ");
    if (conn_name) {
      conn_name->fdrprint();
    } else {
      doPrint("OOPS: null name.");
    }
    addTab();
    newLine();

    SplitLongSets ();
    ProduceRolesAndGlue (NULL);

    // Déclare des canaux pour les événements de rôle. Un canal est nommé
    // après le rôle et paramétré par tous les événements de ce rôle.

    //          role_name : { events w/in role }
    // Si "x" et "y" sont des événements dont le rôle est "R", nous
    // obtenons:      R : {x,y}  -ou- R.x et R.y

    // Parcourir tous les rôles qui déclarent ces canaux
    curRole = (declaration *) roles->startNode();
    while (curRole) {
      doPrint("channel ");
      if (curRole->n->gtype != NAME_T) {
          fprintf(efile,"Internal Error:  mangled declaration name.\n");
          exit(1);
      }
      ((name *)curRole->n)->fdrprintWithNoParams();
      doPrint(": {");
      curRole->totEvents->fdrprint();
      doPrint("}");
      newLine();
      curRole = (declaration *) roles->nextNode();
    } // fin while

    // Déclare le connecteur égal au "Glue" avec le renommage effectué
    // sur les rôles
    if (conn_name) {
      conn_name->fdrprint();
    } else {
      doPrint("OOPS: null name.");
    }
    doPrint(" = (");
    i=0;
    // Effectuer le renommage d'un rôle à la fois; renommer les événements
    // tels que chaque événement "x" dans le rôle de "R" sera renommé en
    // "R.x"

        curRole = (declaration *) roles->startNode();
    while (curRole) {
      ++i;
      doPrint("( ROLE");
```

```
      curRole->n->fdrprint();
      doPrint("[[ x <- ");
      curRole->n->fdrprint();
      doPrint(".x | x <- {");
      if (curRole->totEvents) {
        curRole->totEvents->fdrprint();
      }

      // Mettre les rôles en parallèle
      doPrint(" } ]]");
      continueLine();
      doPrint("[| diff({|");
      curRole->n->fdrprint();
      doPrint("|}, {");
      // Calculer l'ensemble des événements internes du rôle
      role_internal_events = Set::SetMinus (curRole->ParamTotEvents,
                                            glue->totEvents);
      if (role_internal_events) {
      role_internal_events->fdrprint();
      delete role_internal_events;
      }
      doPrint("}) |]");
      continueLine();
      curRole = (declaration *) roles->nextNode();
    } // Fin while

    // S'assurer que le processus de Glue utilise ces renommages
    glue->n->fdrprint();
    ((connector *)glue->higherScope)->conn_name->fdrprintWithNoParams();
    for (;i>0;i--) {
      doPrint(")");
    }
    doPrint(")");

    // Enfin, ajouter un test de non interblocage du connecteur
    if (conn_name) {
      // L'alphabet du connecteur est égal à l'alphabet du Glue
      //(qui a déjà été calculé).
      // Produire une version abstraite du processus de connecteur
      newLine();
      conn_name->fdrprint();
      doPrint("A = ");
      conn_name->fdrprint();
      doPrint(" [[ x <- abstractEvent | x <- ALPHA_");
      conn_name->fdrprint();
      doPrint(" ]]");

      //(Test 2)
      newLine();
      doPrint("assert DFA [FD= ");
      conn_name->fdrprint();
      doPrint("A");
    } else {
      doPrint("OOPS: can't produce assertion for null name.");
    }
    subTab();
    newLine();

}
```

## 4. Evaluation

L'outil Wr2fdr possède plusieurs points forts :

- ☺ Il est modulaire à la C++. En effet, il respecte le principe de séparation interface/implémentation en utilisant les fichiers « .hpp » pour l'interface et les fichiers « .cpp » pour l'implémentation. En outre, la répartition des classes sur les modules C++ (interface et implémentation) est plus ou moins judicieuse.
- ☺ Il réutilise des outils existants et performants Lex et Yacc afin de réaliser un analyseur lexico-syntaxique de l'ADL Wright.
- ☺ Le code source de Wr2fdr est souvent lisible et accompagné parfois par des commentaires pertinents
- ☺ Wr2fdr propose une implémentation orientée objet de concept arbre syntaxique abstrait dédié à l'ADL Wright et à la traduction de Wright vers CSP de l'outil FDR2.

Quant aux points faibles, nous pourrions signaler les critiques suivantes :

- ☹ L'outil Wr2fdr est écrit en C++. Ceci a des conséquences inévitables liées au caractère hybride du langage C++ : utilisation des méthodes statiques, variables externes et sous programmes.
- ☹ Les structures de données proposées par l'outil Wr2fdr telles que *AstList*, *Set*, *Relation*, *LookUpTable* ne sont pas génériques. Elles sont plutôt dédiées à l'application.
- ☹ Sur le plan architectural, l'outil Wr2fdr n'est pas basé sur des patterns de conception tels que les patterns de GoF [**11**]. En effet, il y a presque un vide entre les classes d'analyse (ou métier) et les classes d'implémentation formant l'outil Wr2fdr.
- ☹ La hiérarchie introduite par la classe *AstNode* est plutôt encombrante. La séparation entre les concepts structuraux et comportementaux de Wright n'est pas nette. On pourrait imaginer deux sous-hiérarchies l'une pour les concepts structuraux et l'autre pour les concepts comportementaux. Ces deux sous-hiérarchies pourraient être reliées par des associations appropriées.
- ☹ La sémantique des méthodes n'est pas définie ni d'une façon informelle (commentaires précis et pertinents) ni d'une façon formelle en utilisant une spécification pré/post exprimée à l'aide de la macro-instruction *assert* du langage C++.
- ☹ L'outil Wr2fdr n'est pas doté des points de contrôles dits assertions internes exprimées par la macro-instruction *assert*. Ceci rend le logiciel Wr2fdr peu testable.

## 5. Conclusion

Dans ce chapitre, nous avons mené une activité de rétro-ingénierie concernant l'outil Wr2fdr. Nous avons pu extraire à partir du code source C++ de Wr2fdr sa partie statique sous forme d'un diagramme de classe UML comportant aussi bien des classes d'implémentation, de conception et d'analyse. En outre, nous avons identifié et compris les choix techniques retenus par les auteurs de Wr2fdr. Enfin, nous avons évalué l'outil Wr2fdr en tant que logiciel : ses points forts et faibles. Dans le chapitre suivant, nous allons maintenir le logiciel Wr2fdr aussi bien sur le plan correctif qu'évolutif.

# *Chapitre 3*

# *Modifications apportées à l'outil Wr2fdr*

L'outil Wr2fdr a pour objectif d'automatiser les quatre propriétés décrites dans le chapitre 1 liées à la cohérence des composants, des connecteurs et des configurations Wright. Pour y parvenir, l'outil Wr2fdr traduit une spécification Wright en une spécification CSP dotée des relations de raffinement à vérifier [**3**]. Ces relations de raffinement traduisent des propriétés standards à vérifier sur des architectures décrites en Wright. La spécification CSP engendrée pour l'outil Wr2fdr est soumise à l'outil de Model-cheker FDR.

Suite à des expérimentations avec l'outil Wr2fdr [**26**], nous avons identifié des défaillances liées aux fonctionnalités souhaitées de Wr2fdr (voir chapitre 1 section 8.3). De plus, nous avons mené une activité de rétro-ingénierie (du concret vers abstrait) afin d'avoir une vue d'ensemble, d'identifier les abstractions principales et de connaître les choix techniques de Wr2fdr (voir chapitre 2). Dans ce chapitre, nous allons effectuer une activité de maintenance corrective et évolutive concernant l'outil Wr2fdr. Les trois premières sections localisent, corrigent et retestent les erreurs liées respectivement à la cohérence d'un connecteur, d'un composant et d'une configuration Wright. La dernière section de ce chapitre propose un analyseur de la sémantique statique de Wright.

## 1. Cohérence d'un connecteur

### *1.1. Description informelle*

La cohérence d'un connecteur Wright est définie par deux propriétés codifiées : propriété 2 et propriété 3. Celles-ci sont exprimées informellement comme suit :

**Propriété 2**: Connecteur sans interblocage

*La glu d'un connecteur interagissant avec les rôles doit être sans interblocage.*

**Propriété 3**: Rôle sans interblocage

*Chaque rôle d'un connecteur doit être sans interblocage.*

### *1.2. Description formelle*

Les deux propriétés 2 et 3 reviennent à vérifier si un processus est sans interblocage. D'une façon formelle, un processus P= (A, F, D) est sans interblocage si pour toute trace telle que $(t,A) \in F$, last(t) = $\surd$, avec A représente l'alphabet du processus, F représente ses échecs et D représente ses divergence [**1**], [**3**] et [**4**]. Mais ceci peut être exprimé par une relation de raffinement entre le processus $DF_A$ et P ($DF_A \sqsubseteq P$) avec $DF_A$ est défini comme suit :

$DF_A = (\sqcap\, e : A \bullet e \rightarrow DF_A) \;\sqcap\; \S.$

Le processus $DF_A$ permet toute les traces possibles sur l'alphabet A mais sans jamais avoir la possibilité de refuser tous les événements : il s'agit d'un processus sans interblocage.

### *1.3. Anomalies détectées*

La DT1 (Donnée de Test 1) fournit par la figure 3.1 couvre les deux propriétés 2 et 3 relatives à la cohérence d'un connecteur.

```
Style ClientServer
Connector CSconnector
   Role Client = (request -> result -> Client) |~| TICK
   Role Server = (invoke -> return -> Server) [] TICK
   Glue = (Client.request -> Server.invoke -> Server.return -> Client.result -> Glue) [] TICK
Constraints
   // no constraints
End Style
```

**Figure 3.1.** *DT1 couvrant les deux propriétés 2 et 3*

En exécutant l'outil Wr2fdr sur la DT1, celui-ci produit une spécification à soumettre au Model-checker FDR contenant en partie la sortie observée illustrée par la figure 3.2.

```
ALPHA_CSconnector = {|Server.invoke, Server.return, Client.result, Client.request|}
Glue = ((Client.request -> (Server.invoke -> (Server.return -> (Client.result-> Glue)))) []SKIP)
ALPHA_Client = {request, result}
ROLEClient = ((request -> (result -> Client)) |~| SKIP)
ClientA = ROLEClient [[ x <- abstractEvent | x <- ALPHA_Client ]]
assert DFA [FD= ClientA

ALPHA_Server = {invoke, return}
ROLEServer = ((invoke -> (return -> Server)) [] SKIP)
ServerA = ROLEServer [[ x <- abstractEvent | x <- ALPHA_Server ]]
assert DFA [FD= ServerA


channel Client: {request, result}
channel Server: {invoke, return}
CSconnector = ( (ROLEClient[[ x <- Client.x | x <- {request, result } ]]
      [| diff({|Client|}, {}) |]
   (ROLEServer[[ x <- Server.x | x <- {invoke, return } ]]
      [| diff({|Server|}, {}) |]
```

```
   Glue)) )
CSconnectorA = CSconnector [[ x <- abstractEvent | x <- ALPHA_Glue ]]
assert DFA [FD= CSconnectorA
```

**Figure 3.2.** *Sortie observée liée au DT1*

Tandis que la sortie attendue liée au DT1 est illustrée par la figure 3.3.

```
ALPHA_CSconnector = {|Server.invoke, Server.return, Client.result, Client.request|}
  Glue = ((Client.request -> (Server.invoke -> (Server.return -> (Client.result-> Glue)))) []SKIP)

  ALPHA_Client = {request, result}
  ROLEClient = ((request -> (result -> ROLEClient)) |~| SKIP)
  ClientA = ROLEClient [[ x <- abstractEvent | x <- ALPHA_Client ]]
  assert DFA [FD= ClientA

  ALPHA_Server = {invoke, return}
  ROLEServer = ((invoke -> (return -> ROLEServer)) [] SKIP)
  ServerA = ROLEServer [[ x <- abstractEvent | x <- ALPHA_Server ]]
  assert DFA [FD= ServerA

channel Client: {request, result}
channel Server: {invoke, return}
CSconnector = ( (ROLEClient[[ x <- Client.x | x <- {request, result } ]]
      [| diff({|Client|}, {}) |]
   (ROLEServer[[ x <- Server.x | x <- {invoke, return } ]]
      [| diff({|Server|}, {}) |]
   Glue)) )
CSconnectorA = CSconnector [[ x <- abstractEvent | x <- ALPHA_ CSconnector ]]
assert DFA [FD= CSconnectorA
```

**Figure 3.3.** S*ortie attendue liée au DT1*

En comparant la sortie observée (Voir figure 3.2) à la sortie attendue jugée correcte (Voir figure 3.3), on remarque les divergences suivantes :

- L'équation relative au rôle *client* est syntaxiquement incorrecte car une équation récursive doit avoir la forme suivante : P= x → P. Ceci est également vrai pour le rôle *Server* (**erreur1**).
- L'identificateur *ALPHA_Glue* qui devrait matérialiser un ensemble d'événements est non défini (**erreur2**).

### *1.4. Identification de la partie concernée*

En se basant sur l'activité de rétro-ingénierie effectuée sur le code de l'outil Wr2fdr (voir chapitre 2), nous avons su localiser la méthode qui génère le code FDR2. Celle-ci est appelée *fdrprint* introduite dans la classe abstraite *AstNode* et redéfinie ou implémentée dans les classes descendantes *Connector*, *Composant*, *Name* … (voir figure 3.4).

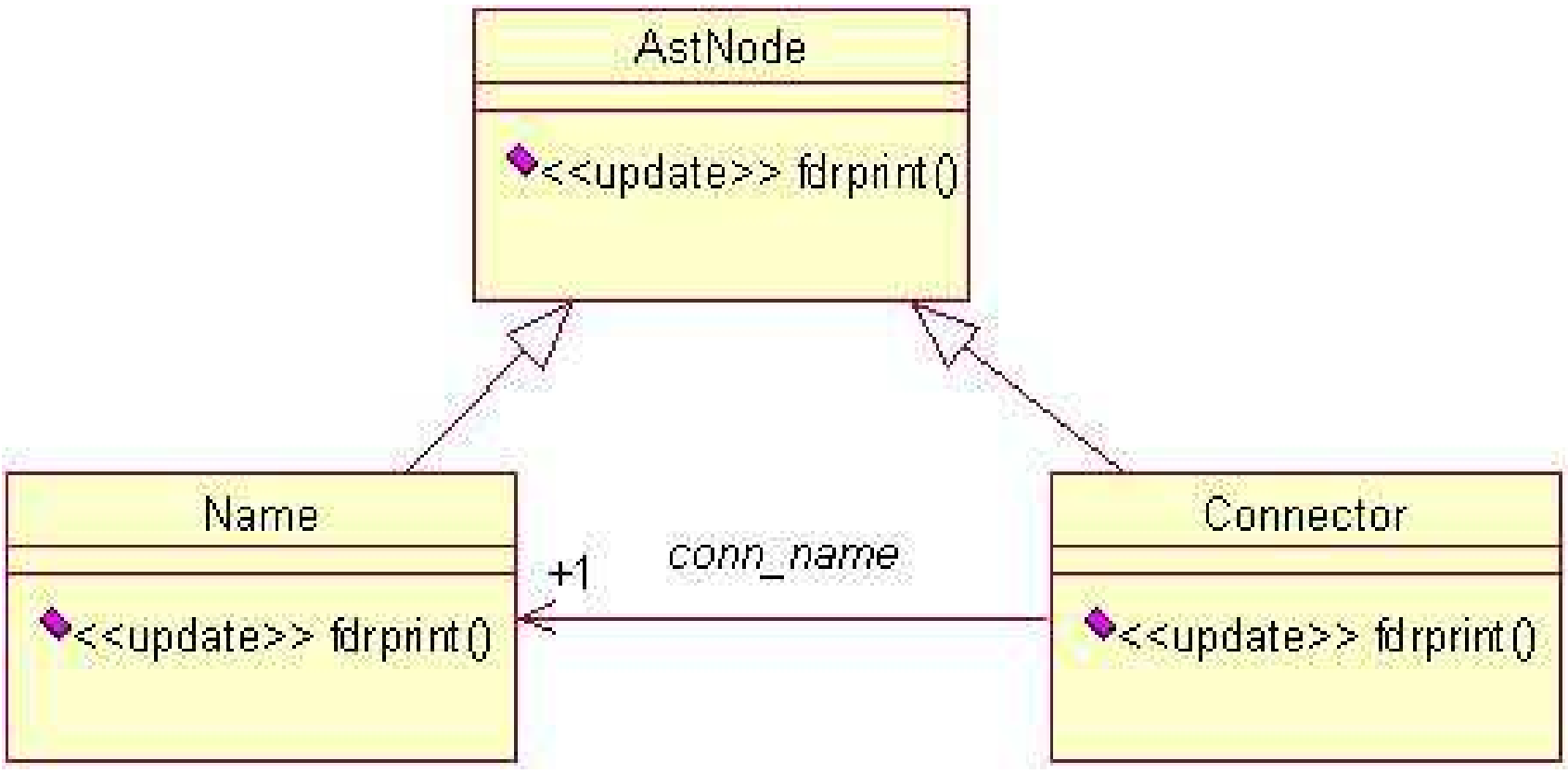

**Figure 3.4.** *Méthode fdrprint*

### *1.5 Correction proposée*

L'*erreur1* est induite par la méthode *fdrprint* de la classe *Name* (voir figure 3.5). La version corrigée de *fdrprint* de la classe *Name* est fournie par la figure 3.6.

```
void name::fdrprint(void) {
   astNode      *param = NULL;
  if (infertype == EVENT_TYPE)
   doPrintWithNoBreak(n); //afficher directement le nom du concept son vérification
  else
   doPrint(n);
  if (params != NULL) {
   if (infertype == EVENT_TYPE) {
         param = params->startNode();
         param->infertype = EVENT_TYPE;
         doPrintWithNoBreak(".");
         params->fdrprint();
   } else {
```

```
        doPrintWithNoBreak("(");
         params->fdrprint();
         doPrint(")");
    }
  }
}
```

**Figure 3.5.** *Version incorrecte de fdrprint de la classe Name*

```
void name::fdrprint(void) {
   astNode      *param = NULL;
   astNode      * higherScope_effectif = NULL;
  if(higherScope && (higherScope->gtype == ROLE_T || higherScope->gtype
  == PORT_T))
         higherScope_effectif = higherScope;
   else if(higherScope && higherScope->higherScope && (higherScope->higherScope->gtype
       == ROLE_T || higherScope->higherScope->gtype== PORT_T))
             higherScope_effectif = higherScope->higherScope;

   if (infertype == EVENT_TYPE)
         doPrintWithNoBreak(n);
   else{

         if(higherScope_effectif){
            switch (higherScope_effectif->gtype){
                  case ROLE_T:
                     if( this->eq(((declaration*)higherScope_effectif)->n))
                           doPrint("ROLE");
                     doPrint(n);
                     if (((connector *)((declaration*)higherScope_effectif )
                   ->higherScope)->ProduceDetRole==1)
                           doPrintWithNoBreak("DET");
                     break;
                  case PORT_T:
                     if( this->eq(((declaration*)higherScope_effectif)->n))
                           doPrint("PORT");
                     doPrint(n);
                     if (((component *)((declaration*)higherScope_effectif )->higherScope)
                   ->ProduceDetPort==1)
                           doPrintWithNoBreak("DET");
                     if (((component *)((declaration*)higherScope_effectif )->higherScope)
                   ->ProduceRestPort==1)
                           doPrintWithNoBreak("R");
                     break;
            }
         }else
```

```
            doPrintWithNoBreak(n);
  }
 if (params != NULL) {
   if (infertype == EVENT_TYPE) {
         param = params->startNode();
         param->infertype = EVENT_TYPE;
         doPrintWithNoBreak(".");
         params->fdrprint();
   } else {
         doPrintWithNoBreak("(");
         params->fdrprint();
         doPrint(")");
   }
  }
}
```

**Figure 3.6.** *Version corrigée de fdrprint de la classe Name*

Lors de la génération de la sortie FDR, la méthode fdrprint de la classe Name de la version actuelle de l’outil Wr2fdr (voir figure 3.5) traite de la même manière les noms associés aux concepts Composant, Connecteur, Configuration, Style, Evénement, Port et Rôle. Nous avons ajouté une structure de contrôle (switch (higherScope_effectif->gtype) voir figure 3.6) pour distinguer les deux concepts Port et Rôle. Dans le cas du concept Rôle, l’injection de son nom est précédé par ‘ROLE’.

L’*erreur2* est induite par la méthode *fdrprint* de la classe *Connector* (Voir figure 3.7). La version corrigée de *fdrprint* de la classe *connector* est fournie par la figure 3.8.

```
void connector::InstantiateWithNoParams (void) {
   declaration      *curRole;
   int               i;
   Set              *role_internal_events = NULL;
   event::BindEventData (NULL);

   doPrint("-- Connector ");//mouti 4
   if (conn_name) {
         conn_name->fdrprint();
   } else {
         doPrint("OOPS: null name.");
   }
   addTab();
   newLine();
```

```
SplitLongSets ();
ProduceRolesAndGlue (NULL);

curRole = (declaration *) roles->startNode();
while (curRole) {
      doPrint("channel ");
      if (curRole->n->gtype != NAME_T) {
         fprintf(efile,"Internal Error:  mangled declaration name.\n");
         exit(1);
      }
      ((name *)curRole->n)->fdrprintWithNoParams();
      doPrint(": {");
      curRole->totEvents->fdrprint();
      doPrint("}");
      newLine();
      curRole = (declaration *) roles->nextNode();
} // end while

if (conn_name) {
      conn_name->fdrprint();
} else {
      doPrint("OOPS: null name.");
}
doPrint(" = (");
i=0;

curRole = (declaration *) roles->startNode();
while (curRole) {
  ++i;
  doPrint("( ROLE");
  curRole->n->fdrprint();
  doPrint("[[ x <- ");
  curRole->n->fdrprint();
  doPrint(".x | x <- {");
  if (curRole->totEvents) {
    curRole->totEvents->fdrprint();
  }

  doPrint(" } ]]");
  continueLine();
  doPrint("[| diff({|");
  curRole->n->fdrprint();
  doPrint("|}, {");
  role_internal_events = Set::SetMinus (curRole->ParamTotEvents, glue->totEvents);
  if (role_internal_events) {
```

```
            role_internal_events->fdrprint();
            delete role_internal_events;
        }
        doPrint("}) |]");
        continueLine();
        curRole = (declaration *) roles->nextNode();
    } // end while
    glue->n->fdrprint();
    ((connector *)glue->higherScope)->conn_name->fdrprintWithNoParams();
    for (;i>0;i--) {
        doPrint(")");
    }
    doPrint(")");

        if (conn_name) {
            newLine();
            conn_name->fdrprint();
            doPrint("A = ");
            conn_name->fdrprint();
            doPrint(" [[ x <- abstractEvent | x <- ALPHA_");
            glue->n->fdrprint();  //on  utilise 'ALPHA_glue' comme nom de l'alphabet du connecteur
            doPrint(" ]]");
            newLine();
            doPrint("assert DFA [FD= ");
            conn_name->fdrprint();
            doPrint("A");
    } else {
            doPrint("OOPS: can't produce assertion for null name.");
    }
    subTab();
    newLine();
}
```

**Figure 3.7.** *Version incorrecte de fdrprint de la classe Connector*

```
void connector::fdrprint (void) {
    declaration      *curRole;
    int               i;
    Set              *role_internal_events = NULL;
    event::BindEventData (NULL);

    doPrint("-- Connector ");
    if (conn_name) {
            conn_name->fdrprint();
    } else {
            doPrint("OOPS: null name.");
```

```
   }
   addTab();
newLine();

   SplitLongSets ();
   ProduceRolesAndGlue (NULL);

   curRole = (declaration *) roles->startNode();
   while (curRole) {
        doPrint("channel ");
        if (curRole->n->gtype != NAME_T) {
            fprintf(efile,"Internal Error:  mangled declaration name.\n");
           exit(1);
        }
        ((name *)curRole->n)->fdrprintWithNoParams();
        doPrint(": {");
        curRole->totEvents->fdrprint();
        doPrint("}");
        newLine();
        curRole = (declaration *) roles->nextNode();
   } // end while
if (conn_name) {
        conn_name->fdrprint();
   } else {
        doPrint("OOPS: null name.");
   }
   doPrint(" = (");

   i=0;
   curRole = (declaration *) roles->startNode();
   while (curRole) {
     ++i;
     doPrint("( ROLE");
     curRole->n->fdrprint();
     doPrint("[[ x <- ");
     curRole->n->fdrprint();
     doPrint(".x | x <- {");
     if (curRole->totEvents) {
       curRole->totEvents->fdrprint();
     }
     doPrint(" } ]]");
     continueLine();
     doPrint("[| diff({|");
     curRole->n->fdrprint();
     doPrint("|}, {");
     role_internal_events = Set::SetMinus (curRole->ParamTotEvents, glue->totEvents);
```

```
      if (role_internal_events) {
            role_internal_events->fdrprint();
            delete role_internal_events;
      }
      doPrint("}) |]");
      continueLine();
      curRole = (declaration *) roles->nextNode();
    } // end while

    glue->n->fdrprint();
    ((connector *)glue->higherScope)->conn_name->fdrprintWithNoParams();
    for (;i>0;i--) {
      doPrint(")");
    }
    doPrint(")");
    if (conn_name) {
            newLine();
            conn_name->fdrprint();
            doPrint("A = ");
            conn_name->fdrprint();
            doPrint(" [[ x <- abstractEvent | x <- ALPHA_");
            conn_name->fdrprint();// utiliser ALPHA_conn_name comme nom de l'alphabet du
                                  //connecteur
            doPrint(" ]]");
            newLine();
            doPrint("assert DFA [FD= ");
            conn_name->fdrprint();
            doPrint("A");
    } else {
            doPrint("OOPS: can't produce assertion for null name.");
    }
    subTab();
    newLine();
}
```

**Figure 3.8.** *Version corrigée de fdrprint de la classe Connector*

Les auteurs de Wright ont décidé d'attribuer l'identificateur *ALPHA_conn_name* à l'alphabet du processus associé au connecteur. Mais la méthode incorrecte *fdrprint* de la classe *Connector* (voir figure 3.7) ne respecte pas cette convention. Elle injecte dans la sortie FDR plutôt *ALPHA_glue*. La solution est de remplacer l'instruction qui injecte 'glue' par une instruction qui injecte le nom du connecteur (*conn_name*) : conn_name->fdrprint() (voir figure 3.8).

### *1.6. Validation*

Après avoir corrigé les anomalies détectées relatives à la cohérence des connecteurs de l’outil Wr2fdr (Voir section 1.3), nous avons retesté avec succès l’outil Wr2fdr [**21**] en utilisant des techniques de couverture issues de test fonctionnel orienté syntaxe.

## 2. Cohérence d’un composant

### *2.1. Description informelle*

Un composant est cohérent si le comportement des ports et le comportement du calcul (computation) sont cohérents. Pour s’assurer de la cohérence entre le comportement des ports et celui du calcul on utilise la notion du projection. Un port est une projection d’un composant si ce dernier agit de la même manière que le port quand nous ignorons tous les évènements n’appartenant pas à l’alphabet de ce port.

La cohérence Port/Calcul Wright est définie par une propriété codifiée : propriété 1. Celle-ci est exprimée informellement comme suit :

**Propriété 1 :** cohérence Port/Calcul

*La spécification d’un port doit être une projection du Calcul, sous l’hypothèse que l’environnement obéisse à la spécification de tous les autres ports.*

### *2.2. Description Formelle*

**Propriété 1** : Cohérence Port/Calcul

*Pour un composant avec un processus de Calcul C et des ports P,* $P_1, \ldots P_n$ *; C est cohérent avec P si* $P \sqsubseteq (C \parallel \forall\ i : 1..n \parallel det(Pi \upharpoonright \alpha oPi)) \upharpoonright \alpha P)$.

<u>*det*(P)</u>: le processus est déterministe, il a les mêmes traces que P, mais avec moins de refus. Ainsi, n’importe quel évènement qui a lieu à tout point est entièrement contrôlable par l’environnement.

<u>*αoP*</u> : le sous ensemble de *αP* correspondant au événements observés

<u>*L’opérateur* ⌈</u>: Pour tout processus P et un ensemble d’évènements E :

$P \upharpoonright E = (A \cap E, F', D')$ *où* $F' = \{(t', r') \mid \ (t, r) \in F \mid t' = t \upharpoonright E \ \wedge \forall \ \ r' = r \cap E\}$ *et* $D' = \{t' \mid \exists\ t \in D \mid t' = t \upharpoonright E\}$.

La projection d'une trace (t ⌈ E) est une trace qui contient tous les éléments de t qui sont dans E, dans le même ordre, sans tous les éléments qui ne sont pas dans E.

*Exemple:*

*<acadbcabc>⌈{a, b} = <aabab>*

### *2.3. Anomalies détectées*

La DT2 (Donnée de Test2) fournit par la figure 3.9 couvre la propriété 1 relative à la cohérence d'un composant.

```
Style ClientServer
  Component Client
    Port p = request -> reply -> p |~| TICK
    Computation = internalCompute -> p.request -> p.reply -> Computation |~| TICK
constraints
    //no constraints
End Style
```

**Figure 3.9.** *DT2 couvrant la propriété 1*

En exécutant l'outil Wr2fdr sur la DT2, celui-ci entraîne une erreur à l'exécution (Voir figure 3.10)

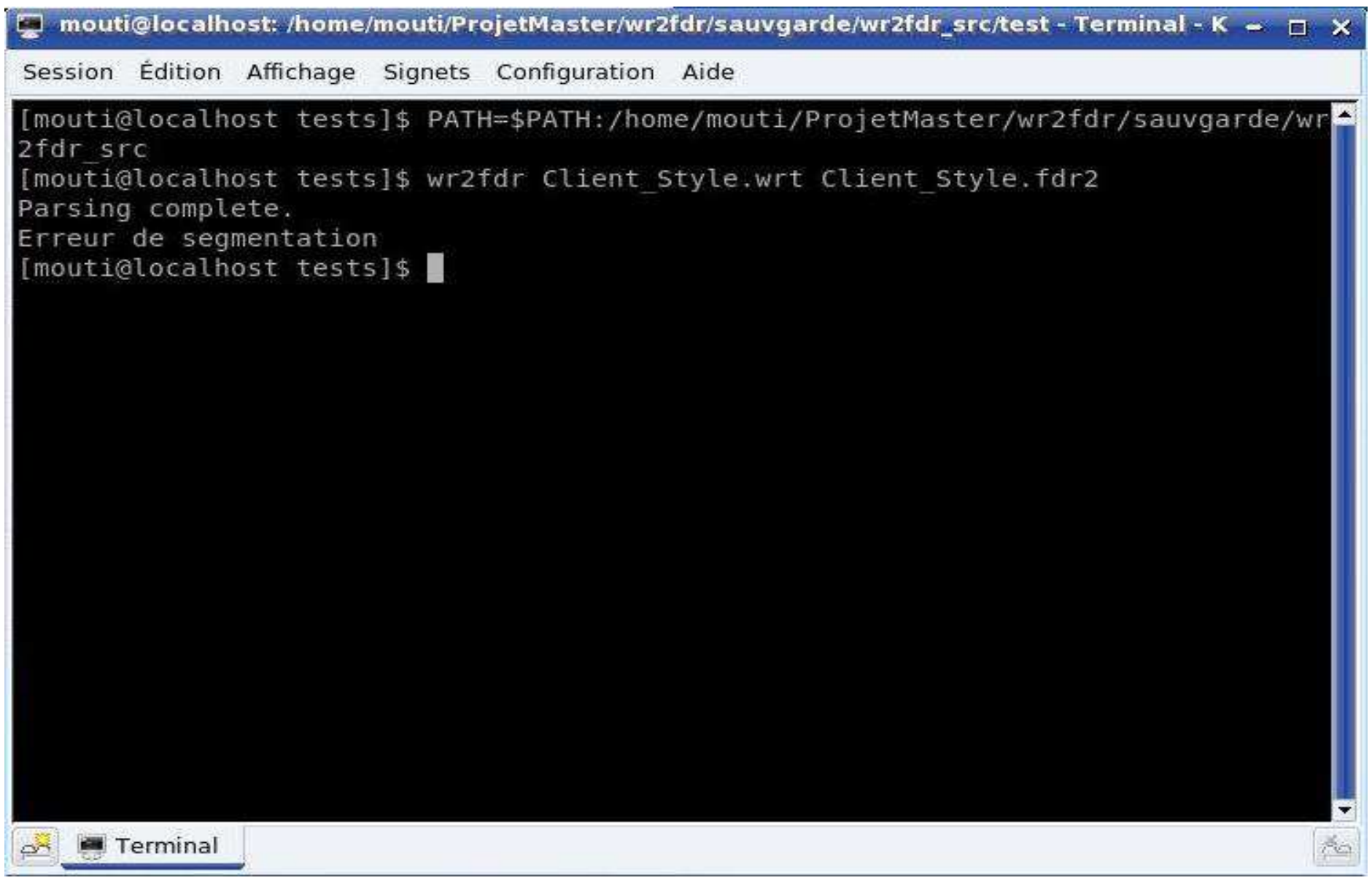

**Figure 3.10.** *Erreur d'exécution liée au DT2*

### *2.4. Identification de la Partie concernée*

La structure macro-scopique extraite du code de l'outil Wr2fdr (Voir Chapitre 2 figure 2.19) nous a permis de localiser la partie coupable: c'est-à-dire la méthode *calculationPass* de la classe *Component* (Voir figure 3.11). En effet, l'attribut polymorphe *higherscope* de type *AstNode* n'est pas initialisé au sein de cette méthode. Ceci provoque tôt ou tard une erreur d'exécution lorsqu'on tente d'appliquer une méthode sur cette entité (*higherscope*) non attachée.

```
void component::CalculationPass(void) {
   // l'attribut higherScope n'est pas initialisé.
   ports->CalculationPass();
   computation->CalculationPass();
}
```

**Figure 3.11.** *Définition de la méthode CalculationPass*

De plus, la méthode *fdrprint* de la classe *Component* comparée à celle de la classe *Connector* (voir figure 3.8) ne comporte pas de traitement lié à la génération de la propriété 1 cohérence Port/Calcul (Voir figure 3.12).

```
void component::fdrprint(void) {
   doPrint("-- Component ");
   if (comp_name)
         comp_name->fdrprint();
   else
         doPrint("OOPS: null name");
   if (params) {
         doPrint("(");
         params->fdrprint();
         doPrint(")");
   }
   addTab();newLine();
   if (ports)
         ports->fdrprint();          //pas de tests à injecter dans le fichier de sortie pour les
                                      //ports
   newLine();
   doPrint("Computation = ");
   if (computation)
         computation->fdrprint();  //pas de tests à injecter dans le fichier de sortie pour la
                                    //computation
```

```
    else
        doPrint("ERROR:  no computation found");
    subTab();newLine();
}
```

**Figure 3.12.** *Version incorrecte de fdrprint de la classe Component*

En conclusion, la fonctionnalité relative à la cohérence Port/Calcul connue sous le code Propriété 1 n’est pas implémentée.

### *2.5. Correction proposée*

Dans cette section, nous allons proposer une réalisation de la propriété 1: cohérence Port/Calcul. Dans un premier temps, nous allons apporter des modifications de la méthode *calculationPass* de *Component* en s’inspirant de celle de *Connector*. Ceci nous a permis de résoudre le problème d’initialisation de l’attribut polymorphe *higherscope* (Voir figure 3.13) sachant que dans l’outil Wr2fdr, les processus CSP associés aux ports et à la partie calcul d’un composant Wright sont modélisés par des arbres binaires.

```
void component::CalculationPass(void) {
    astNode     *current_port;
    Set         *tmp = NULL;
    processAlphabet = new Relation;

    // parcourir la liste des ports
    current_port = ports->startNode();
    while (current_port) {
        // initialisation de l’attribut higherScope pour la liste des ports
        current_port->higherScope = this;
        current_port = ports->nextNode();
    }

    // calculer l’alphabet de chaque port
    ports->CalculationPass();

    // calculer l’alphabet de la computation
    computation->higherScope = this;
    computation->CalculationPass();

    // calculer l’alphabet du composant
    current_port = ports->startNode();
    while (current_port) {
```

```
        totEvents = Set::Union (tmp, current_port->totEvents);
        if (tmp)
           delete tmp;
        tmp = totEvents;
        current_port = ports->nextNode();
   }
   totEvents = Set::Union (tmp, computation->totEvents);
   if (tmp)
        delete tmp;
}
```

**Figure 3.13.** *Modification de la méthode CalculationPass*

Dans un deuxième temps, nous avons réalisé les opérations au sens strict relatives à la propriété 1. Ces opérations sont :

- L'adaptation des processus CSP d'un composant pour qu'ils soient reconnus par l'outil FDR.
- L'implantation de la déterminisation des processus.
- L'implantation la restriction des processus (⌈) à un ensemble d'événements.
- La génération des relations de raffinement de la cohérence d'un composant.

   - **Adaptation des processus d'un composant**

Afin de réaliser cette tâche, nous avons réutilisé avec adaptation le code relatif à l'adaptation des processus CSP associés aux rôles et glu d'un connecteur.

   - **Implantation de la déterminisation (det)**

Nous l'avons implantée dans le générateur de sortie FDR (méthode *fdrprint*) par la simple substitution de l'opérateur du choix non déterministe ⊓ par l'opérateur du choix déterministe □, mais il faut que tous les processus de la forme P = e → Q ⊓ e → S (ou e est un événement, P, Q et S des processus) seront remplacés a priori par P = e → (Q ⊓ S) [**23**].

   - **Implantation de la restriction des processus (⌈)**

Comme on peut remarquer, l'opérateur ⌈ est utilisé deux fois dans l'expression formelle vérifiant la cohérence du calcul C avec un port P: restreindre chaque processus des ports restant à l'ensemble de ses événements observés, restreindre le processus obtenu après la composition parallèle de ces processus à l'ensemble des événements du port traité (voir section 2.2).

Dans l'implantation de cet opérateur, nous avons rencontré des difficultés, puisque dans le premier type d'utilisation de l'opérateur est appliqué directement sur un processus, mais dans le deuxième type d’utilisation l’opérateur vient s’appliquer sur un processus issu de la composition parallèle de plusieurs processus.

Etant donné que l’opérateur ⌈ n’est pas implanté dans l’outil FDR, nous avons choisi d’utiliser l’opérateur de hiding (\) [**28**], qui d’après la définition de l’opérateur ⌈, peut générer un processus équivalant en passant par la réécriture suivante:

$$P \lceil A = P \setminus (\alpha P\text{-}A).$$

Mais cette solution basée sur l’opérateur hiding (\) de FDR est incompatible avec la solution retenue afin d’implémenter la déterminisation d’un processus.

*Exemple:*

*Soit le processus P = a →b →P ⊓ b →P. L’alphabet du processus P est {a, b}.*

*On veut calculer et déterminiser le processus Q tel que Q = P ⌈{b}. la réécriture de l’opérateur ⌈ en passant par l’opérateur (\) de FDR donne : Q = P\ (αP – {b}). Mais l’opérateur (\) ne permet pas de calculer d’une façon explicite l’expression analytique de Q.*

Dans la suite, nous allons proposer une technique d’implémentation de l’opérateur ⌈ basée sur l’élimination syntaxique des branches d’un processus CSP.

*Technique proposée*

Un processus simple en CSP (pas de composition | |, ni de paramètre) est composé de 3 opérateurs binaire →, □ et ⊓. Un processus CSP est modélisé par un arbre binaire où les nœuds représentent les opérateurs et les feuilles représentent les opérandes (évent, nameProcess, processus prédéfini) [**4**].

Pour l’opérateur de préfixage → :

Le fils gauche doit être un évent.

Le fils droit peut être:

- nameProcess.
- processus prédéfini
- opérateur de préfixage →

Pour les opérateurs □ et ⊓:

Le fils gauche peut être:

- processus prédéfini.
- opérateur de préfixage →
- opérateur □
- opérateur ⊓

De même le fils droit peut être:

- processus prédéfini.
- opérateur de préfixage →
- opérateur □
- opérateur ⊓

Reprenons maintenant la définition de notre opérateur ⌈: P⌈A revient à cacher dans la trace de P tous les événements n'appartenant pas à A.

*Règles d'élimination d'événement dans un processus:*

D'après les possibilités de construction d'un processus dégagées plus haut, un événement ne peut être que le fils gauche l'opérateur de préfixage (→). Cacher événement aura comme conséquence l'élimination de cet opérateur de préfixage (**Règle 1**), dans ce cas si le fils droit est un *nameProcess* et que c'est le nom du processus lui-même il faut l'éliminer sinon il est conservé (**Règle 2**). Pour les deux opérateurs □ et ⊓ si l'un de ses fils gauche ou droit est éliminé il faut le supprimer (**Règle 3**).

*Exemple:*

*Soit les trois processus P1, P2 et P3 :*

*P1= a → b → P1*

*P2 = b → P2*

*P3 = a → P3 □ b → P3*

*L'élimination de l'événement b des trois processus donnera*

*P1= a → P1(application Règle 1).*

*P2= §     (application Règle 1 et Règle 2).*

*P3= a → P3(application Règle1, Règle 2 et Règle 3).*

Maintenant après avoir terminé l'analyse des opérateurs det et ⌈ on arrive à l'étape de réalisation. L'opérateur det est implanté par la vérification lors de l'injection de l'opérateur ⊓ dans le processus d'un port, si on est en train d'appliquer l'opérateur 'det' sur ce port (voir figure 3.14)

```
void binaryOp::fdrprint(void) {
…
case TOK_CHOOSEKW:          //cas d'injection d'opérateur de choix non déterministe
        if ((higherScope_effectif && higherScope_effectif->gtype==PORT_T && ((component
*)((declaration     *)higherScope_effectif)->higherScope)->ProduceDetPort     ==     1)     ||
(higherScope_effectif && higherScope_effectif->gtype==ROLE_T &&
((connector *)((declaration *)higherScope_effectif)->higherScope)->ProduceDetRole == 1))
             doPrint(" [] "); // dans le cas d'un port, l'attribut ProduceDetPort de la
                          //classe component indiquera si on applique l'opérateur 'det'
                          //sur ce port.
        else
             doPrint(" |~| ");
        break;
…
}
```

**Figure 3.14.** *Déterminisation d'un processus*

L'implantation de l'opérateur ⌈, dans le cas où le processus traité sera modifié par l'opérateur det, nécessite que la représentation syntaxique du processus soit modifiée. Vu qu'un processus est représenté par un arbre binaire dans sa forme la plus simple (pas de composition ||, ni de paramètre) et que restreindre un processus à un ensemble d'événements revient à cacher les événements qui n'appartient pas à cet ensemble. Dans ce cas il faut commencer par calculer l'ensemble d'événements initialisé (voir figure 3.15) et le passer comme paramètre à une méthode qui va parcourir l'arbre représentant le processus concerné permettant de supprimer les événements initialisés selon les trois règles définies plus haut. Ce parcours est un parcours d'un arbre binaire, il peut être préfixé, infixé ou postfixé. Pour identifier le type de parcours à choisir il faut déterminer quand le traitement de chaque nœud sera réalisé par rapport à son sous-arbre gauche et à son sous-arbre droit. Selon les règles d'élimination, un opérateur n'est supprimé qu'après la vérification de l'opérande gauche et droite ce qui amène à choisir un parcours postfixé [**5**] et [**27**](voir figure 3.16).

```
Set * Set::InitiatedEvent(){
Set * s;
sList *c;
s = new Set();
// parcourir l'ensemble des événements
for (c=names; c!=NULL; c=c->next) {
        //vérifier si c'est un événement initialisé
        if (c->val != NULL && c->val->gtype == EVENT_T && ((event *)c->val)->event_type ==
        INITIATED_T) {
                        s->add(c->val);
        }
}
        return s;
}
```

**Figure 3.15.** *Calcul des événements initialisés*

```
void Hiden (astNode ** racine, astNode * noeud, Set * events, Set * names, int afd){
   astNode * restant;
// cas général
if(noeud && noeud->gkind == BINARY_IND){
        // traiter le sous arbre droit
        HidenTest(racine, ((binaryOp *)noeud)->first, events, names, afd);
        if(((binaryOp *)noeud)->first && noeud->gtype == TOK_THENKW)
                afd = 1;//pour ne pas supprimer les noms des processus en cas de préfixage
        // traiter le sous arbre gauche
        HidenTest(racine,((binaryOp *)noeud)->second, events, names, afd);
        // si l'un des sous arbre n'est pas supprimé
        if(!((binaryOp *)noeud)->first || !((binaryOp *)noeud)->second){
                //si les deux sous arbre sont supprimés
                if(!((binaryOp *)noeud)->first && !((binaryOp *)noeud)->second)
                                restant = NULL;
                        // si le sous arbre gauche est supprimé
                else    if(!((binaryOp *)noeud)->first)
                                restant = ((binaryOp *)noeud)->second;
                        // si le sous arbre droit est supprimé
                        else
                                restant = ((binaryOp *)noeud)->first;
                // pour lier le sous arbre restant
                if( noeud->pere ){
                        if(noeud == ((binaryOp *)noeud->pere)->first)
                                ((binaryOp *)noeud->pere)->first = restant;
                        else
                                ((binaryOp *)noeud->pere)->second = restant;
                }else
```

```
                    *racine = restant;
                //supprimer le noeud
                delete noeud;
                }
        // cas trivial
}else{
        // nœud terminal de type événement initialisé
        if(noeud->gtype == EVENT_T && events->IsElement((event *)noeud)){
                if(noeud->pere){
                        if(noeud == ((binaryOp *)noeud->pere)->first)
                                ((binaryOp *)noeud->pere)->first = NULL;
                        else
                                ((binaryOp *)noeud->pere)->second = NULL;

                }else
                        *racine = NULL;
                delete noeud;
        }
        // nom de processus qui vient après l'opérateur → dont l'opérande gauche est supprimé
        if(noeud->gtype == NAME_T && names->IsElement((name *)noeud) && afd == 0){
                if(noeud->pere){
                        if(noeud == ((binaryOp *)noeud->pere)->first)
                                ((binaryOp *)noeud->pere)->first = NULL;
                        else
                                ((binaryOp *)noeud->pere)->second = NULL;
                }else
                        racine = NULL;
                delete noeud;
        }
}}
```

**Figure 3.16.** *Elimination des événements initialisés*

- **Génération des relations de raffinement de la cohérence d'un composant**

Après avoir réussi à implémenter les opérateurs manquants (det, $\lceil$), nous avons réutilisé ces opérateurs afin de réaliser les relations de raffinement des processus CSP liées à la cohérence d'un composant.

Soit un composant Comp dont la liste des ports sont P, P1, P2, …, Pn et un calcul C. Vérifier que P respecte la première propriété revient à vérifier que :

$P \sqsubseteq (C \parallel \forall\ i : 1..n \parallel det(Pi \lceil \alpha oPi)) \lceil \alpha P)$. Afin de réaliser cette relation de raffinement, nous avons suivi la démarche comportant les étapes suivantes :

- Première étape : donner une présentation de chaque processus respectant la syntaxe du model-cheker FDR [**29**].
- Deuxième étape : renommer tous les processus des ports pour passer d'une représentation locale à une représentation globale. Ce renommage est réalisé de deux manières : soit par l'association du nom du port à chacun de ses événements au niveau de la méthode *fdrprint,* soit en utilisant l'opérateur de préfixage du model-cheker FDR. Vu que les processus sont déjà présents, on utilisera le préfixage ([[ ← ]]).
- Troisième étape : production de la version déterministe de chaque port après sa restriction à ses événements observés (*det(Pi ⌈ αoPi)*): la restriction du processus à ses événements observés est effectuée par l'élimination des événements initialisés de l'arbre d'analyse représentatif du processus puisque la déterminisation est implémentée par une modification syntaxique (voir section 2.5). Donc il faut commencer par le calcul des événements initialisée de chaque port (voir figure 3.15). Ensuite, il faut utiliser la méthode *copy()* pour créer une copie de l'arbre représentatif du processus du port. En outre, on applique la méthode *Hiden()* en utilisant le sous-ensemble d'événements initialisés pour obtenir un arbre représentatif du processus (*Pi ⌈ αoPi*). Finalement le processus est injecté (*Pi ⌈ αoPi*) dans la sortie FDR en appliquant la déterminisation.
- Quatrième étape : la composition parallèle des ports $P_1$, $P_2$, …, $P_n$ et du calcul : *C || ∀ i : 1..n || det(Pi ⌈ αoPi).* L'opérateur de composition parallèle est implémenté par le model-cheker FDR. Pour utiliser cet opérateur on doit préciser la liste des événements en commun [**28**]. Ceci correspond à l'alphabet du port privé des événements internes.
- Cinquième étape : la restriction du processus *C || ∀ i : 1..n || det(Pi ⌈ αoPi)* à l'alphabet du port P : *(C || ∀ i : 1..n || det(Pi* C *αoPi)) ⌈αP).* Vu que l'outil n'admet pas un modèle d'analyse de ce processus, cette restriction sera réalisée en utilisant l'opérateur hiding (\) de FDR : Pour un processus Q et un ensemble d'événement E, Q ⌈ E = Q \ (αQ – E). Il suffit donc de calculer la différence entre l'alphabet du composant et celle du port considéré. Cette opération est implantée dans le model-cheker FDR (diff) [**29**].
- Sixième étape : préparer le test qui vérifie la relation de raffinement : *P ⊑ (C || ∀ i : 1..n || det(Pi ⌈ αoPi)) ⌈αP).*

### *2.6. Validation*

Une série de test était réalisée pour vérifier que les additions et les modifications apportées à l’outil produiront les résultats attendus. Prenons par exemple le style ayant un composant qui prend en entrée une valeur, calcule une formule et produit le résultat. La description d’un tel système est donnée par la figure 3.17.

```
Style calculFormul
        Component calcul
                Port In = read -> In [] close -> TICK
                Port Out = _write -> Out |~| _close -> TICK
                Computation = In.read -> _Out.write -> Computation [] In.close -> _Out.close
                                -> TICK
        Constraints
                //no constraints
End Style
```

**Figure 3.17.** *Style calculFormul*

Après avoir enregistré la spécification du style dans un fichier (CalculFormule.wrt) on génère les propriétés relatives à cette spécification à l’aide de l’outil Wr2fdr modifié et puisque la description ne contient que la spécification d’un composant, le fichier de sortie n’explicitera que les propriétés relatives à la cohérence d’un composant : Port/Calcul (voir figure 3.18).

```
-- Style CalculFormule
-- Types declarations
  -- events for abstract specification
  channel write, close, read

  -- Component Calcul
    ALPHA_Calcul = {|Out.close, Out.write, In.read, In.close|}
    ComputationCalcul = ((In.read -> (Out.write -> ComputationCalcul)) [] (In.close
          -> (Out.close -> SKIP)))
--Port Process
    ALPHA_In = {close, read}
    ALPHA_InI = {}
    PORTIn = ((read -> PORTIn) [] (close -> SKIP))
    InG = PORTIn[[ x <-In.x | x <- ALPHA_In ]]

    ALPHA_Out = {close, write}
    -- no events observed!
    PORTOut = ((write -> PORTOut) |~| (close -> SKIP))
    OutG = PORTOut[[ x <-Out.x | x <- ALPHA_Out ]]
```

```
channel In: {close, read}
   channel Out: {close, write}

   --Deterministic Process restricted to the observed event
   PORTInDETR = ((read -> PORTInDETR) [] (close -> SKIP))
   PORTOutDETR = SKIP

   COMPIn = (( PORTOutDETR
      [| diff({}, {}) |]
      ComputationCalcul))\ diff(ALPHA_Calcul, {|In|})
   assert InG [FD= COMPIn

   COMPOut = (( PORTInDETR [[ x <- In.x | x <- {close, read } ]]
      [| diff({In.close, In.read}, {}) |]
      ComputationCalcul))\ diff(ALPHA_Calcul, {|Out|})
   assert OutG [FD= COMPOut

-- No constraints
-- End Style
```

**Figure 3.18.** *Code FDR relatif au style CalculFormule en Wright*

La vérification des propriétés obtenues -signalées par assert- pour le style *calculFormule* avec le model-cheker FDR confirme que la description du style respecte la propriété 1 : cohérence Port/Calcul (Voir figur 3.19).

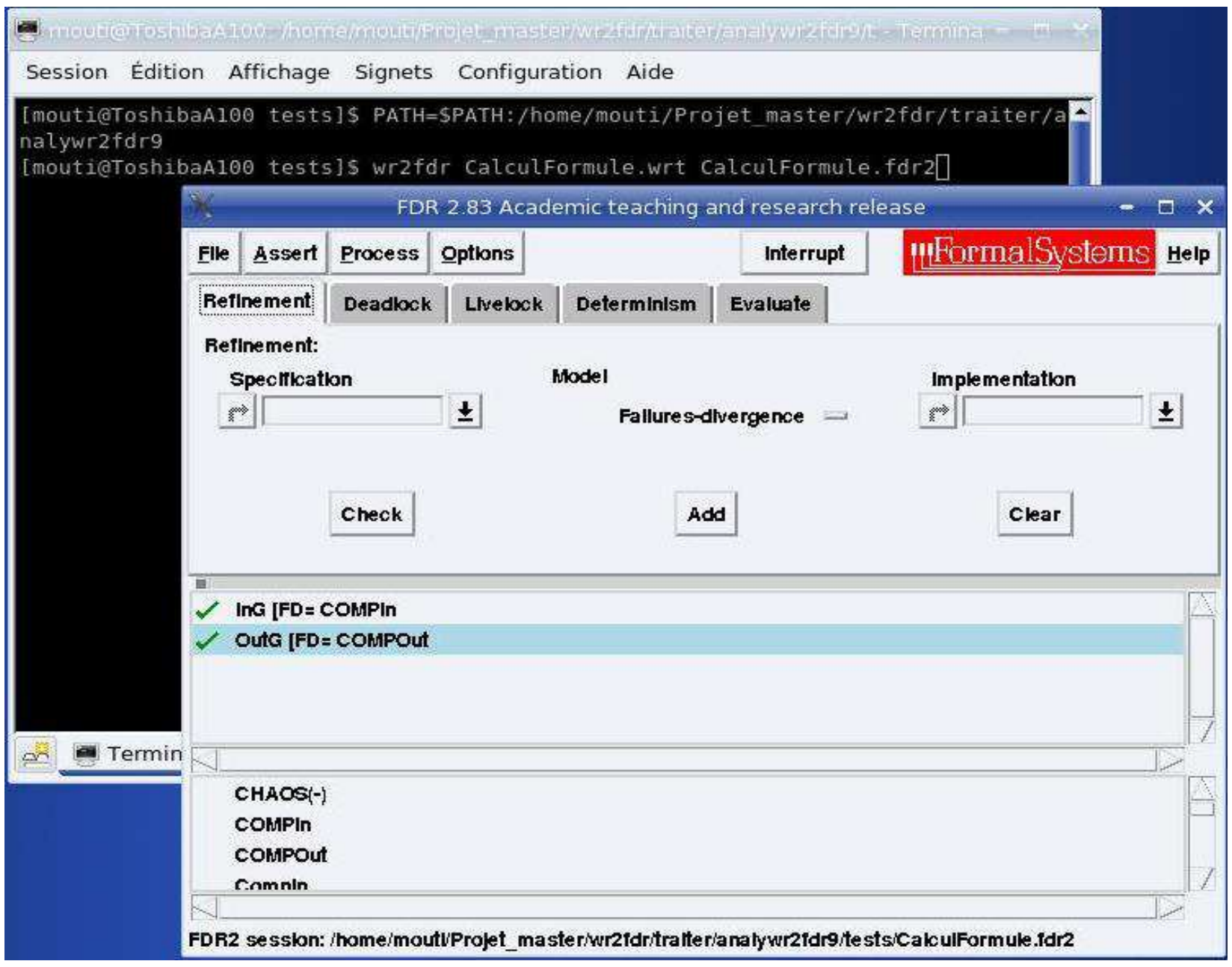

**Figure 3.19.** *Vérification du style CalculFormule avec le model-cheker FDR*

## 3. Compatibilité port/rôle

### *3.1. Description informelle*

Pour vérifier la cohérence d'un système, ce n'est pas suffisant de s'assurer uniquement de la cohérence de ses composants mais il faut vérifier le bon déroulement de l'interaction entre ces composants. D'une par les composants d'un système défini dans le langage de description d'architecture Wright communique à travers les connecteurs, d'autre par la liaison entre un composant et un connecteur est définie par l'attachement d'un port à un rôle. Donc vérifier l'interaction des composants d'un système revient à vérifier la compatibilité entre les ports et les rôles attachés. Comme il était précisé précédemment la vérification de compatibilité entre port et rôle ne doit pas se limiter au cas où le protocole du port soit identique à celui du rôle (voir chapitre1 section 6.1.1.3), elle doit accepter d'attacher le port qui continue son protocole dans une direction que le rôle peut avoir.

**Propriété 8 :** compatibilité Port/Rôle

*Tout port attaché à un rôle doit toujours continuer son protocole dans une direction que le rôle peut avoir.*

### *3.2. Description formelle*

La propriété 8 peut être traduite formellement par une relation de raffinement où le port augmenté par les événements spécifiques composé parallèlement à une version déterministe du rôle, raffine le rôle en question augmenté par les événements spécifiques du port.

D'où un port P est compatible avec un rôle R, si

$R_{+(\alpha P - \alpha R)} \sqsubseteq P_{+(\alpha R - \alpha P)} \mid\mid det(R)$

$P_{+A}$ : augmenter l'alphabet du processus P par les événements de A. Il est définie par le processus $P \mid\mid STOP_A$ [**24**].

*det*(R) : est la version déterministe du processus R.

### *3.3. Anomalies détectées*

La DT3 (Données de test 3) fournit par la figure 3.20 couvre la propriété 8 relative la propriété 8 relative à la compatibilité Port/Rôle. Pour y parvenir, on fiat appel au concept configuration de Wright.

```
Configuration ABC
  Component Atype
    Port Output = _a -> Output |~| TICK
    Computation = _Output.a -> Computation |~| TICK
  Component Btype
    Port Input = c -> Input [] TICK
    Computation = Input.c -> _b -> Computation
  connector Ctype
    Role Origin = _a -> Origin |~| TICK
    Role Target = c -> Target [] TICK
    Glue = Origin.a -> _Target.c -> Glue [] TICK
  Instances
    A : Atype
    B : Btype
    C : Ctype
  Attachments
    A.Output As C.Origin
    B.Input As C.Target
End Configuration
```

**Figure 3.20.** *DT3 couvrant la propriété 8*

En exécutant l'outil Wr2fdr sur la DT3, celui-ci produit une spécification à soumettre au Model-checker FDR contenant en partie la sortie observée illustrée par la figure 3.21.

```
-- FDR compression functions
transparent diamond
transparent normalise
-- Wright defined processes
channel abstractEvent
DFA = abstractEvent -> DFA |~| SKIP
quant_semi({},_) = SKIP
quant_semi(S,PARAM) = |~| i:S @ PARAM(i) ; quant_semi(diff(S,{i}),PARAM)
power_set({}) = {{}}
power_set(S) = { union(y,{x}) | x <- S, y <- power_set(diff(S,{x}))}
-- Configuration ABC
  -- Types declarations
```

```
-- events for abstract specification
  channel b, c, a

  -- Connector Ctype
    -- generated definitions (to split long sets)
    ALPHA_Ctype = {|Target.c, Origin.a|}
    GlueCtype = ((Origin.a -> (Target.c -> GlueCtype)) [] SKIP)
    ALPHA_Origin = {a}
    ROLEOrigin = ((a -> ROLEOrigin) |~| SKIP)
    OriginA = ROLEOrigin [[ x <- abstractEvent | x <- ALPHA_Origin ]]
    assert DFA [FD= OriginA
    ALPHA_Target = {c}
    ROLETarget = ((c -> ROLETarget) [] SKIP)
    TargetA = ROLETarget [[ x <- abstractEvent | x <- ALPHA_Target ]]
    assert DFA [FD= TargetA
    channel Origin: {a}
    channel Target: {c}
    Ctype = (( ROLEOrigin[[ x <- Origin.x | x <- {a } ]]
        [| diff({|Origin|}, {}) |]
        ( ROLETarget[[ x <- Target.x | x <- {c } ]]
        [| diff({|Target|}, {}) |]
        GlueCtype)))
    CtypeA = Ctype [[ x <- abstractEvent | x <- ALPHA_Ctype ]]
    assert DFA [FD= CtypeA
-- Currently, the rest of the configuration is not shown in FDR.
  -- End Configuration
```

**Figure 3.21.** *Sortie observée liée au DT3*

La sortie observée liée au DT3 est incorrecte. En effet, elle ne comporte pas des relations de raffinements – exprimées par assert – relatives à la propriété 8. Ainsi, la génération des relations de raffinement liées à la propriété 8 n'est pas implémentée par la version actuelle de l'outil Wr2fdr.

### *3.4. Identification de la partie concernée*

En se basant sur l'activité de rétro-conception menée sur le code de l'outil Wr2fdr (voir chapitre 2), la partie coupable est la classe *Configuration* descendante d'*AstNode* et précisément la méthode *fdrprint* redéfinie dans la classe *Configuration* (Voir figure 3.22 et 3.23).

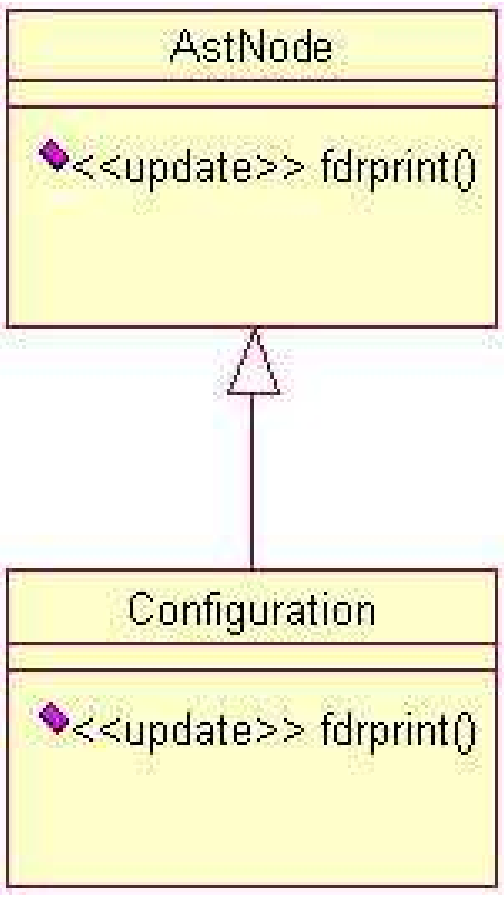

**Figure 3.22.** *Redéfinition de la méthode fdrprint par la classe configuration*

```
void configuration::fdrprint(void) {
    astNode      *current_instance;
    name         *instance_typename;
    connector    *connector_type;       // the type to be instantiated
    newLine();
    doPrint("-- Configuration ");
    if (conf_name)
        conf_name->fdrprint();
    addTab();
    newLine();
    current_instance = instances->startNode();
    while (current_instance) {
        if (current_instance->gtype != TOK_INSTANCESKW) {
            fprintf(efile, "Error:  internal structures mangled.\n");
            exit(1);
        }
        instance_typename = ((name *)((declaration *)current_instance)
->defn);
        connector_type = FindTypeIfConnector (instance_typename);
        if (connector_type) {
            connector_type->Instantiate (instance_typename->params);
            newLine();
        }
        current_instance = instances->nextNode();
    }
    subTab();
    doPrint("-- Currently, the rest of the configuration is not shown in FDR.");//
    newLine();
    doPrint("-- End Configuration");
    newLine();
} // configuration::fdrprint
```

**Figure 3.23.** *Implémentation de la méthode fdrprint par la classe configuration*

### *3.5. Correction proposée*

La réalisation de la propriété 8 exige l'utilisation des deux opérateurs :déterminisation et augmentation d'un processus. En ce qui concerne la déterminisation, nous avons réutilisé avec profit l'implémentation de l'opérateur det proposée dans 2.5.

En ce qui concerne l'augmentation, nous avons réutilisé l'opérateur diff offert par FDR. En outre, nous avons conçu et réalisé deux modules AttachmentPortName et AttachmentRoleName C++ permettant respectivement de récupérer le port et le rôle d'un attachement donné.

La génération des relations de raffinement liées à la propriété 8 exige les traitements suivants :

- Déterminisation du processus associé au Rôle.
- Augmentation du processus associé au Port par les événements spécifiques au processus associé au Rôle (Voir figure 3.24).
- Augmentation du processus associé au Rôle par les événements spécifiques du processus associé au Port (Voir figure 3.25)
- Composition parallèle du processus augmenté associé au Port et de la déterminisation du processus associé au Rôle (Voir figure 3.26).
- Génération de la relation de raffinement entre les deux processus déjà construits (Voir figure 3.27).

```
portInstance->first->fdrprint();        // nom de l'instance du composant
doPrint("_");
portInstance->second->fdrprint();       // nom du port
doPrint("PLUS = PORT");
portInstance->second->fdrprint();
continueLine();
doPrint("[| diff( ALPHA_");
roleInstance->second->fdrprint();
doPrint(" , ALPHA_");
portInstance->second->fdrprint();
doPrint(" ) |] STOP");
newLine();
```

**Figure 3.24.** *Augmentation du processus Port par les événements du processus Rôle*

```
roleInstance->first->fdrprint();           // nom de l'instance du connecteur
doPrint("_");
roleInstance->second->fdrprint();          // nom du rôle
doPrint("PLUS = ROLE");
roleInstance->second->fdrprint();
continueLine();
doPrint("[| diff( ALPHA_");
portInstance->second->fdrprint();
doPrint(" , ALPHA_");
roleInstance->second->fdrprint();
doPrint(" )|] STOP");
newLine();
```

**Figure 3.25.** *Augmentation du processus Rôle par les événements du processus Port*

```
….
portInstance->first->fdrprint();
doPrint("_");
portInstance->second->fdrprint();
doPrint("PLUSDET = ");
portInstance->first->fdrprint();
doPrint("_");
portInstance->second->fdrprint();
doPrint("PLUS");
continueLine();
doPrint("[| union(ALPHA_");
portInstance->second->fdrprint();
doPrint(" , ALPHA_");
roleInstance->second->fdrprint();
doPrint(" ) |]");
continueLine();
doPrint("ROLE");
roleInstance->second->fdrprint();
doPrint("DET");
newLine();
….
```

**Figure 3.26.** *Port augmenté et Rôle déterminisé composés en parallèle*

```
doPrint("assert ");
            roleInstance->first->fdrprint();
            doPrint("_");
            roleInstance->second->fdrprint();
            doPrint("PLUS  [FD= ");
            portInstance->first->fdrprint();
            doPrint("_");
            portInstance->second->fdrprint();
            doPrint("PLUSDET ");
            newLine();
            newLine();
            current_attachment = (binaryOp *) attachments->nextNode();
```

**Figure 3.27.** *Génération de la relation de raffinement*

### *3.6. Validation*

En suivant une approche orientée test syntaxique, nous avons testé avec succès notre implémentation de la propriété 8 : compatibilité Port/Rôle. Par exemple, l'exécution de notre version de l'outil Wr2fdr sur la DT3 (Voir section 3.3) fournit la sortie observée illustrée par la figure 3.28. La figure 3.29 récapitule les propriétés 1, 2, 3 et 8 vérifiés par le model-checker FDR concernant la configuration Wright DT3.

```
-- Configuration ABC
  -- Types declarations
  -- events for abstract specification
  channel b, c, a
  …..
    ROLEOriginDET = ((a -> ROLEOriginDET) [] SKIP)
    ROLETargetDET = ((c -> ROLETargetDET) [] SKIP)
……

  --Attachment Test
    A_OutputPLUS = PORTOutput
        [| diff( ALPHA_Origin , ALPHA_Output ) |] STOP
    C_OriginPLUS = ROLEOrigin
        [| diff( ALPHA_Output , ALPHA_Origin )|] STOP
    A_OutputPLUSDET = A_OutputPLUS
        [| union(ALPHA_Output , ALPHA_Origin ) |]
        ROLEOriginDET
    assert C_OriginPLUS  [FD= A_OutputPLUSDET
```

```
B_InputPLUS = PORTInput
        [| diff( ALPHA_Target , ALPHA_Input ) |] STOP
    C_TargetPLUS = ROLETarget
        [| diff( ALPHA_Input , ALPHA_Target )|] STOP
    B_InputPLUSDET = B_InputPLUS
        [| union(ALPHA_Input , ALPHA_Target ) |]
        ROLETargetDET
    assert C_TargetPLUS  [FD= B_InputPLUSDET
  -- End Configuration
```

**Figure 3.28.** *Relation de raffinement FDR relatives au DT3*

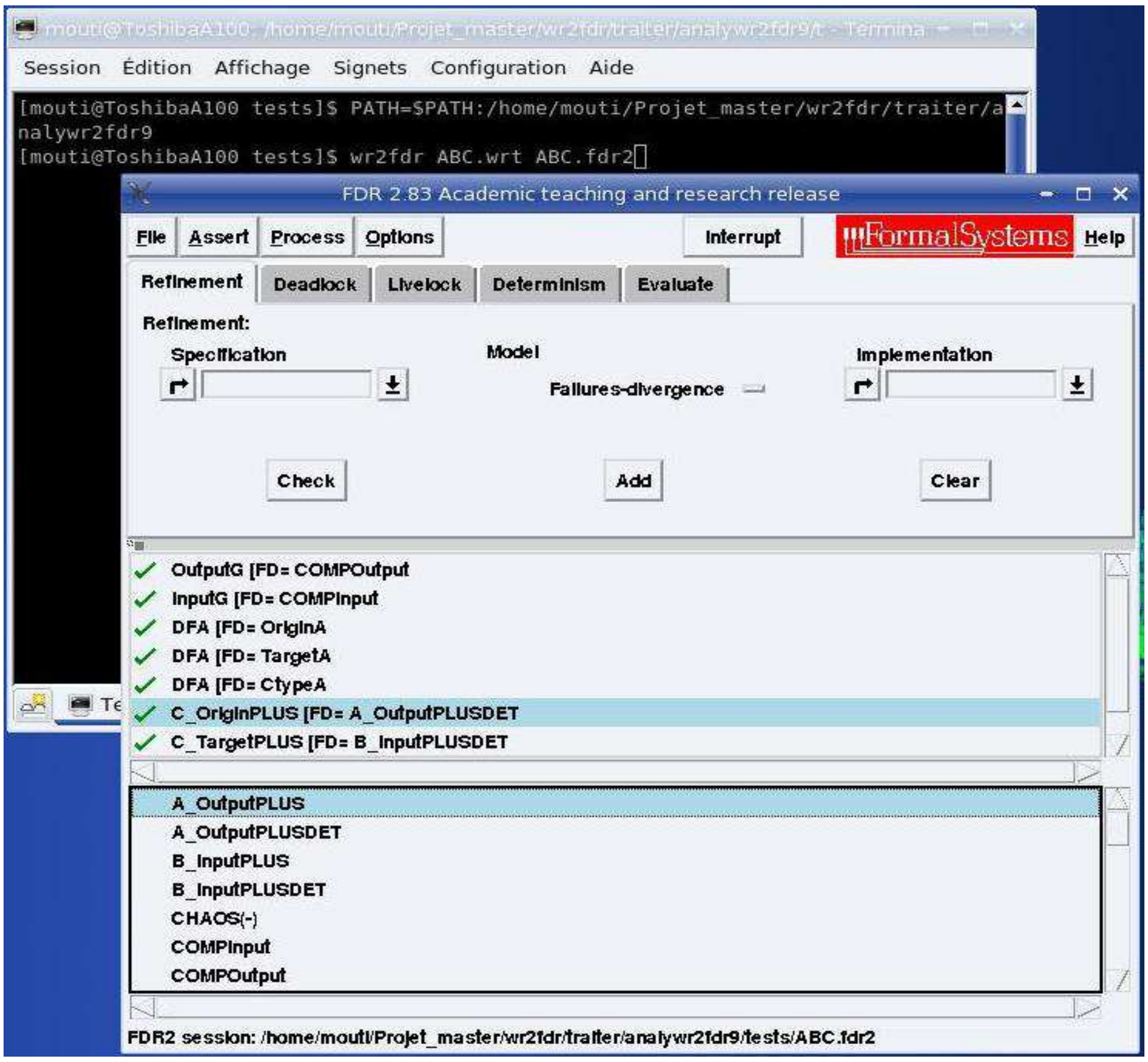

**Figure 3.29.** *Vérification des propriétés 1, 2, 3 et 8 par l'outil FDR de la configuration Wright DT3*

## 4. Un analyseur de la sémantique statique de Wright

### *4.1. Règles proposées*

Nous avons enrichi l'outil Wr2fdr par un analyseur sémantique statique de Wright. Ceci permet d'avoir des constructions cohérentes aussi bien sur le plan syntaxique que sémantique. Nous avons établi et implémenté les six règles suivantes :

- **Règle1:** Un identificateur doit désigner un seul élément architectural (component, connector, port, role, configuration et style)
- **Règle2:** Le type d'une instance (component, connector) doit être précédemment déclaré.
- **Règle3:** Toute instance doit être déclaré (clause instances) avant d'être utilisée dans les attachements (clause attachments).
- **Règle4:** Une interface d'un composant ou d'un connecteur doit être de la forme *instance.port ou instance.role*. Chaque port (respectivement role) doit figurer au sein du type composant (respectivement connecteur) utilisé pour définir l'instance.
- **Règle5:** Un attachement (clause attachement) doit être de la forme *instance.port as instance.role.*
- **Règle6:** Chaque port (respectivement rôle) d'un composant (respectivement d'un connecteur) doit être attaché à un et un seul rôle (respectivement port) d'un connecteur (respectivement d'un composant).

### *4.2. Partie concernée*

Nous avons implémenté les règles proposées précédemment relatives à la sémantique statique de Wright en augmentant l'analyseur lexico-syntaxique de Wr2fdr par des actions sémantiques appropriées. En effet, l'analyseur syntaxique Yacc [**10**], [**12**] et [**17**] prend en entrée un fichier « .y » qui contient les règles de production de la grammaire non contextuelle de Wright. Ces règles peuvent être augmentées par des actions sémantiques.

Une action sémantique est séquence d'instructions C écrite entre accolades à droite d'une production. Cette séquence est recopiée par Yacc [**15**] de telle manière qu'elle sera exécutée lorsque la production correspondante aura été employée pour faire une réduction.

### *4.3. Solution*

Nous avons enrichi l'analyseur lexico-syntaxique de Wright par une table de symboles permettant de regrouper des informations utiles à la vérification des règles proposées précédemment.

La technique utilisée pour implémenter cette table de symboles est appelée Adressage dispersé ouvert qui est structuré autour d'une table de hachage (voir figure 3.30) et d'une fonction d'adressage (de hachage).

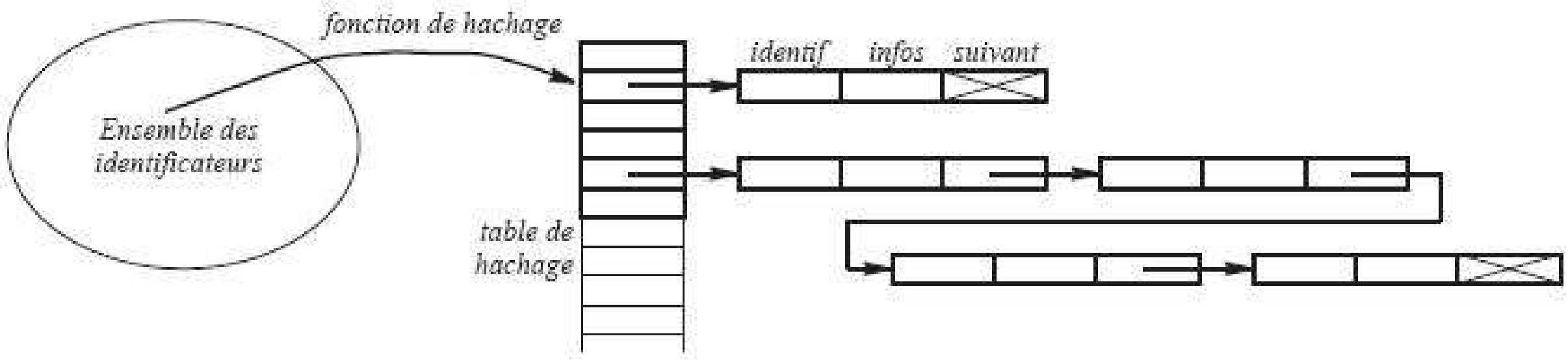

**Figure 3.30.** *Représentation d'une table de hachage*

Nous avons conçu et réaliser un module C++ [**5**], [**21**] et [**27**] (interface et implémentation) afin de matérialiser la structure de données table de symboles. Nous avons utilisé avec profit la construction *union* afin d'implémenter la table de symboles. En outre, nous avons utilisé une fonction de hachage appropriée favorisant la distribution plus ou moins équitable des symboles sur la table de hachage.

En se basant sur le module table de symboles [**14**], nous avons pu implémenter les règles syntaxo-sémantiques de Wright proposées dans 4.1. Par exemple la règle 5 (un attachement doit être de la forme *instance.port as instance.role*) est implémentée comme suit: on doit rechercher les deux insttances dans la table de symboles. En cas de succès, on doit vérifier que le type de la première instance est un type composant et le type de la deuxième instance est un type connecteur (Voir figure 3.31)

```
Attachment : Interface TOK_ASKW Interface
        {
                li1=$1;
                li2=$3;
                //rechercher le type de la première instance
                p=rechercher(ts,li1->sym);
                p=p->s.attributs.casInstance.type;
                // rechercher le type de la deuxième instance
                q=rechercher(ts,li2->sym);
                q=q->s.attributs.casInstance.type;
                //vérifier le type des deux instances
                if(p->s.nature!=Composant || q->s.nature!=Connecteur )
                {
                        yyerror("***Attachement: Composant.Port as Connecteur.Role***");
                        YYABORT;
                }
                li1=dernier($1);
                li1->suivant=li2;
                $$=$1;
        }
;
```

**Figure 3.31.** *Implémentation de la règle 5*

## 5. Validation

Nous avons testé avec succès notre analyseur sémantique en adoptant une approche basée sur le test fonctionnel orienté tests syntaxiques.
Soit DT4 (Donnée de Test 4) qui représente la spécification d'une configuration décrite en Wright (voir figure 3.32). Le concept configuration définit la notion d'attachement qui est l'objet de la règle 5. La génération de la sortie correspondante à DT4 aboutit avec succès (voir figure 3.34). Mais la DT5 (Donnée de Test 5) fournit par la figure 3.33 engendre une erreur sémantique (voir figure 3.35).

```
Configuration ABC
  Component Atype
    Port Output = _a -> Output |~| TICK
    Computation = _Output.a -> Computation |~| TICK
  Component Btype
    Port Input = c -> Input [] TICK
    Computation = Input.c -> _b -> Computation [] TICK
  connector Ctype
    Role Origin = _a -> Origin |~| TICK
    Role Target = c -> Target [] TICK
    Glue = Origin.a -> _Target.c -> Glue [] TICK

  Instances
    A : Atype
    B : Btype
    C : Ctype
  Attachments
    A.Output As C.Origin
    B.Input As C.Target
End Configuration
```

**Figure 3.32.** *DT4 respectant la règle 5*

```
Configuration ABC
  Component Atype
    Port Output = _a -> Output |~| TICK
    Computation = _Output.a -> Computation |~| TICK
  Component Btype
    Port Input = c -> Input [] TICK
    Computation = Input.c -> _b -> Computation [] TICK
  connector Ctype
    Role Origin = _a -> Origin |~| TICK
    Role Target = c -> Target [] TICK
    Glue = Origin.a -> _Target.c -> Glue [] TICK
  Instances
    A : Atype
    B : Btype
    C : Ctype
  Attachments
    C.Origin As A.Output
    B.Input As C.Target
End Configuration
```

**Figure 3.33.** *DT5 violant la règle 5*

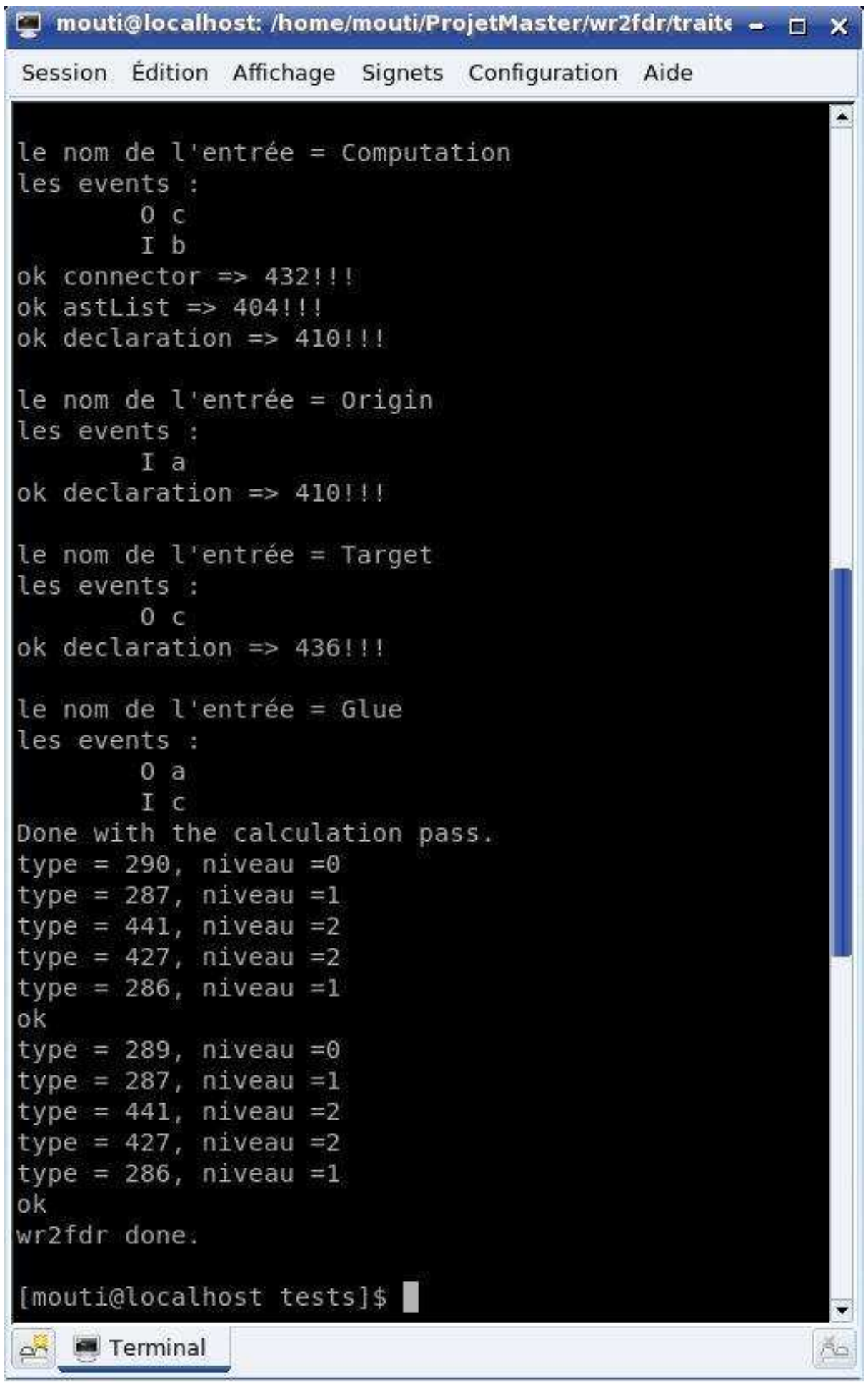

**Figure 3.34.** *Règle 5 respectée*

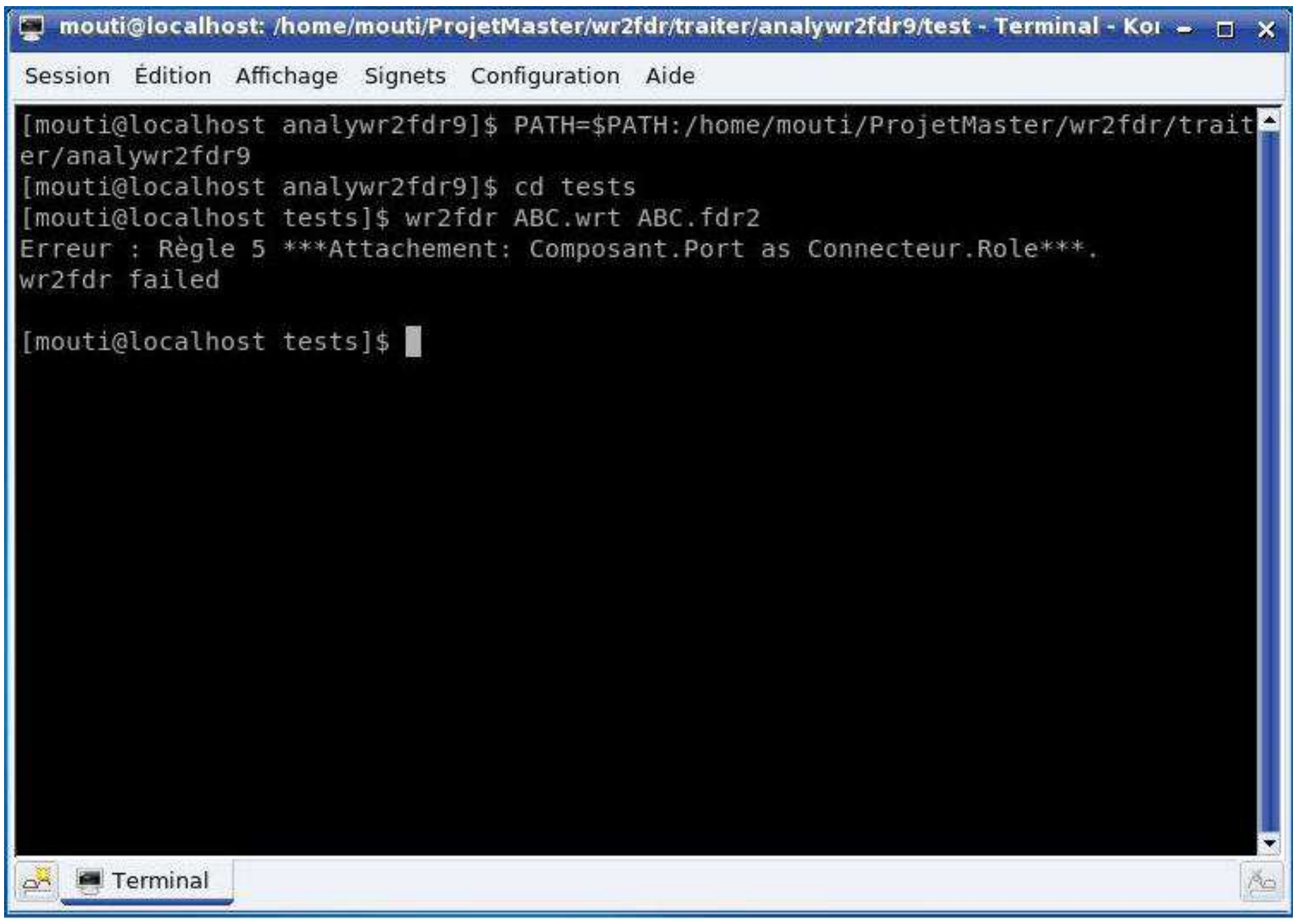

**Figure 3.35.** *Règle 5 violée*

## 6. Conclusion

Dans ce chapitre, nous avons réalisé une activité de maintenance de l’outil Wr2fdr qui accompagne Wright. Nous avons corrigé les erreurs liées aux deux propriétés 2 et 3. En outre, nous avons proposé une implémentation des deux propriétés 1 et 8. Enfin, nous avons enrichi l’outil Wr2fdr avec un analyseur sémantique de Wright.

# *Conclusion et Perspectives*

Dans ce mémoire, nous avons mené une activité de maintenance de l'outil Wr2fdr qui accompagne l'ADL formel Wright. Pour y parvenir, nous avons exercé une activité de test fonctionnel orienté tests syntaxiques afin de vérifier si l'outil Wr2fdr est conforme à sa spécification. Ceci nous a permis d'identifier les défaillances de l'outil Wr2fdr. Afin de localiser et corriger les erreurs détectées par l'activité de test fonctionnel, nous avons suivi une démarche de rétro-conception de l'outil Wr2fdr. Ainsi toutes les erreurs liées à l'automatisation des propriétés relatives à la cohérence d'un connecteur : propriétés 2 et 3 ont été bel et bien corrigées. En outre, nous avons implémenté les deux propriétés 1 et 8 relatives à la cohérence d'un Composant et à la compatibilité Port/Rôle. Egalement nous avons enrichi l'outil Wr2fdr par un analyseur sémantique de Wright en augmentant l'analyseur lexico-syntaxique de l'outil Wr2fdr par des actions sémantiques appropriées. Enfin nous avons testé les modifications apportées à l'outil Wr2fdr en utilisant le test fonctionnel. Pour mener à bien la maintenance corrective et évolutive de l'outil Wr2fdr, nous avons fourni des efforts importants afin de comprendre les 16000 lignes C++ formant Wr2fdr. A partir de la source de l'outil Wr2fdr, nous avons pu extraire son architecture à objet sous forme d'un diagramme de classes. Ceci nous a permis de localiser et corriger les erreurs de cet outil et de le faire évoluer. En conclusion, nous avons su travailler sur le code source de l'outil Wr2fdr et nous l'avons enrichi avec 800 Lignes C++.

Quant aux perspectives de ce travail, nous pourrions envisager les prolongements suivants :

- Documenter davantage l'outil Wr2fdr en utilisant la conception par contrat (Design by Contract). [**25**] : équiper les classes par des propriétés invariantes et les méthodes par une spécification pré/post ;
- Implémenter les facilités syntaxiques offertes par CSP telles que quantificateurs sur les processus CSP ;
- Améliorer la structure interne de l'outil Wr2fdr en utilisant les patterns de conception de GoF [11]. Ceci est connu sous le nom de refactoring.
- Réécrire Wr2fdr en Eiffel afin de tirer profit des possibilités orthogonales offertes par Eiffel telles que : généricité, héritage simple et multiple, conception par contrat et des bibliothèques de classes liées aux structures de données fondamentales et à la génération des analyseurs lexico-syntaxiques.

# *Annexes*

# Annexe A

Cette annexe présente l'implémentation corrigée de la méthode fdrprint de la classe *Connecteur*. La méthode fdrprint de cette classe génère les relations de raffinement relatives aux propriétés 2 et 3.

```
void connector::fdrprint(void) {
   declaration      *curRole;
   int               i;
   Set              *role_internal_events = NULL;
   event::BindEventData (NULL);

   // first output the basic role and glue processes which will be used
   //   to build up the connector definition
   doPrint("-- Connector ");
   if (conn_name) {
         conn_name->fdrprint();
   } else {
         doPrint("OOPS: null name.");
   }
   addTab();
   newLine();

   SplitLongSets ();
   ProduceRolesAndGlue (NULL);

   // declare channels for the role events.  The channel is named after
   //   the role and parameterized by all events from that role.
   //        role_name : { events w/in role }
   // so if "x" and "y" are events within role "R", we get:
   //        R : {x,y}  -OR- R.x and R.y
   // iterate through all roles declaring these channels
   curRole = (declaration *) roles->startNode();
   while (curRole) {
         doPrint("channel ");
         if (curRole->n->gtype != NAME_T) {
            fprintf(efile,"Internal Error:  mangled declaration name.\n");
            exit(1);
         }
         ((name *)curRole->n)->fdrprintWithNoParams();
         doPrint(": {");
         curRole->totEvents->fdrprint();
         doPrint("}");
         newLine();
         curRole = (declaration *) roles->nextNode();
```

```
} // end while

// declare the connector equal to "Glue" with renaming performed
//    over the roles
if (conn_name) {
       conn_name->fdrprint();
} else {
       doPrint("OOPS: null name.");
}
doPrint(" = (");
i=0;

// perform the renaming one role at a time; rename the events such
//   that each event "x" in role "R" will be renamed to "R.x"
curRole = (declaration *) roles->startNode();
while (curRole) {
  ++i;
  doPrint("( ROLE");
  curRole->n->fdrprint();
  doPrint("[[ x <- ");
  curRole->n->fdrprint();
  doPrint(".x | x <- {");
  if (curRole->totEvents) {
    curRole->totEvents->fdrprint();
  }

  // put the roles in parallel
  doPrint(" } ]]");
  continueLine();
  doPrint("[| diff({|");
  curRole->n->fdrprint();
  doPrint("|}, {");
  // calculate the set of role internal events
  role_internal_events = Set::SetMinus (curRole->ParamTotEvents,glue->totEvents);
  if (role_internal_events) {
       role_internal_events->fdrprint();
       delete role_internal_events;
  }
  doPrint("}) |]");
  continueLine();
  curRole = (declaration *) roles->nextNode();
} // end while

// make sure that the Glue process uses these renamings
glue->n->fdrprint();
((connector *)glue->higherScope)->conn_name->fdrprintWithNoParams();
```

```
    for (;i>0;i--) {
      doPrint(")");
    }
    doPrint(")");

    // finally, add a check for connector deadlock freedom
    if (conn_name) {
          // the alphabet of the connector is equal to the alphabet of the
          //   glue (which has already been calculated), so we'll just use
          //   that.

          // then produce an abstract version of the connector process
          newLine();
          conn_name->fdrprint();
          doPrint("A = ");
          conn_name->fdrprint();
          doPrint(" [[ x <- abstractEvent | x <- ALPHA_");
          conn_name->fdrprint();
          doPrint(" ]]");

          // finally throw in the check for Connector Deadlock Freedom (Test 2)
          newLine();
          doPrint("assert DFA [FD= ");
          conn_name->fdrprint();
          doPrint("A");
    } else {
          doPrint("OOPS: can't produce assertion for null name.");
    }
    subTab();
    newLine();
}  // end connector::fdrprint
```

# Annexe B

Cette annexe présente l'implémentation corrigée de la méthode *fdrprint* de la classe *Component*. La méthode *fdrprint* de cette classe génère les relations de raffinement relatives à la propriété 1.

```
void component::fdrprint(void) {
    astNode     *dec =NULL;
    Set * s;
    name * n;
    declaration     *current_port =NULL;
    declaration     *current_portO =NULL;
    astList    *copyPorts;
    char *string;
    int             i;

    Set             *compo_alphabet = NULL;         // alphabet of the component
    Set             *tmp = NULL;
    event           *current_event = NULL;
    Set             *prefixed_alphabet = NULL;
    event           *prefixed_event = NULL;         // events w/ port names prepended

    Set             *port_internal_events = NULL;
    // first output the basic ports and computation processes which will be used
    //   to build up the component definition
    doPrint("-- Component ");
    if (comp_name)
          comp_name->fdrprint();
    else
          doPrint("OOPS: null name");
     addTab();
    newLine();

    if (params) {
          doPrint("(");
          params->fdrprint();
          doPrint(")");
    }
    newLine();
    newLine();

    tmp = Set::Union (NULL, computation->totEvents);
    current_event = (event *) tmp->start();
    while (current_event) {
          if (current_event->gtype != EVENT_T) {
              fprintf(efile, "Internal Error:  not a valid event.\n");
              exit(1);
```

```
        }

        // must check for prefixed events
        if ( current_event->prefix_name ) {
            tmp->remove (current_event);

            // NOTE:  do the check here to make sure the event is in a
            //        proper role & add the event to the role alphabet

        }
        current_event = (event *) tmp->next();
}

// calculate the Component's alphabet as the union of the alphabets of all
//   ports and the alphabet of the computaion expression
current_port =(declaration *) ports->startNode();
while (current_port) {
        // if it's a port, we need to prefix its events with the port name
        prefixed_alphabet = new Set;

        current_event = (event *) current_port->totEvents->start();
        while (current_event) {
            if (current_event->gtype != EVENT_T) {
                fprintf(efile,"Internal Error:  bogus event.\n");
                exit(1);
            }
            if (current_port->n->gtype != NAME_T) {
                fprintf(efile,"Internal Error:  mangled declaration name.\n");
                exit(1);
            }
            prefixed_event = current_event->AddPrefixToEvent ( (name *)
                                            current_port->n);
            prefixed_alphabet->add (prefixed_event);
            delete prefixed_event;                  // this was copied
            current_event = (event *) current_port->totEvents->next();
        } // end while

        // now bind all event subscripts in the new set
        BindEventsInSet (prefixed_alphabet, NULL);

        // save the Port's alphabet in this new, channeled form
        current_port->ParamTotEvents = prefixed_alphabet->
                                                                copy();
        compo_alphabet = Set::Union(tmp, prefixed_alphabet);
        if (tmp)
            delete tmp;
```

```
        tmp = compo_alphabet;
        delete prefixed_alphabet;        // this was copied in the Union
        prefixed_alphabet = NULL;
        current_port =(declaration *) ports->nextNode();
    }

    if ( ELEMENTS_PER_SET > compo_alphabet->Cardinality()) {
        doPrint("ALPHA_");
        if (comp_name)
           comp_name->fdrprint();
        doPrint(" = {|");
        compo_alphabet->fdrprintAbstract();
        doPrint("|}");
        newLine();
    } else {
        // We have to break it up into smaller sets that FDR can handle.
        //Break up Component Alphabet (compo_alphabet);
    }
    newLine();

    // Check to see if the (Component alphabet \ the Computation alphabet) is not the
    //   null set.  If it's not, provide a warning!  The ports really shouldn't
    //   have internal events.
    tmp = NULL;
    tmp = Set::SetMinus (compo_alphabet, computation->totEvents);
    if ( NULL != tmp->start() ) {
        fprintf(efile,"WARNING:  Ports really shouldn't have internal events.");
        fprintf(efile,"\n           Consider this carefully.\n");
    }
    delete tmp;
    // declare "Computation" equal to the computation process expression
    if (computation) {
        computation->fdrprint();
        subTab();
    } else {
        doPrint("ERROR:  no computation expression in Computation");
    }
    newLine();
    newLine();

    // produce the Ports after the Compution since the computation alphabet calculation
    //   may need to add events to a Port's alphabet
    if (ports) {
        doPrint("--Port Process");
        newLine();
        ports->fdrprint();
```

```
        newLine();
//****pour generer les channels*************
        current_port = (declaration *) ports->startNode();
        while (current_port) {
           doPrint("channel ");
           if (current_port->n->gtype != NAME_T) {
                fprintf(efile,"Internal Error:  mangled declaration name.\n");
                exit(1);
           }
           ((name *)current_port->n)->fdrprintWithNoParams();
           doPrint(": {");
           current_port->totEvents->fdrprint();
           doPrint("}");
           newLine();
           current_port = (declaration *) ports->nextNode();
        } // end while

        doPrint("--Deterministic Process restricted to the observed event");
        newLine();
        ProduceDetPort = 1;
        ProduceRestPort = 1;
//**********le test de la suppression physique des noeud de lexpr. d'un process ******
//mouti pour la production des PORTDET(les ports deterministe)
        newLine();
        current_port = (declaration *) ports->startNode();
        while (current_port) {
           s = current_port->totEvents->InitiatedEvent();
           dec = current_port->copy();
           ((declaration *)dec)->Hiden(s);
           ((declaration *)dec)->fdrprint();
           newLine();
           delete s;
           s = new Set();
           for(current_event=(event *)current_port->totEvents->start(); current_event!=NULL;
           current_event= (event *)current_port->totEvents->next()){
                if(current_event->gtype == EVENT_T){
                   prefixed_event = current_event->AddPrefixToEvent((name *) current_port->n);
                   s->add(prefixed_event);
                }else{
                   fprintf(efile,"Internal Error:  totEvents contient que des event.\n");
                   exit(1);
                }
           }
           delete dec;
           dec = computation->copy();
           dec->higherScope = dec;
```

```
            dec->gtype = DECLARATION_T;
            string = new char[strlen((((name *)current_port->n)->n)) +1];
            strcpy(string,((name *)current_port->n)->n);
            ((declaration *)dec)->renomage(string);
            tmp = Set::SetMinus (compo_alphabet,s);
            ((declaration *)dec)->Hiden(tmp);
            AfficheProcess(((declaration *)dec)->defn, 0);
            printf("ok\n");
            //newLine();
            //computation->fdrprint();
            ((declaration *)dec)->fdrprint();
            newLine();
            delete tmp;
            delete dec;
            delete s;
            delete string;
            current_port = (declaration *) ports->nextNode();
        } // end while
        delete compo_alphabet;
        newLine();
        newLine();

//*********pour generer les test sur les ports**************
        current_port =(declaration *) ports->startNode();
        while (current_port) {
            doPrint("COMP");
            ((name *)current_port->n)->fdrprintWithNoParams();
            doPrint(" = (");
            i=0;
            current_portO =(declaration *) ports->startNode();
            while (current_portO) {
                if(!((name *)current_portO->n)->eq((name *)current_port->n)){
                    ++i;
                    doPrint("( PORT");
                    current_portO->n->fdrprint();
                    doPrint("DETR ");
                    if (current_portO->totEvents->haveObservedEvent()) {
                        doPrint("[[ x <- ");
                        current_portO->n->fdrprint();
                        doPrint(".x | x <- {");
                        current_portO->totEvents->fdrprintObservedEvent(NULL);
                        doPrint(" } ]]");
                    }
                    continueLine();
                    doPrint("[| diff({");
                    current_portO->totEvents->fdrprintObservedEvent(((name *)current_portO->n)->n);
```

```
                doPrint("}, {");
                // calculate the set of port internal events
                port_internal_events = Set::SetMinus (current_portO->ParamTotEvents,
                                                computation->totEvents);
                if (port_internal_events) {
                    port_internal_events->fdrprint();
                    delete port_internal_events;
                }
                doPrint("}) |]");
                continueLine();
            }
            current_portO = (declaration *) ports->nextNode();
        } // end while
        computation->n->fdrprint();
        comp_name-> fdrprintWithNoParams();//mouti
        for (;i>0;i--)
            doPrint(")");
        doPrint(")\\ diff(ALPHA_");
        comp_name-> fdrprintWithNoParams();
        doPrint(", {|");
        current_port->n->fdrprint();
        doPrint("|})");
        newLine();
        doPrint("assert ");
        ((name *)current_port->n)->fdrprintWithNoParams();
        doPrint("G [FD= COMP");
        ((name *)current_port->n)->fdrprintWithNoParams();
        newLine();
        doPrint("assert ");
        ((name *)current_port->n)->fdrprintWithNoParams();
        doPrint("G [FD= Comp");
        ((name *)current_port->n)->fdrprintWithNoParams();
        newLine();
        newLine();
        //pour construire le test du port suivant
        for (current_portO =(declaration *) ports->startNode();current_portO != current_port;
        current_portO =(declaration *) ports->nextNode());
            current_port =(declaration *) ports->nextNode();
      }//end while
      subTab();
  }
}
```

# Annexe C

Cette annexe présente l'implémentation corrigée de la méthode *fdrprint* de la classe *Configuration*. La méthode *fdrprint* de cette classe génère les relations de raffinement relatives à la propriété 8.

```
void configuration::fdrprint(void) {
   binaryOp      *current_attachment;
   namePair      *portInstance;
   namePair      *roleInstance;

   newLine();
   doPrint("-- Configuration ");
   if (conf_name)
         conf_name->fdrprint();
   addTab();
   newLine();

   doPrint("-- Types declarations");
   newLine();
   if (types) {
         types->fdrprint();
         newLine();
         newLine();
   }

//générer le relation de raffinement de la propriété 8
   if(attachments){
         doPrint("--Attachment Test");
         addTab();
         newLine();
         newLine();
         current_attachment = (binaryOp *) attachments->startNode();
         while (current_attachment) {
            portInstance = AttachmentPortName(current_attachment);
            roleInstance = AttachmentRoleName(current_attachment);

         //generer PortPlus
            portInstance->first->fdrprint();
            doPrint("_");
            portInstance->second->fdrprint();
            doPrint("PLUS = PORT");
            portInstance->second->fdrprint();
            continueLine();
            doPrint("[| diff( ALPHA_");
```

```
    roleInstance->second->fdrprint();
    doPrint(" , ALPHA_");
    portInstance->second->fdrprint();
    doPrint(" ) |] STOP");
    newLine();

//generer RolePlus
    roleInstance->first->fdrprint();
    doPrint("_");
    roleInstance->second->fdrprint();
    doPrint("PLUS = ROLE");
    roleInstance->second->fdrprint();
    continueLine();
    doPrint("[| diff( ALPHA_");
    portInstance->second->fdrprint();
    doPrint(" , ALPHA_");
    roleInstance->second->fdrprint();
    doPrint(" )|] STOP");
    newLine();
//générer PortPlusDet
    portInstance->first->fdrprint();
    doPrint("_");
    portInstance->second->fdrprint();
    doPrint("PLUSDET = ");
    portInstance->first->fdrprint();
    doPrint("_");
    portInstance->second->fdrprint();
    doPrint("PLUS");
    continueLine();
    doPrint("[| union(ALPHA_");
    portInstance->second->fdrprint();
    doPrint(" , ALPHA_");
    roleInstance->second->fdrprint();
    doPrint(" ) |]");
    continueLine();
    doPrint("ROLE");
    roleInstance->second->fdrprint();
    doPrint("DET");
    newLine();
// générer le test
    doPrint("assert ");
    roleInstance->first->fdrprint();
    doPrint("_");
    roleInstance->second->fdrprint();
    doPrint("PLUS  [FD= ");
    portInstance->first->fdrprint();
```

```
            doPrint("_");
            portInstance->second->fdrprint();
            doPrint("PLUSDET ");
            newLine();
            newLine();
            current_attachment = (binaryOp *) attachments->nextNode();
        } // end while
   }
   subTab();
   newLine();
   doPrint("-- End Configuration");
   newLine();
} // configuration::fdrprint
```

# Annexe D

Cette annexe présente le module *Table_Symbole* et les actions sémantiques utilisées par l'analyseur de sémantique statique de Wright pour vérifier les règles sémantiques.

Fichier entête  Table_Symbole :

```
#include<stdio.h>
#include<conio.h>
#include<malloc.h>
#include<string.h>
#include<assert.h>

#define taille 10

//different type d'identificateur
typedef enum typeIdentificateur
	{
		Interface=1, Composant=2, Port=3, Connecteur=4, Role=5, Instance=6
	};

// c'est la structure de de données d'un symbole
typedef struct symbole
	{
		char nom[10];
		enum typeIdentificateur nature;
		union
		{
			struct
			{
				struct noeudTableSymbole * composantRelatif;
			}casPort;
			struct
			{
				struct noeudTableSymbole * connecteurRelatif;
			}casRole;
			struct
			{
				struct noeudTableSymbole * type;
			}casInstance;
		}attributs;
	};

//c'est la structure de de données d'un noeud dans la table de symbole
typedef struct noeudTableSymbole
```

```
        {
                struct symbole s;
                struct noeudTableSymbole * suivant;
        };

//pour determiner l'indice dans la table de symbole
unsigned adresse(char *);
//pour initialser les cases du tableau à NULL
void creer(struct noeudTableSymbole * []);
//pour chercher un identificateur apartir de son nom, elle renvoie son adresse sil existe ,NULL sinon
struct noeudTableSymbole * rechercher(struct noeudTableSymbole * [],char *);
//pour verifier l'existance d'un identificateur
unsigned present(struct noeudTableSymbole * [],char *);
//pour ajouter un identificateur
void inserer(struct noeudTableSymbole * [],char *, enum typeIdentificateur,struct noeudTableSymbole *);

typedef struct listeIdentificateur
        {
                char sym[10];
                struct listeIdentificateur * suivant;
        };
struct listeIdentificateur * dernier(struct listeIdentificateur *);
unsigned couplExist (struct listeIdentificateur * , char *, char *);
void destructeurListe(struct listeIdentificateur *);
```

## Fichier des déclarations C++ Table_Symbole

```
#include"Table_Symbole.h"

unsigned adresse(char* id)
{
        unsigned i,j;
        char ch[20];
        strcpy(ch,id);
        for(j=0,i=0;ch[i]!='\0';i++)
                j=(j*256+(ch[i]))%taille;
        return j;
}

void creer(struct noeudTableSymbole * tabsym[])
{
        unsigned i;
        for(i=0; i<taille; i++)
                tabsym[i]=NULL;
}
```

```
struct noeudTableSymbole * rechercher(struct noeudTableSymbole * tabsym[],char *id)
{
	struct noeudTableSymbole * p;
	unsigned i;
	i=adresse(id);
	p=tabsym[i];
	while(p!=NULL && strcmp(p->s.nom,id)!=0)
	{
		p=p->suivant;
	}
	return p;
}

unsigned present(struct noeudTableSymbole * tabsym[],char * id)
{
	return NULL != rechercher(tabsym,id);
}

void inserer(struct noeudTableSymbole * tabsym[],char * id, enum typeIdentificateur t, struct
noeudTableSymbole * q)
{
	struct noeudTableSymbole * p;
	unsigned i;
	i=adresse(id);
	p=(struct noeudTableSymbole*)malloc(sizeof(struct noeudTableSymbole));
	strcpy(p->s.nom,id);
	p->s.nature=t;
	if(t==Port)
		p->s.attributs.casPort.composantRelatif=q;
	if(t==Role)
		p->s.attributs.casRole.connecteurRelatif=q;
	if(t==Instance)
		p->s.attributs.casInstance.type=q;
	p->suivant=tabsym[i];
	tabsym[i]=p;
}

struct listeIdentificateur * dernier(struct listeIdentificateur * tete)
{
	struct listeIdentificateur * l;
	l=tete;
	while(l->suivant)
		l=l->suivant;
	return l;
}
```

```
unsigned couplExist (struct listeIdentificateur * liste, char * ch1,char * ch2)
{
	struct listeIdentificateur * l;
	l=liste;
	while(l && l->suivant && (strcmp(l->sym,ch1)||strcmp(l->suivant->sym,ch2)))
		l=l->suivant->suivant;
	return l!=NULL;
}
void destructeurListe(struct listeIdentificateur *l)
{
	struct listeIdentificateur * p,* q;
	p=l;
	while(p)
	{
		q=p;
		p=p->suivant;
		free(q);
	}
}
```

Dans la suite, nous donnons les action sémantiques associées aux productions de l’analyseur lexico-synatxique de Wright écrit en Lex et Yacc.

```
NameList : SimpleName
{	li1=(struct listeIdentificateur*)malloc(sizeof(struct listeIdentificateur));
	strcpy(li1->sym,(char*)$1);
	li1->suivant=NULL;
	 $$=li1;
}
	| NameList ',' SimpleName
{	li1=(struct listeIdentificateur*)malloc(sizeof(struct listeIdentificateur));
	strcpy(li1->sym,(char*)$3);
	li1->suivant=$1;
	$$=li1;
}

Component : TOK_COMPONENTKW SimpleName
		PortList
		TOK_COMPUTATIONKW BehaviorDescription
{	if(!present(ts,(char*)$2))
	{
		inserer(ts,(char*)$2,Composant,NULL);
		p=rechercher(ts,(char*)$2);
		li1=$3;
		while(li1)
```

```
                {
                        q=rechercher(ts, (char*)li1->sym);
                        q->s.attributs.casPort.composantRelatif=p;
                        li1=li1->suivant;
                }
                $$=$2;
        }
        else
        {
                yyerror("***Identificateur Redondant***");
                YYABORT;
        }
}

TypeUse : SimpleName
{       if(present(ts,(char*)$1))
                $$=rechercher(ts,(char*)$1);
        else
        {
                yyerror("***Type non Declarer***");
                YYABORT;
        }
}

AttachmentList : AttachmentList Attachment
{       li1=$2;//couple (composant,port)
        li2=li1->suivant->suivant;//couple (connecteur,role)
        if(couplExist((struct listeIdentificateur*)$1,li1->sym,li1->suivant->sym))
        {
                yyerror("***Port deja relier***");
                YYABORT;
        }
        else    if(couplExist((struct listeIdentificateur*)$1,li2->sym,li2->suivant->sym))
                {
                        yyerror("***Role deja relier***");
                        YYABORT;
                }
                else
                {
                        li1=dernier((struct listeIdentificateur*)$2);
                        li1->suivant=$1;
                        $$=$2;
                }
}
```

```
Attachment : Interface TOK_ASKW Interface
{
        li1=$1;
        li2=$3;
        p=rechercher(ts,li1->sym);
        p=p->s.attributs.casInstance.type;
        q=rechercher(ts,li2->sym);
        q=q->s.attributs.casInstance.type;
        if(p->s.nature!=Composant || q->s.nature!=Connecteur )
        {
                yyerror("***Attachement: Composant.Port as Connecteur.Role***");
                YYABORT;
        }
        li1=dernier($1);
        li1->suivant=li2;
        $$=$1;
}

Interface : SimpleName '.' ActualPRName
{       if(present(ts,(char*)$1))
        {
                p=rechercher(ts,(char*)$1);
                if(p->s.nature==Instance)
                {
                        p=p->s.attributs.casInstance.type;
                        if(present(ts,(char*)$3))
                        {
                                q=rechercher(ts,(char*)$3);
                                if(q->s.nature==Port)
                                        q=q->s.attributs.casPort.composantRelatif;
                                else    if(q->s.nature==Role)
                                                q=q->s.attributs.casRole.connecteurRelatif;
                                        else
                                        {
                                                yyerror("***La deusieme partie doit etre soit un Port soit
                                                                                                un Role***");
                                                YYABORT;
                                        }
                        }
                        else
                        {
                                yyerror("***Identificateur non declarer***");
                                YYABORT;
                        }
                }
                else
```

```
        {
                yyerror("***La premiere partie doit etre une Instance***");
                YYABORT;
        }
    }
    else
    {
        yyerror("***Identificateur non declarer***");
        YYABORT;
    }
    if(p && q && strcmp(p->s.nom,q->s.nom)==0)
    {
        li1=(struct listeIdentificateur*)malloc(sizeof(struct listeIdentificateur));
        strcpy(li1->sym,(char*)$1);
        li1->suivant=(struct listeIdentificateur*)malloc(sizeof(struct listeIdentificateur));
        strcpy((li1->suivant)->sym,(char*)$3);
        li1->suivant->suivant=NULL;
        $$=li1;
    }
    else
    {
        yyerror("***L'Instance et l'Interface non pas le meme Type***");
        YYABORT;
    }
}
```

# *Bibliographie*

**[1]**Ali Abdallah, Cliff Jones, Jeff Sanders, *Communicating Sequential Processes The First 25 Years*, Symposium on the Occasion of 25Years of CSP, London UK, July 7-8, Springer 2004.

**[2]***Architecture Based Languages and Environments*.
http://www.cs.cmu.edu/afs/cs/project/able/www/able.html.

**[3]**Charles Antony Richard Hoare, *Communicating Sequential Processes,* Prentice Hall, Englewood Califfs NJ, 1985.

**[4]**Charles Antony Richard Hoare, *Process Algebra: A Unifying Approach*, 25 Years Communicating Sequential Processes 2004: 36-60.

**[5]**Claude Delannoy, *Programmer en langage C++*, 6ème édition Eyrolles, ISBN10: 2-212-11500-8, Mai 2004.

**[6]***Communicating Process Architectures 2004 (Concurrent Systems Engineering Series, Vol. 62)*, ISBN: 1-58603-458-8, Oxford, UK, 6-8 September 2004.

**[7]**David Garlan, *What is Style?*, Proceedings of the Dagstuhl Workshop on Software Architecture, Saarbruecken, Germany, February, 1995.

**[8]**David Garlan, Mary Shaw, *An Introduction to Software Architecture,* CMU-CS-94-166, School of Computer Science, Carnegie Mellon University, Pittsburgh, PA 15213-3890.

**[9]**David Garlan, Robert Monroe, David Wile, *Acme: Architectural Description of Component-Based Systems*, Foundations of Component-Based Systems, Gary T. Leavens and Murali Sitaraman (eds), Cambridge University Press, 2000, pp. 47-68.

**[10]**Didier Donsez, *Outils pour la Compilation Lex et Yacc,* ISTV /UVHC, 1995-2000.
www-adele.imag.fr/~donsez/cours/csyacclex.pdf.

**[11]**Erich Gamma, Richard Helm, Ralph Johnson, John Vlissides, *Design Patterns: Elements of Reusable Object-Oriented Software*, Addison Wesley - 1994 - ISBN 0-201-63361-2.

**[12]**Etienne Bernard. *Manuel d'utilisation de Lex et Yacc*, English version SGML, 30 Mars 1997.
www.linux-france.org/article/devl/lexyacc/.

**[13]**Gilles Clavel, Nicolas Fagart, David Grenet, Jorge Miguéis, *C++ La Synthèse Concepts objet standard ISO et modélisation UML*, 1ère édition DUNOD, ISBN10 : 2100081926, Mai 2003.

**[14]**Henri Garreta, *Techniques et outils pour la compilation*, Cours: Génie Logiciel & Théorie des Langages, Faculté des Sciences de Luminy, Université de la Méditerranée, Janvier 2001.

**[15]***Interface graphique pour la compilation sous Linux*.
www.archilinux.org.

**[16]**Jacques-Louis Lions, *Echec du vol Ariane 501*, Rapport de la Commission d'enquête Ariane 501, Communiqué de presse conjoint ESA-CNES, Paris, Juillet 1996.

**[17]**John Levine, Tony Mason, Doug Brown, *Lex & Yacc*, O'Reilly & Associates 2$^{nd}$ edition, ISBN: 1565920007, October 1992.

**[18]**Mohamed Graiet, Contribution à une démarche de vérification formelle d'architectures logicielles, thèse en co-tutelle de l'UJF et de l'université de Sfax, 25 Octobre 2007.

**[19]**Pierre Alain Muller, *Modélisation objet avec UML*, pam Avril 1997.
yasr2002.free.fr/soft/uml/uml-pam.pdf.

**[20]**Projet ACCORD, *Etat de l'art sur les Langages de Description d'Architecture (ADLs)*, Juin 2002.
www.infres.enst.fr/projets/accord/lot1/lot_1.1-2.pdf.

**[21]**Renaud Blanch, *HOWTO GNU Make & makefile*, 2005-2007.
http://iihm.imag.fr/blanch/howtos/GNUMake.html.

**[22]**Robert Allen, *A Formal Approach to Software Architecture*, CMU-CS-97-144, School of Computer Science, Carnegie Mellon University, Pittsburgh, PA 15213, May 1997.

**[23]**Robert Allen, David Garlan, *A Formal Basis For Architectural Connection*, ACM Transactions on Software Engineering and Methodology, July 1997.

**[24]**Robert Allen, David Garlan, *The wright architectural specification language*, Rapport technique CMU-CS-96-TBD, Carnegie Mellon University, School of Computer Science, Pillsburgh, PA, septembre 1996.

**[25]**Spyros Xanthakis, Pascal Regnier, Constantin Karapoulios, *Le test des logiciels*, Hermès science publication : Etudes et logiciels informatiques, 1999.

**[26]***The Wright Architecture Description Language*.
http://www.cs.cmu.edu/~able/wright/.

**[27]**Timothy Budd, Cay Horstmann, *La Bible C++*, 1$^{ère}$ édition Editions Micro Application, ISBN10: 2-7429-3717-X, 2004.

**[28]**User Manual and Tutorial, *Failures Divergence Refinement*, Formal Systems (Europe) Ltd, Oxford, England, 1.3 edition, August 1993.
www.fsel.com/documentation/fdr2/fdr2manual.pdf.

**[29]**William Roscoe, *Model-checking CSP*, A Classical Mind: Essays in Honour of CARHoare Prentice-Hall, 1997.